\documentclass[11pt,a4paper]{article}
\pdfoutput=1
\usepackage{jheppub}

\usepackage{amsmath,amsfonts,amsbsy,amssymb,array,accents,dsfont}
\usepackage{enumerate,array,latexsym,graphicx,mathrsfs,verbatim,psfrag}
\usepackage{enumerate}
\usepackage{graphicx} 
\usepackage{caption} 
\usepackage{subcaption} 

\usepackage{bm}
\usepackage{amssymb}
\usepackage{amsmath}
\usepackage{cancel}
\usepackage{epigraph}

\newcommand{\captionfonts}{\small}

\makeatletter  
\long\def\@makecaption#1#2{%
  \vskip\abovecaptionskip
  \sbox\@tempboxa{{\captionfonts #1: #2}}%
 \ifdim \wd\@tempboxa >\hsize
    {\captionfonts #1: #2\par}
  \else
    \hbox to\hsize{\hfil\box\@tempboxa\hfil}%
  \fi
  \vskip\belowcaptionskip}
\makeatother   

\usepackage{fancyhdr}
\usepackage{datetime}

\usepackage{mciteplus}

\usepackage[all]{hypcap}     

\makeatletter
\renewcommand\section{\@startsection {section}{1}{\z@}%
                                   {-3.5ex \@plus -1ex \@minus -.2ex}
                                   {2.3ex \@plus.2ex}%
                                   {\normalfont\Large\bfseries}}
\renewcommand\subsection{\@startsection{subsection}{2}{\z@}%
                                     {-3.25ex\@plus -1ex \@minus -.2ex}%
                                     {1.5ex \@plus .2ex}%
                                     {\normalfont\bfseries}}
\renewcommand\subsubsection{\@startsection{subsubsection}{3}{\z@}%
                                     {-2.5ex\@plus -1ex \@minus -.2ex}%
                                     {1.25ex \@plus .2ex}%
                                     {\normalfont\textit}}
\makeatother

\newcommand{\rhoo}{\ensuremath{\!\!\!\; \rho \;\! }}

\newcommand{\cAb}{\ensuremath{\bar{\mathcal{A}}}}

\DeclareMathSymbol{\medhatsym}{\mathord}{largesymbols}{"62} 
\newcommand\lowermedhatsym{
  \text{\smash{\raisebox{-1.28ex}{%
    $\medhatsym$}}}}
\newcommand\medhat[1]{
  \mathchoice
    {\accentset{\displaystyle\lowermedhatsym}{#1}}
    {\accentset{\textstyle\lowermedhatsym}{#1}}
    {\accentset{\scriptstyle\lowermedhatsym}{#1}}
    {\accentset{\scriptscriptstyle\lowermedhatsym}{#1}}
}

\DeclareMathSymbol{\medtildesym}{\mathord}{largesymbols}{"65}

\DeclareMathAlphabet{\mathpzc}{OT1}{pzc}{m}{it}

\def\sl{\text{sl}}
\def \su{\text{su}}

\def\dlam{{\delta\hskip -1pt\lambda}}
\def\lam{\lambda}

\newcommand{\msl}{m_{\sl}}
\newcommand{\msu}{m_{\su}}
\newcommand{\wsl}{w_{\sl}}
\newcommand{\wsu}{w_{\su}}

\newcommand{\bmsl}{\bar{m}_{\sl}}
\newcommand{\bmsu}{\bar{m}_{\su}}
\newcommand{\bwsl}{\bar{w}_{\sl}}
\newcommand{\bwsu}{\bar{w}_{\su}}
\newcommand{\jsl}{j_{\sl}}
\newcommand{\jsu}{j_{\su}}

\newcommand{\eq}[1]{\eqref{#1}}

\newenvironment{centermath}
 {\begin{center}$\displaystyle}
 {$\end{center}}

\def\mathbi#1{\textbf{\em #1}}
\def\tight#1{\! #1 \!}

\def\({\left(}
\def\){\right)}
\def\[{\left[}
\def\]{\right]}

\def\naive{na\"ive}
\def\sltwo{{SL(2,\IR)}}

\def\sutwo{{SU(2)}}
\def\uone{U(1)}

\def\mbar{\bar m}

\def\ie{{i.e.}}
\def\eg{{e.g.}}

\def\etc{{etc}}

\def\eff{{\rm eff}}

\def\lstr{\ell_{\textit{str}}}

\def\gstr{g_{\textit s}^{\;}}
\def\gstrsq	{g_{\textit s}^{2}}

\def\rads{R_{\textit AdS}}

\def\suplabel{m}
\def\supera{{\sst (\suplabel)}}

\def\nfive{{n_5}}
\def\nfivetil{{\tilde n_5}}
\def\nfivehat{{\hat n_5}}
\def\none{{n_1}}

\def\gosc{\ensuremath{\mathfrak{g}}}
\def\g{{\mathbi g}}

\def\R{{\sst \text{R}}}
\def\NS{{\sst \text{NS}}}
\def\LM{{\sst \text{LM}}}

\def\sst#1{\scriptscriptstyle{#1}}

\renewcommand{\P}{\mathrm{\sst{P}}}

\def\sfa{{\mathsf a}}
\def\xx{{\bf x}}
\def\xx{{\bf  x}}
\def\flabel{{\sst (m)}}

\def\bA{{\bf A}}
\def\bB{{\bf B}}
\def\A{{\mathsf A}}
\def\B{{\mathsf B}}
\def\C{{\mathsf C}}

\def\F{{\mathsf F}}
\def\H{{\mathsf H}}
\def\K{{\mathsf K}}
\def\sfH{{\mathsf H}}
\def\sfU{{\mathsf U}}
\def\sfV{{\mathsf V}}

\def\sfX{{\mathsf X}}
\def\sfY{{\mathsf Y}}
\def\sfZ{{\mathsf Z}}
\def\sfv{{\mathsf v}}

\def\mfw{{\cU}}

\def\x1x2{$x^1$-$x^2$}

\def\ytil{{\tilde y}}

\def\Ry{R_y}
\def\Rytil{R_\ytil}

\def\alphab{{\boldsymbol\alpha}}
\def\lamb{{\boldsymbol\lambda}}

\def\mub{{\boldsymbol\mu}}

\def\Phihat{\medhat\Phi}
\def\Psihat{\medhat\Psi}

\def\jj{{\mathsf J}}
\def\P{{\mathsf P}}

\def\sst{\scriptscriptstyle}
\def\half{\frac12}
\def\coeff#1#2{{\textstyle \frac{#1}{#2}}}
\def\hf{\coeff12}
\def\tr{{\rm Tr}}
\def\One{{\hbox{1\kern-1mm l}}}

\def\barray{\begin{array}}
\def\earray{\end{array}}
\def\be{\begin{equation}}
\def\ee{\end{equation}}
\def\bea{\begin{eqnarray}}
\def\eea{\end{eqnarray}}
\def\bal{\begin{align}}
\def\eal{\end{align}}
\def\nn{\nonumber}

\newcommand{\bC}{{\mathbb C}}

\newcommand{\bR}{{\mathbb R}}
\newcommand{\bS}{{\mathbb S}}
\newcommand{\bT}{{\mathbb T}}
\newcommand{\bZ}{{\mathbb Z}}

\def\IN{\mathbb{N}}

\def\IR{\mathbb{R}}

\def\IZ{\mathbb{Z}}

\definecolor{cardinal}{rgb}{0.6,0,0}
\definecolor{darkgreen}{rgb}{0,0.4,0}
\definecolor{green}{rgb}{0,0.4,0}
\definecolor{golden}{rgb}{0.92, 0.7, 0}
\definecolor{midnight}{rgb}{0, 0, 0.5}
\definecolor{darkblue}{rgb}{0, 0, 0.7}

\numberwithin{equation}{section}

\mathchardef\mhyphen="2D

\def\cA{\mathcal {A}}  \def\cC{\mathcal {C}}
\def\cD{\mathcal {D}} \def\cE{\mathcal {E}} \def\cF{\mathcal {F}}
\def\cG{\mathcal {G}} \def\cH{\mathcal {H}} \def\cI{\mathcal {I}}
\def\cJ{\mathcal {J}}  \def\cL{\mathcal {L}}
\def\cM{\mathcal {M}} \def\cN{\mathcal {N}} \def\cO{\mathcal {O}}
\def\cP{\mathcal {P}} \def\cQ{\mathcal {Q}} \def\cR{\mathcal {R}}
\def\cS{\mathcal {S}}  \def\cU{\mathcal {U}}
\def\cV{\mathcal {V}} \def\cW{\mathcal {W}} 
\def\cY{\mathcal {Y}}

\def\mR{\mathbb{R}}

\def\one{{\hbox{\kern+.5mm 1\kern-.8mm l}}}
\def\zero{{\hbox{0\kern-1.5mm 0}}}

\newcommand{\ket}[1]{{\,| {#1} \rangle}}

\newcommand{\T}[3]{\ensuremath{ #1{}^{#2}_{\phantom{#2} \! #3}}}		

\begin{document}

\title{Stringy Structure at the BPS Bound} 

\author{Emil J. Martinec$^a$, Stefano Massai$^{a,b}$ {\it and}\,
  David Turton$^c$}

\vspace{0.85 cm}

\affiliation[a]{
Enrico Fermi Institute and Dept. of Physics \\
5640 S. Ellis Ave.,
Chicago, IL 60637-1433, USA 
}

\affiliation[b]{
Institut f\"ur Theoretische Physik, ETH Zurich\\
CH-8093 Z\"urich, Switzerland
}

\affiliation[c]{
Mathematical Sciences and STAG Research Centre, University of Southampton, \\
Highfield, Southampton, SO17 1BJ, UK
}

 \emailAdd{ejmartin@uchicago.edu}
 \emailAdd{smassai@phys.ethz.ch}
 \emailAdd{d.j.turton@soton.ac.uk}

\vspace{0.7cm} 

\abstract{ 
We explore the stringy structure of 1/2-BPS bound states of NS fivebranes carrying momentum or fundamental string charge,
in the decoupling limits leading to little string theory and to $AdS_3/CFT_2$ duality.  
We develop an exact worldsheet description of these states using null-gauged sigma models, and illustrate the construction by deriving the closed-form solution sourced by an elliptical NS5-F1 supertube.
The Calabi-Yau/Landau-Ginsburg correspondence maps this geometrical worldsheet description to a non-compact LG model whose superpotential is determined by the fivebrane source configuration. 
Singular limits of the 1/2-BPS configuration space result when the fivebrane worldvolume self-intersects, as can be seen from both sides of the CY/LG duality~-- on the Landau-Ginsburg side from the degeneration of the superpotential(s), and on the geometrical side from an analysis of D-brane probes.  
These singular limits are a portal to black hole formation via the condensation of the branes that are becoming massless, and thus exhibit in the gravitational bulk description the central actors in the non-gravitational dual theory underlying black hole thermodynamics.
}


\setcounter{tocdepth}{2}   
\maketitle


\baselineskip=15pt
\parskip=3pt


\setcounter{footnote}{0}


\section{Introduction}
\label{sec:intro}

The Neveu-Schwarz fivebrane occupies a special place in the assortment of extended objects in string theory.  It is a soliton of the closed string sector, and thus much heavier and fatter than D-branes.  D-branes can end on fivebranes, and thus serve as both the W-objects of NS5-brane vector or tensor gauge theory, as well as precise probes of the stringy geometry in the vicinity of fivebranes.  

Fivebrane gauge theory provides intriguing examples of gauge/gravity duality.  The fivebrane decoupling limit $\gstr\to0$ leads to {\it little string theory}, a non-local and non-gravitational theory whose holographic dual has a null rather than timelike conformal boundary and thus may provide a tantalizing stepping stone toward understanding holography in asymptotically flat spacetimes (see~\cite{Kutasov:2001uf} for a review).  

Furthermore, when little string theory of $\nfive$ fivebranes is compactified on $\bT^4\times\bS^1$ or $K3\times\bS^1$, the superselection sector of large fundamental string winding charge $\none$ along $\bS^1$ admits an $AdS_3$ decoupling limit upon taking $R_{\bS^1}\to\infty$, holographically dual to a 2d CFT of central charge $6\none\nfive$.  This CFT has a rich set of 1/2-BPS excitations. These NS5-F1 bound states are in a U-duality orbit that includes D1-D5, F1-P, and NS5-P bound states, where P denotes momentum along $\bS^1$. The NS5-P duality frame will play an important role~-- the interplay between
the NS5-F1 and NS5-P descriptions will be a primary tool in our analysis.

The general family of supergravity solutions describing 1/2-BPS states in the NS5-P and other duality frames 
was constructed and studied in~\cite{Lunin:2001fv,Lunin:2001jy,Lunin:2002iz,Mathur:2005zp,Taylor:2005db,Kanitscheider:2007wq} (see also~\cite{Black:2010uq,Mathur:2018tib}).  We review their properties in section~\ref{sec:fivebranes}, focusing on solutions with pure NS-NS flux.  
The solutions are characterized by a set of profile functions $\F^I(v)$ which, in the NS5-P frame, specify the fivebrane source configuration; here $I$ are polarization labels, and $v$ parametrizes a left-moving wave on $\bS^1$.   The resulting {\it Lunin-Mathur supertube geometries} admit a pair of null Killing vectors, bilinear in the Killing spinors associated to the preserved supersymmetries.  Among the profile functions are those that describe the shape gyrations of the fivebranes in the ambient spacetime.  When the fivebranes are well separated, perturbative strings self-consistently avoid the strong-coupling region near the fivebranes, and the Lunin-Mathur geometries are nonsingular; conversely, when fivebranes come together, strings can freely propagate into the strong coupling region and perturbation theory breaks down.  The development of a strong coupling region when branes coalesce signals the liberation of W-branes and the onset of formation of a near-extremal black hole.

There are several related worldsheet descriptions of the spacetimes dual to little string theory.  The best studied example is that of fivebranes separated on their Coulomb branch in a circular array, which can be mapped onto a nonlinear sigma model on the coset orbifold 
$\bigl(\frac\sltwo\uone\tight\times\frac\sutwo\uone\bigr)/\bZ_\nfive$~\cite{Sfetsos:1998xd,Giveon:1999px}.  
A refinement of this presentation employs the gauging of a pair of null isometries
$\bigl(\frac{\sltwo\times\sutwo}{\uone_L\times\uone_R}\bigr)$~\cite{Israel:2004ir,Itzhaki:2005zr,Martinec:2017ztd}.  These descriptions are both related by mirror symmetry to a Landau-Ginsburg orbifold involving $\cN\tight=2$ Liouville field theory~\cite{Ooguri:1995wj,Hori:2001ax}, a phenomenon known as FZZ duality~\cite{FZZref,Kazakov:2000pm}.  This duality results from operator identifications in the $\sltwo$ WZW model~\cite{Maldacena:2000hw,Giveon:2016dxe}.

In recent work, the gauging of null isometries of nonlinear sigma models has been applied to study configurations of fivebranes, as well as NS5-F1 and NS5-P supertubes, that are based on circular source profiles~\cite{Martinec:2017ztd,Martinec:2018nco,Martinec:2019wzw}. The null gauging of the $\sltwo\tight\times\sutwo$ WZW model is the simplest example, one in which the current algebra symmetries enable the exact solution of the worldsheet theory. 
By tilting the null vector into a second null direction along the fivebrane worldvolume, one can construct supertube backgrounds from Coulomb branch configurations~\cite{Martinec:2017ztd}.

In this paper we generalize this construction to all Lunin-Mathur geometries with pure NS-NS flux, exploiting the fact that they possess asymptotically $AdS_3\tight\times\bS^3$ limits with a pair of null Killing vectors which can be gauged. 
We then apply the formalism to explore the worldsheet dynamics of the 1/2-BPS configuration space of little string theory.
We show in section~\ref{sec:nullgauging} that gauge projection of this family of geometries fills out the Coulomb branch moduli space, as well as its 1/2-BPS supertube configuration space.  
We illustrate the construction by finding the closed-form nonlinear deformation of the circular source profile into an ellipse. 
We study the gaugings that lead to an elliptical array of fivebranes as well as to an elliptical supertube solution. 
We also analyze the limit in which the ellipse degenerates.

We present in section~\ref{sec:CoulombfromLM} the 1/2-BPS vertex operators that describe the linearized deformations along the Coulomb branch moduli space away from the circular array, as well as the BPS deformations of circular supertubes, and match them to the deformations of the profile functions $\F^I$ away from this special point.  Each of these linearized deformations has an FZZ dual, which we exhibit in section~\ref{sec:FZZduality}.  We argue that FZZ duality is not specific to the circular array and circular supertube, but rather should be a property of the generic point in the Coulomb branch moduli space, and of the generic point in the supertube configuration space.  

The key to the structure of the duality is the source profile $\F^I(v)$, as we discuss in section~\ref{sec:locatingbranes}.  In the supergravity approximation embodied in the NS5-F1 Lunin-Mathur geometries, the contour $\F^I(v)$ specifies the location of KK monopole singularities in the effective geometry.  On the Landau-Ginsburg side, $\F^I(v)$ specifies the data of the worldsheet superpotential and twisted superpotential.  The 1/2-BPS perturbations of the background are deformations of the profile $\F^I(v)$ and can be thought of equally as perturbations of the geometry or as perturbations of the Landau-Ginsburg superpotential and twisted superpotential.

The full worldsheet theory knows about the discrete locations of fivebranes, information that is lost in the supergravity approximation which only sees a smeared source.  One can see this structure on both sides of FZZ duality.  In the geometrical description as a null-gauged sigma model on a Lunin-Mathur background, D-branes ending on the fivebranes localize the fivebranes at discrete (relative) positions determined by the quantization of worldvolume flux.  This phenomenon was first seen in our analysis of D-branes in the null-gauged WZW model in~\cite{Martinec:2019wzw}; we verify that this structure continues to hold in the deformation of the circular fivebrane array to an elliptical array, and propose that it holds in general Lunin-Mathur geometries.  On the Landau-Ginsburg side, the fivebrane locations are directly coded by the zeros of the holomorphic worldvolume superpotential; we exhibit a precise map between the parameters of the superpotential and those that determine the shape profile $\F^I(v)$ of the supertube source in the supergravity description.  In both descriptions, one can see a strong-coupling singularity develop when fivebranes approach one another -- through coalescing zeros of the superpotential in the Landau-Ginsburg description, and through the development of vanishing cycles in the Lunin-Mathur geometry when the source profile develops a self-intersection.  The ellipse example exhibits this vanishing cycle nicely; we analyze D-branes that wrap it using the DBI approximation, and show that their mass vanishes when the ellipse degenerates.

We conclude in section~\ref{sec:discussion} with a discussion of the consequences of this stringy structure for the process of near-extremal black hole formation that results when these 1/2-BPS geometries are perturbed.  The singularities that appear in the 1/2-BPS configuration space are akin to those that develop at any vanishing cycle (see \eg~\cite{Strominger:1995cz})~-- W-branes that wrap the vanishing cycle become massless and condense, signalling the transition to a new phase.  Here that new phase is the black hole phase, and the phase transition is associated to the formation of a near-extremal black hole horizon in the corresponding low-energy effective theory.



\section{Fivebrane backgrounds}
\label{sec:fivebranes}

There are a variety of backgrounds involving NS5-branes that are related through the general formalism of null gauging.
These include NS5-branes separated on their Coulomb branch moduli space, as well as NS5-P and NS5-F1 supertubes.  In this section we give a brief overview of these backgrounds.

\subsection{The Coulomb branch of fivebranes}
\label{sec:coulomb}

Static NS5 branes on their Coulomb branch source the (string-frame) supergravity solution (we work in units in which $\alpha'=1$, and take from the beginning the fivebrane decoupling limit $\gstr\to 0$)
\begin{align}
\label{eq:CB1}
ds^2 = ds^2_{\bR^{5,1}} + \H \, d\xx\!\cdot\! d\xx 
~~,~~~~ 
H^{(3)} = dB = *_\perp d\H 
~~,~~~~ 
e^{2\Phi} = \H \,,
\end{align}
where the harmonic function $\H $ is given by
\be
\label{eq:CB2}
\H  = \sum_{m=1}^\nfive \frac{1}{|\xx-\F_\flabel|^2} \,,
\ee
and where $*_\perp$ indicates Hodge duality in the transverse $\bR^4$ parametrized by $\xx$. 

In the fivebrane decoupling limit, the separation between the fivebrane sources is often substringy.  Supergravity is then not sensitive to the locations of individual fivebranes, but rather sees a smeared source distribution.  For instance, in the situations we consider here, the fivebranes are uniformly distributed along a one-dimensional contour, and the harmonic function $\H$ is smeared over this contour.

\subsection{Lunin-Mathur geometries}
\label{sec:LMgeom}

Our main interest will be two-charge configurations in the NS5-F1 and NS5-P frames. We first consider the onebrane-fivebrane system, where the fivebranes are wrapped on the four-manifold $\cM=\bT^4$ or $K3$. The 1/2-BPS supergravity solutions of this system can be put in the standard form~\cite{Lunin:2001fv,Kanitscheider:2007wq} in the NS5-F1 duality frame (for now we restrict to solutions with pure NS-NS fluxes)
\begin{align}
\begin{aligned}
\label{LMgeom}
ds^2 &\;=\; \K  ^{-1}\bigl[ -(d\tau+\A)^2 + (d\sigma + \B)^2 \bigr] + \H \, d\xx\!\cdot\! d\xx + ds^2_\cM \,,
\\[.2cm]
B &\;=\; 
\K  ^{-1}\bigl(d\tau + \A\bigr)\wedge\bigl(d\sigma+\B\bigr)  + \C_{ij}\, dx^i\wedge dx^j \,,
\\[.2cm]
e^{2\Phi} &\;=\; 
{\gstrsq}\,\frac{\H}{\K}  \,, \qquad~~~ d\C =  *_{\sst\perp} d\H \;, 
\qquad~~ d\B = *_{\sst\perp} d\A \,,
\end{aligned}
\end{align}
where again $\xx$ are Cartesian coordinates on the transverse space to the fivebranes.

The harmonic forms and functions appearing in this solution can be written in terms of a Green's function representation,
which in the $AdS_3$ decoupling limit takes the form
\begin{align}
\begin{aligned}
\label{greensfn}
\H  \,=\, \frac{1}{2\pi}\sum_{m=1}^\nfive\int\limits_0^{2\pi} \frac{d\tilde{v}}{|\xx-\F_\flabel(\tilde{v})|^2} ~, & \qquad~
\K   \,=\,  
\frac{1}{2\pi}\sum_{m=1}^\nfive\int\limits_0^{2\pi} \frac{d\tilde{v} \, \dot\F_\flabel \tight\cdot \dot\F_\flabel}{|\xx-\F_\flabel(\tilde{v})|^2} \;,\\[.2cm]
\A \,=\, \A_i dx^i \,, \qquad
\A_i \,=\, & \frac{1}{2\pi}\sum_{m=1}^\nfive\int\limits_0^{2\pi} \frac{d\tilde{v} \,\dot \F^\flabel_i(\tilde{v})}{|\xx-\F_\flabel(\tilde{v})|^2}\;,
\end{aligned}
\end{align}
involving source profile functions $\F_\flabel(\tilde{v})$, $m=1,2,\dots,\nfive$, that describe the locations of the fivebranes in their transverse space (overdots denote derivatives with respect to $\tilde{v}$).
These geometries extend into a linear dilaton fivebrane throat region upon adding a constant to the F1 harmonic function $\K  $ (and to asymptotically flat spacetime by in addition adding a constant to the NS5 harmonic function $\H $).

\begin{figure}[h!]
\centering
\includegraphics[width=.58\textwidth]{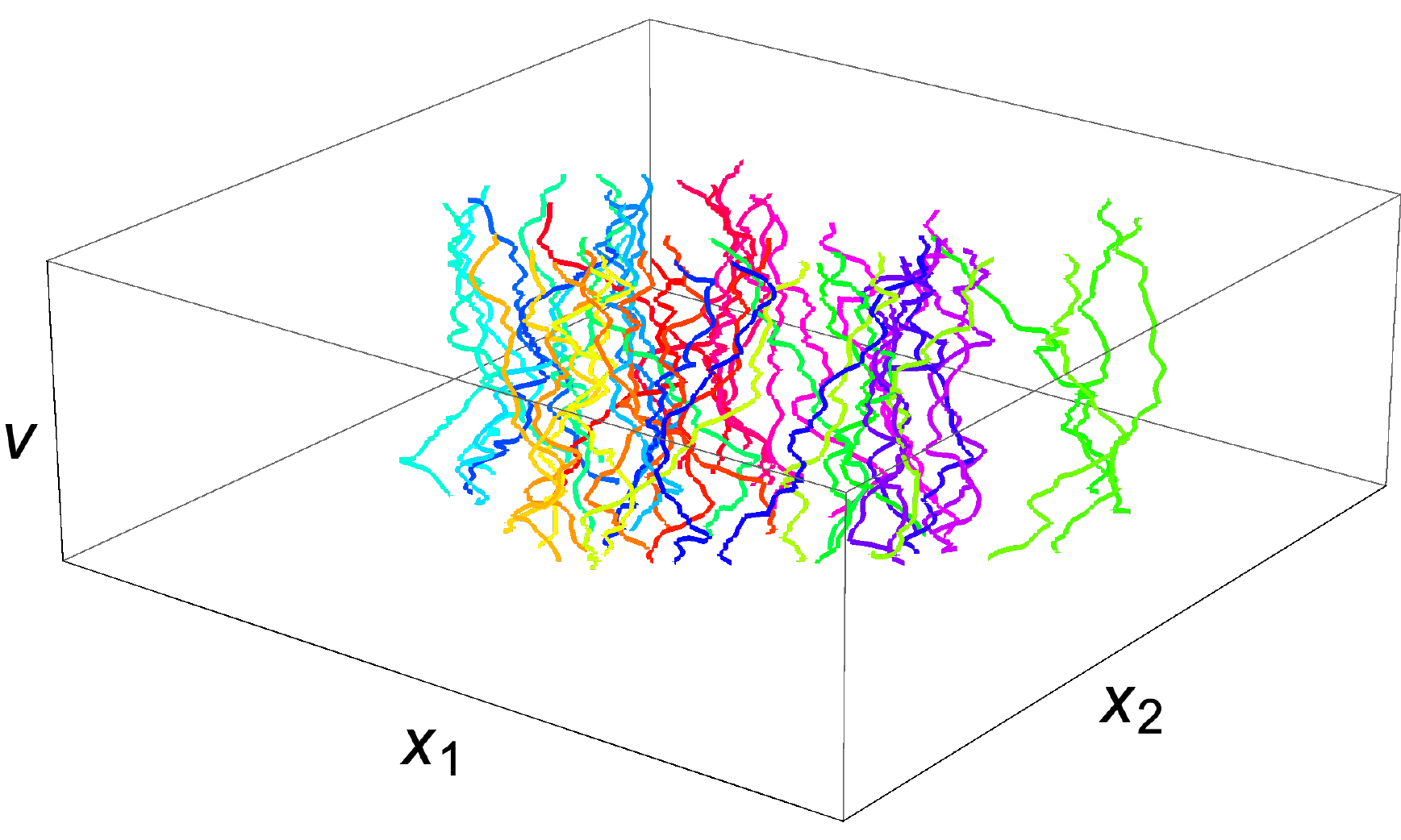}
\caption{\it Typical source profile $\,\F(\tilde{v})$, with successive windings color coded with evolving hue around the color wheel to indicate connectivity.}
\label{fig:Thermalprimary}
\end{figure}

It is convenient to introduce the null coordinates
\be
\hat{v} \,=\, \tau+\sigma \,, \qquad \hat{u} \,=\, \tau-\sigma 
\ee
(note that we distinguish the spacetime coordinate $\hat v$ from the variable $\tilde v$ which parametrizes the source profile).
The above NS5-F1 solutions are T-dual to corresponding NS5-P solutions
\begin{align}
\label{NS5pgeom}
\begin{aligned}
ds^2 &= -d\hat{u}\,d\hat{v} + \K  \, d\hat{v}^2 + 2\, \A_i\, dx^i \,d\hat{v} + \H \, d\xx\!\cdot\! d\xx + ds^2_\cM
\\[.3cm]
H_{vij} &= \epsilon_{ijkl}\,\partial^k\A^l
~~,~~~~
H_{ijk} = \epsilon_{ijkl}\partial^l \H 
~~,~~~~
e^{2\Phi} =  \H 
\end{aligned}
\end{align}
which, because the harmonic functions~\eqref{greensfn} have been smeared over $\tilde{v}$, are straightforwardly obtained by using the standard Buscher transformation~\cite{Buscher:1987qj}.  In the NS5-P duality frame, one has a direct geometric interpretation of the profile function $\F(\tilde{v})$~-- it is the expectation value of the four scalars in the fivebrane worldvolume theory describing its gyrations in the transverse $\bR^4$ parametrized by $\xx$.
%
  We choose twisted boundary conditions for the source profile functions,
\be
\label{twistedbc}
\F_\flabel(\tilde{v}+2\pi)=\F_{\sst (m+1)}(\tilde{v}) 
\ee
that bind all the fivebranes together.  We can then bundle all the fivebrane profile functions together into a single profile $\F^I(\tilde{v})$ that extends over the range $[0,2\pi\nfive)$.  An example profile is depicted in figure~\ref{fig:Thermalprimary}.

\vspace{0.5mm}
\vspace{1mm}

In the above expressions, we have ignored four additional polarizations of fivebrane excitation that correspond to internal modes on the fivebranes; in type IIB, these are the four polarizations of the fivebrane gauge field $\cA_\mu(\tilde{v})$, while in type IIA they are the excitations of the self-dual antisymmetric tensor $\cA_{\mu\nu}(\tilde{v})$ and associated scalar $\cA(\tilde{v})$ (which characterizes the location of the fivebrane along the M-theory circle).  Note that these internal excitations are RR fields -- for instance, D-branes that end on NS5-branes source the fivebrane gauge field. 

The space of 1/2-BPS states of momentum excitations on fivebranes is thus labelled by the mode excitation numbers $\{ N^{(i)}_{k} \}$ for the profile functions $\F_{i}(\tilde{v})$ describing fivebrane gyrations in the transverse space; and additionally by the mode excitation numbers $\{ N^{(\mu)}_{k} \}$ for the gauge field profile $\cA_\mu(\tilde{v})$ in type IIB, or $\{ N^{[\mu\nu]}_{k},N_{k} \}$ for the tensor gauge field $\cA_{\mu\nu}$ and scalar $\cA$ in type IIA.  A standard notation employs bispinor parametrizations $N^{(i)}\to N^{\alpha\dot\beta}$ of the transverse space; and on the internal space $\cM$, one has $(N^{[\mu\nu]}, N) \to N^{AB}$ for type IIA fivebranes, or $N^{(\mu)}\to N^{A\dot B}$ for type IIB fivebranes.%
\footnote{Note that these states are written in the NS5-P frame; for the NS5-F1 frame, one should interchange the IIA/IIB designations due to the T-duality involved.}
States are then written in the basis (say for type IIA)
\be
\label{halfBPS}
\prod_{k,\alpha,\dot\beta} \Bigl(\bigl|{\alpha\dot\beta}\bigr\rangle^{~}_k\Bigr)^{N_k^{\alpha\dot\beta}}
\prod_{k',A, B} \Bigl(\bigl|{A B}\bigr\rangle^{~}_{k'}\Bigr)^{N_k^{\!A B}} 
\ee
(where $\alpha,\dot\beta,A,B=\pm$),
with classical supergravity backgrounds corresponding to coherent states built from this mode number basis.
In the worldsheet theory, $\alpha$ and $A$ are associated to the left-moving sector, while $\dot\beta$ and $B$ come from the right-moving sector, as we will see below when we analyze the perturbative string spectrum.  Classically, there is a continuous moduli space of BPS states characterized by the arbitrary choice of source profile $\F(\tilde{v})$, whose Fourier modes are the coherent state parameters; quantum mechanically, the underlying basis states~\eqref{halfBPS} comprise a discretuum of ${ \exp}[2\pi\sqrt{2\none\nfive}\,]$ states~\cite{Sen:1994eb} (see also~\cite{Rychkov:2005ji}).  
The explicit description of this exponential number of states in the bulk has been interpreted as resolving the singularity of a string-sized black hole at the correspondence point and demonstrating the absence of a horizon in the microscopic bulk description~\cite{Lunin:2001jy,Mathur:2005zp}, though this interpretation remains a matter of debate~\cite{Sen:2009bm,Mathur:2018tib}.

Introducing the one-forms
\be
\label{b-o-A-B}
\beta \,=\, (\A+\B) \,,\qquad \omega \,=\, \A-\B \,,
\ee
the line element and B-field in Eq.\;\eq{LMgeom} become
\begin{align}
\begin{aligned}
\label{eq:uv-form}
ds^2_{\LM} &\;=\; -\frac{1}{\K  }(d\hat v+\beta)(d\hat u + \omega )+ \H \, d\xx\!\cdot\! d\xx + ds^2_\cM \,,\\
B &\;=\; \frac{1}{2\K  }(d\hat u + \omega )\wedge(d\hat v+\beta)  +\C_{ij}\, dx^i\wedge dx^j \,.
\end{aligned}
\end{align}

Let us introduce spherical bipolar coordinates for the flat four-dimensional base,
whose relation to the standard Cartesian coordinates $\xx$ for $\bR^4$ is given by
\begin{align}
\begin{aligned}
\label{sphericalbipolar}
&\qquad\qquad~~ x^1+ix^2 \,=\, a \cosh\rho \, \sin\theta \, e^{i\phi_{\R}}
\;,~~~~
x^3+ix^4 \,=\, a \sinh\rho \,\cos\theta \, e^{i\psi_{\R}} \;,  \\
& d\xx\!\cdot\! d\xx \;=\; a^2 \Big[ (\sinh^2\rhoo + \cos^2\theta) \left( d\rho^2 + d\theta^2\right) + \cosh^2\rhoo \sin^2\theta \,  d\phi_{\R}^2 + \sinh^2\rhoo \cos^2\theta  \, d\psi_{\R}^2 \Big] \,.
\end{aligned}
\end{align}
The simplest configurations in the family~\eqref{LMgeom} are those specified by a circular profile function~\cite{Balasubramanian:2000rt,Maldacena:2000dr}, with only a single mode $N_k^{\alpha\dot\beta}$ populated macroscopically (conventionally taken to be the one with $\alpha=\dot\beta=+$) and all the other mode numbers set to zero.

\noindent
This circular supertube solution is given by%
\footnote{Our conventions for this solution are related to those of~\cite{Bena:2016ypk,Bena:2017xbt} by $\phi_{\text{there}} = - \phi_{\R}$, $\,\psi_{\text{there}} = - \psi_{\R}$, $\,r_{\text{there}}=a \sinh \rho$, $v_{\text{there}}=\hat v$, $u_{\text{there}}=\hat u$. The dimensionful factors $a$ and $R_y$ in~\cite{Bena:2017xbt} are rescaled into the coordinates here, and we recall that we work in units in which $\alpha'=1$. The smoothness condition on the dimensionful brane charges, $Q_1 Q_5 = a^2 R_y^2$ in~\cite[Eq.~(3.5)]{Bena:2017xbt}, translates into the coefficient in the F1 harmonic function $\K$ being $1/n_5$ in Eq.\;\eq{roundST}. Our group theory conventions also differ from those of~\cite{Bena:2016ypk,Bena:2017xbt}, as described in Appendix~\ref{app:conventions}.}
\be
\label{profile1}
\F^1_{\!\supera}+i\F^2_{\!\supera} 
= a \, \exp\Bigl[ \frac{ik}{\nfive} \tilde{v} + \frac{2\pi i \suplabel}{\nfive}\Bigr]
\,, \qquad~~ \F^3_{\!\supera} \,=\, \F^4_{\!\supera} \,=\,0 \,,
\ee
which, from \eq{greensfn}, gives rise to \cite{Lunin:2001fv,Lunin:2001jy} 
\begin{align}
\begin{aligned}
\label{roundST}
\H &\,=\, \frac{\nfive}{a^2\Sigma_0} \,, \qquad 
\K \,=\,  \frac{k^2}{\nfive \Sigma_0} \,, \qquad 
\Sigma_0 \,=\,  \sinh^2\rhoo + \cos^2\theta  \,,
\\[.3cm]
\A &\,=\, -\frac{k \:\! \sin^2\theta}{\Sigma_0} \,d\phi_{\R} \,,~~~~
\B \,=\, \frac{k \:\! \cos^2\theta}{\Sigma_0}\, d\psi_{\R} \,.
\end{aligned}
\end{align}

For $k=1$, the six-dimensional part of this solution is related to global $AdS_3\times\bS^3$ by the  spectral flow (large) coordinate transformation 
\be
\label{bkgd specflow}
\phi_{\R} \,=\, \phi_{\NS} - \tau \,,\quad~~
\psi_{\R} \,=\, \psi_{\NS} - \sigma \,, 
\ee
under which the metric becomes
\be \label{eq:metric-0}
ds^2 \,=\,   
\nfive \Big[ \bigl(  - \cosh^2 \rhoo \;\! d\tau^2 + d\rho^2 +  \sinh^2 \rhoo \;\!  d\sigma^2 \bigr)
+ \bigl(d\theta^2 + \sin^2\!\theta \;\! d\phi_{\NS}^2 +  \cos^2\!\theta \;\! d\psi_{\NS}^2 \big)\Big]+ ds^2_\cM\,.
\ee
For $k>1$, the six-dimensional metric is a particular $\bZ_k$ orbifold of $AdS_3\times\bS^3$, see e.g.~\cite{Maldacena:2000dr,Mathur:2012tj}.

Provided that the defining profile function does not self-intersect, the configurations~\eqref{halfBPS} are everywhere smooth~\cite{Lunin:2001jy,Lunin:2002fw}. In the next section we will discuss an ellipse profile function; the circular profile function defining the above configuration is a special case of that ellipse profile, given in Eq.~\eqref{Fellipse}. In special examples where the profile runs over the same path multiple times, with successive iterations separated in $\hat{v}$ (as in the $k>1$ case above), the supergravity solution has an orbifold singularity at the profile (being otherwise smooth); however the configuration is completely nonsingular in low-energy string theory, due to the separation of the fivebrane windings in $\hat{v}$. 

Note that the solutions~\eqref{LMgeom}, \eq{eq:uv-form} in general admit the Killing vector fields $\partial_{\hat u}$ and $\partial_{\hat v}$.  Rescaling these by an overall factor of two for convenience, we introduce the notation
\begin{align}
\label{eq:kvecs-r}
\xi_1^{\R} & \; =\;  \partial_\tau  - \partial_\sigma \;=\; 2\partial_{\hat u} \,, \qquad \quad
\xi_2^{\R}  \;=\, \partial_\tau +\partial_\sigma \; =\; 2\partial_{\hat v}  \,.
\end{align}
The Killing spinors that square to these Killing vectors have periodic (Ramond) boundary conditions.
In the circular configuration, we have additional Killing vector fields $\partial_{\psi_{\R}}$ and $\partial_{\phi_{\R}}$. More generally, in the $AdS_3\times\bS^3$ decoupling limit, we have \emph{asymptotic} symmetries $\partial_{\psi_{\R}}$ and $\partial_{\phi_{\R}}$, which are generically not symmetries of the full configuration. 

The smooth configurations can be mapped to a BPS supergravitational wave deformation of global $AdS_3$ by the spectral flow coordinate transformation \eq{bkgd specflow}. Note that the coordinates $\psi_{\R}$, $\phi_{\R}$ are adapted to configurations close to the circular one; one could consider coordinates that approach the above $\psi_{\R}$, $\phi_{\R}$ near the asymptotic boundary of $AdS_3$ but which differ in the interior, and we shall do so later for the ellipse. Note also that the behaviour near the asymptotic boundary of $AdS_3$ defines the large diffeomorphism \eq{bkgd specflow}; different extensions into the bulk differ only by small diffeomorphisms, which are redundancies of the bulk description, and which thus do not change the state.

We note for use in the following section that under the spectral flow transformation \eqref{bkgd specflow}, the Killing vector fields in Eq.~\eq{eq:kvecs-r} transform to 
\begin{align}
\begin{aligned}
\label{eq:kvecs-ns}
\xi_1^{\NS} &\,=\, (\partial_\tau + \partial_{\phi_{\NS}}) - (\partial_\sigma + \partial_{\psi_{\NS}}) \,,\\[2mm]
\xi_2^{\NS} &\,=\, (\partial_\tau + \partial_{\phi_{\NS}}) +(\partial_\sigma + \partial_{\psi_{\NS}}) \,. \\[1mm]
\end{aligned}
\end{align}

In the next section we shall describe a family of gauged sigma models that involve gauging the null isometries of a set of (10+2)-dimensional solutions containing the general class of Lunin-Mathur solutions with pure NS-NS flux~\eqref{LMgeom}.

\section{Null gauging: general results and examples}
\label{sec:nullgauging}

In this section we describe the formalism of obtaining fivebrane backgrounds via null-gauged sigma models. These models consist of an ``upstairs'' (10+2)-dimensional spacetime with a pair of null isometries that are gauged, removing (1+1) directions, resulting in a (9+1)-dimensional ``downstairs'' target space. The primary examples studied to date are the point on the Coulomb branch moduli space where the fivebranes are arranged in a circular array~\cite{Giveon:1999px,Israel:2004ir}, and the closely related round supertube profile~\cite{Martinec:2017ztd}.  In both these examples, an underlying current algebra symmetry allows the worldsheet dynamics to be solved exactly.
In this section, we review these two examples, and in each case we describe how to generalize the construction to the full family of Lunin-Mathur solutions. We then present a fully explicit and novel example of this generalization, where the source profile is an ellipse.

We note that the nonlinear sigma model on a general Lunin-Mathur background~\eqref{LMgeom}-\eqref{greensfn} falls within the class of {\it chiral null models} studied in~\cite{Horowitz:1994ei,Horowitz:1994rf,Tseytlin:1995fh}, where it was shown that they are exact solutions to all orders in $\alpha'$.  The basic idea of these works is to show that if the ``transverse'' sigma model
\be
\label{ns5coul}
ds^2 = \H\,  d\xx\tight\cdot d\xx
~~,~~~~
B = \C_{ij} \, dx^i\wedge dx^j ~~,~~~~ e^{2\Phi} = \H
\ee
is conformally invariant, then the corresponding chiral null models~\eqref{NS5pgeom}, \eqref{eq:uv-form} are conformally invariant.  In our case, the transverse geometry is just that of a collection of parallel fivebranes on their Coulomb branch (with the source smeared over a contour).  The sigma model for the transverse geometry is $(4,4)$ supersymmetric and therefore exactly conformally invariant, and therefore the results on chiral null models apply.  As we show below, the transverse geometry and the chiral null model are indeed directly related -- one can gauge the null isometries of the chiral null model and obtain the transverse geometry.

\subsection{The null-gauging formalism}
\label{sec:null-gaug-gen}

The kinetic terms in the sigma model action involve the covariant derivative 
\be
\cD\varphi^i = \partial\varphi^i - \cA^a \xi_a^i(\varphi)
\ee
with a gauge potential $\cA^a$ for gauging each Killing vector $\xi^a$.
We have two independent gauge fields $(\cA^1,\cAb^1)$ and $(\cA^2,\cAb^2)$, each associated to a null Killing vector; the sigma model kinetic term is given by 
\be
\label{gauged KE}
\cL_{\rm K} = \mathcal{D} \varphi^i \, G_{ij} \, \overline{\mathcal{D}} \varphi^j
\;=\; (\partial \varphi^i - \cA^a \:\! \xi_a^i ) \, G_{ij} \,  (\bar\partial \varphi^j - \cAb^a \:\! \xi_a^j) \;,
\ee
while the Wess-Zumino term is given in terms of target-space one-forms $\theta_a$ (we follow the notation of~\cite{Figueroa-OFarrill:2005vws}), pulled back to the worldsheet:
\be
\label{eq:g-wz-term}
\cL_{\rm WZ} = B_{ij} \partial \varphi^i \bar\partial\varphi^j + \cA^a \theta_{a,i} \bar\partial\varphi^i + \cAb^a \theta_{a,i} \partial\varphi^i
+ \xi_{[a}^i \theta^{~}_{b],i} \cA^a \cAb^b \;.
\ee
For the pair of null Killing vectors $\xi_a$, the target-space one-forms $\theta_a$ are given by 
\be \label{eq:theta-xi}
\qquad\quad    \theta_{a} \;=\; (-1)^{a+1}  \xi_{a} \cdot d\varphi  \;\equiv\; (-1)^{a+1} \xi_a^i G_{ij} d\varphi^j \:\!      \qquad\quad (a=1,2)~;
\ee
this causes half the gauge field components to decouple. 
Due to these cancellations, the coefficient of the term quadratic in gauge fields ends up being proportional to the quantity
\be
\Sigma \;\equiv\; -\frac12\xi_1^i G_{ij} \xi_2^j   \,.
\ee 
For a consistent gauging, we have the conditions (see e.g.~\cite{Figueroa-OFarrill:2005vws}) 
\be
\imath_a H \;=\;  d\theta_a \,,\qquad \imath_a\theta_b \;=\; - \imath_b \theta_a   \,.
\ee
Overall, the gauge field terms in the action reduce to
\be
-2\cA^2 \xi_2^i G_{ij} \bar\partial\varphi^j -2 \cAb^1 \xi_1^i G_{ij} \partial\varphi^j - 4 \cA^2 \cAb^1 \Sigma  ~.
\ee
In what follows, we denote $\cA\equiv\cA^2$, $\cAb\equiv\cAb^1$. 

We define worldsheet currents $\cJ$, $\bar{\cJ}$ to be pull-backs of target-space one-forms as follows,\footnote{Note that $\cJ,\bar\cJ$ differ from the worldsheet current operators that measure momenta by a factor of $i$;  for instance the holomorphic operator that measures $y$ momentum is $\hat P_y=i\partial y$, and the holomorphic $SU(2)$ current operator is $\hat J^3_\su=i\jj^3_\su$.  Thus we have the null current operator $\widehat \cJ=i\cJ$.  We hope this does not cause confusion.  See Appendix~\ref{app:conventions} for details.} 
\be
\label{eq:J-def}
\cJ \,\equiv\, - \theta_{1} \cdot \partial\varphi \,\equiv\, - \theta_{1,i} \:\! \partial\varphi^i \;, \qquad \bar\cJ \,\equiv\, \theta_{2} \cdot \bar\partial\varphi \,\equiv\, \theta_{2,i} \:\!\bar\partial\varphi^i \,.
\ee
Using \eq{eq:theta-xi}, we can then rewrite the gauge terms as
\begin{align}
\label{eq:g-terms-gen}
2\cA \;\! \theta_{2,i} \:\!\bar\partial\varphi^i -2 \cAb \;\! \theta_{1,i} \:\! \partial\varphi^i -4 \cA \cAb \:\! \Sigma 
~\,\equiv\,~ 2 \cA \bar \cJ + 2 \bar\cA \cJ - 4 \cA \bar\cA \Sigma\,.
\end{align}
On integrating out the gauge fields, the gauge terms in the action, \eqref{eq:g-terms-gen}, in general become
\be \label{eq:g-terms-int-out}
\frac{\cJ \bar\cJ}{\Sigma} ~=~ - \frac{1}{\Sigma} \big( \theta_{1} \cdot \partial\varphi  \big) \big( \theta_{2} \cdot \bar\partial\varphi \big) 
~=~ \frac{1}{\Sigma} \big( \xi_{1} \cdot \partial\varphi  \big) \big( \xi_{2} \cdot \bar\partial\varphi \big) \,,
\ee
where similarly $\xi_{1} \cdot \partial\varphi \equiv\xi_1^i G_{ij} \partial\varphi^j$.
Thus we see that the null gauging procedure in effect adds the terms \eq{eq:g-terms-int-out} to the sigma model lagrangian.

Recent work on null-gauged sigma models for fivebrane backgrounds has focused on cosets where the ``upstairs'' (10+2)-dimensional spacetime contains global $AdS_3\times\bS^3$~\cite{Martinec:2017ztd,Martinec:2018nco,Martinec:2019wzw}. 
The simplest example of the family of models that can be obtained in this way is the Coulomb branch circular array of fivebranes~\cite{Giveon:1999px,Israel:2004ir,Martinec:2017ztd}, where the isometry that is gauged corresponds to the Killing vectors in Eq.~\eq{eq:kvecs-ns}. (Compare~\cite[Eqs.\,(4.7), (4.14), (4.15)]{Martinec:2019wzw} with Eqs.~\eq{eq:metric-0},~\eq{eq:kvecs-ns}). We shall exhibit this example in section \ref{sec:circular} below, however for now we shall keep the discussion general. 

\subsection{General Coulomb branch configurations from null gauging}

We now observe that the null gauging structure generalizes to the whole class of Lunin-Mathur geometries, demonstrating how to obtain the general Coulomb branch configurations described in section \ref{sec:coulomb} as the result of the gauging procedure. We start by considering the set of (10+2)-dimensional geometries that consist of adding additional directions $\bR^{1}_t\times \bS^1_y$ where $y\sim y+2\pi R_y$\,,
\begin{align}
\begin{aligned}
\label{eq:uv-form-10-2}
ds^2 = ds^2_{\LM} -dt^2 + dy^2 &= -\frac{1}{\K  }(d\hat v+\beta)(d\hat u + \omega )+ \H \, d\xx\!\cdot\! d\xx  -dt^2 + dy^2 + ds^2_\cM \,,\\
\end{aligned}
\end{align}
with the $B$-field and dilaton as given above in~\eqref{eq:uv-form}.

We work in the R-R sector, where the Killing vectors are as given in Eq.\;\eq{eq:kvecs-r}: 
\be
\label{eq:kv-1}
\xi_1 \,=\, 2\partial_{\hat u} \,, \qquad \xi_2 \,=\, 2\partial_{\hat v} \,.
\ee
From the general null form of the metric~\eqref{eq:uv-form}, using Eq.\;\eq{eq:theta-xi} we read off 
\be
\label{thetasigma to KAB}
\theta_1 = -\frac{1}{\K  }(d\hat{v}+\beta)
~~,~~~~
\theta_2 = \frac{1}{\K  }(d\hat{u}+\omega)
~~,~~~~
\Sigma = \frac{1}{\K  } ~.
\ee
The gauge terms that are added to the worldsheet lagrangian, after integrating out the gauge fields, from Eq.\;\eq{eq:g-terms-int-out} thus become
\be \label{eq:g-terms-circ-array}
\frac{1}{\K  }\Big( \partial \hat v + \beta_i \partial\varphi^i \Big) \Big( \bar\partial \hat u + \omega_j\bar\partial\varphi^j\Big)
\ee
which exactly cancel the first term in the metric~\eqref{eq:uv-form} and the corresponding terms in the $B$-field. 
We are then left with the downstairs fields 
\be
\label{dsperp}
ds^2 = -dt^2 + dy^2 + \H\,  d\xx\tight\cdot d\xx + ds^2_\cM
~~,~~~~
B = \C_{ij} \, dx^i\wedge dx^j 
\ee
of fivebranes on the Coulomb branch, in a (smeared) array of sources determined by the harmonic function $\H $ in Eq.~\eqref{greensfn}.  Along the way, the dilaton shifts by $\log(\K  )$ due to the spatially varying size of the gauge orbits, so that it becomes
\be
\label{dil-circ-array}
e^{2\Phi} = \H \;.
\ee
Thus, starting from a Lunin-Mathur geometry in 5+1d and gauging its null isometries yields a geometry in four Euclidean dimensions which is the transverse geometry of fivebranes on the Coulomb branch, with the harmonic function $\H $ given in Eq.~\eqref{greensfn}. Thus we have generalized the null gauging procedure to the full family of Lunin-Mathur solutions, and arbitrary arrays of fivebranes evenly distributed on a curve in $\mR^4$. We will demonstrate explicit examples of circular and elliptical profiles later in this section.

\subsection{General supertubes from tilted null gauging}
\label{sec:STfromCoulomb}

We now generalize the above construction to obtain supertube configurations after the null gauging procedure. 

Again working in the R-R sector, we tilt the null gauging procedure by modifying Eq.~\eqref{eq:kv-1} to gauge instead the following Killing vectors. Defining for convenience $\alphab=kR_y$ with $k$ a positive integer, we gauge
\be
\label{tiltedKV}
 \xi_1 = (\partial_\tau-\partial_\sigma)-\alphab\bigl( \partial_t+\partial_y \bigr)
~~,~~~~
 \xi_2 = (\partial_\tau+\partial_\sigma)-\alphab\bigl( \partial_t-\partial_y \bigr) .
\ee 
Introducing the notation
\be
v \;=\; t+y \,, \qquad u \;=\; t-y \,,
\ee
the one-forms $\theta_a$ and function $\Sigma$ become
\be
\label{thetasigma to KAB-sup}
\theta_1 = -\frac{1}{\K  }(d\hat{v}+\beta) + \alphab du
~~,~~~~
\theta_2 = \frac{1}{\K  }(d\hat{u}+\omega) - \alphab dv
~~,~~~~
\Sigma = \frac{1}{\K  } + \alphab^2 .
\ee
The terms that get added to the action after integrating out the gauge fields, \eq{eq:g-terms-int-out}, evaluate to 
\begin{align}
\label{eq:g-terms-circ-suptube}
& \frac{1}{\K(1+\alphab^2 \K ) }(\partial \hat v + \beta )( \bar\partial \hat u + \bar\omega)- \frac{\alphab}{1+\alphab^2 \K} \Big[ \partial u (\bar \partial \hat u + \bar \omega) + \bar\partial v (\partial \hat v + \beta) \Big] + \frac{\alphab^2 \K}{1+\alphab^2 \K}\partial u \bar\partial v \,.
\end{align}
Thus the downstairs effective action after integrating out the gauge fields evaluates to
\begin{align}
\begin{aligned}
\cS_\eff &\;=\; -\frac{1}{1+\alphab^2 \K}\Big[ (\partial u + \alphab \partial \hat v) + \alphab \beta \Big] 
\Big[ ( \bar\partial v + \alphab \bar\partial \hat u) + \alphab \bar\omega \Big] 
+\Bigl( \H \;\! \delta_{ij}+\C_{ij} \Bigr) \partial x^i \bar\partial x^j + \cS_{\cM} \,.
\end{aligned}
\end{align}
Choosing the gauge $\tau=\sigma=0 ~\Rightarrow \hat u = \hat v = 0$, this becomes
\begin{align}
\begin{aligned}
\label{eq:Seff-suptube}
\cS_\eff &\;=\; -\frac{1}{1+\alphab^2 \K}\big (\partial u + \alphab \beta \big)
\big(  \bar\partial v  + \alphab \bar\omega \big) 
+\Bigl( \H \;\! \delta_{ij}+\C_{ij} \Bigr) \partial x^i \bar\partial x^j + \cS_{\cM} \,.
\end{aligned}
\end{align}
The dilaton is given by
\be \label{dil-suptube}
e^{2\Phi} = \frac{\gstrsq \alphab^2 \H }{(1+\alphab^2 \K)}~.
\ee
Note that if one sends $\alphab \to 0$ and rescales the dilaton appropriately, Eqs.\;\eq{eq:g-terms-circ-suptube}--\eq{dil-suptube} reduce to \eq{eq:g-terms-circ-array}--\eq{dil-circ-array}. By contrast, for non-zero $\alphab$, the F1 harmonic function $(1+\alphab^2 \K)$ (which replaces $\K$ in \eq{eq:uv-form}) now asymptotes to a constant, so the procedure has resulted in an asymptotically linear dilaton solution.
 
To see this in more detail, note that from Eq.\;\eq{eq:Seff-suptube} one can take the $AdS_3$ decoupling limit, $\Ry\to\infty$ with $\tilde t=t/\alphab$, $\tilde y=y/\alphab$ fixed.  This recovers the asymptotically $AdS_3$ Lunin-Mathur geometry~\eqref{LMgeom}, with $\tau=\tilde t, \sigma=\tilde y$.  Thus if we start with an asymptotically $AdS_3$ Lunin-Mathur geometry, null gauging simply adds back an asymptotically linear dilaton region.  Naively one might have thought this a bit redundant -- after all, why not simply start with the sigma model with the harmonic function appropriate to the fivebrane decoupling limit without the further $AdS_3$ decoupling limit, and eliminate the null gauging procedure which simply adds and subtracts two additional coordinates in the sigma model target space?
The point is that the geometry at the supergravity level is rather singular at the fivebrane source locus; but in the null gauging construction, this singularity in the effective geometry downstairs in 9+1d arises from the gauge projection of a completely nonsingular geometry upstairs in 10+2d.  This allows us to use semiclassical calculations in the upstairs geometry to see what are otherwise highly stringy features of the background.  For instance one can see the discrete fivebrane locations through the quantization of flux on the worldvolume of D-brane probes in the upstairs geometry~\cite{Martinec:2019wzw}; this quantization is not apparent in a semiclassical analysis of the sigma model downstairs in (9+1) dimensions.

Another useful feature of the null gauging procedure is the way it neatly separates the structure of the $AdS_3$ cap from the asymptotically linear dilaton region.  In particular, one starts with a sigma model geometry with two timelike directions; these are time as measured by observers in the cap ($\tau$), and by asymptotic observers ($t$).  Of course, there is only one physical time.  The role of null gauging is to eliminate the appropriate combination of these two times, in effect synchronizing the clocks of asymptotic and cap observers.  This synchronization is accomplished by the null current being gauged, which involves mostly the time $\tau$ at large radius, and mostly the time $t$ at small radius.  Thus the physical (gauge invariant) time coordinate smoothly transitions from $t$ at large radius to $\tau$ in the cap.

\subsection{Circular source profiles}
\label{sec:circular}

We now review a family of explicit examples of null-gauged models, based on circular source profiles.
In this subsection we will take the upstairs geometry to be that corresponding to circular source profile \eq{profile1} with $k=1$. 
We reviewed in section \ref{sec:LMgeom} that after the spectral flow coordinate transformation \eq{bkgd specflow}, the six-dimensional metric becomes that of global $AdS_3\times\bS^3$ \eq{eq:metric-0}.

The supergravity backgrounds for circular source profiles have an exact worldsheet description as gauged WZW models, where one gauges left and right null isometries on the group manifold%
~\cite{Israel:2004ir,Itzhaki:2005zr,Martinec:2017ztd}
\be
\label{Gupstairs2}
\cG =
SL(2,\mathbb{R}) \times SU(2) \times \bR_t^{~}\times\bS^1_y\times \bT^4 ~;
\ee
in this way one can realize exactly solvable worldsheet descriptions for a circular array of fivebranes on the Coulomb branch, as well as for round NS5-F1 and NS-P supertubes.


We recall our conventions for the worldsheet sigma models on $\sutwo$ and $\sltwo$; for more details see~\cite{Martinec:2017ztd,Martinec:2018nco,Martinec:2019wzw}.
We use Euler angle group parametrizations
\be
\label{Eulerangles-1}
g_{\sl} \;=\; e^{\frac{i}{2}(\tau-\sigma)\sigma_3}e^{\rho \sigma_1}e^{\frac{i}{2}(\tau + \sigma)\sigma_3} 
 ~, ~~~~~~~ g_{\su} \;=\; e^{\frac{i}{2}(\psi-\phi)\sigma_3}e^{i \theta \sigma_1}e^{\frac{i}{2}(\psi + \phi)\sigma_3} ~.
\ee
The bosonic Wess-Zumino-Witten model on these groups has level
$\nfivehat=\nfive+2$ for $\sltwo$, and $\nfivetil=\nfive-2$ for $\sutwo$; fermion contributions to the currents shift both levels to $\nfive$.  We are going to ignore these shifts in our exposition of the null gauging formalism in order to avoid clutter; when it becomes important, for instance in the exposition of FZZ duality in section~\ref{sec:FZZduality}, we will carefully keep track of the distinction.
The line element for $\sltwo\times\sutwo\times\bR_t\times\bS^1_y$ is then
\be \label{eq:metric}
ds^2 \,=\,   
\nfive \bigl(  - \cosh^2 \rhoo d\tau^2 + d\rho^2 +  \sinh^2 \rhoo d\sigma^2 \bigr)
+ \nfive\bigl(d\theta^2 + \sin^2\!\theta \,d\phi^2 +  \cos^2\!\theta d\psi^2 \big)-dt^2+dy^2\,,
\ee
and comparing with \eq{eq:metric-0} we identify $\psi=\psi_{\NS}$, $\phi=\phi_{\NS}$. We will suppress the subscript $\NS$ in much of the following, for ease of notation.
The $H$-flux is
\be \label{eq:Hslsu}
H_{(3)} ~=~ \nfive \bigl( \sinh 2\rho \,d\rho \wedge d\tau \wedge d\sigma\bigr) + \nfive\bigl(\sin 2\theta \,d\theta \wedge d\psi \wedge d\phi \bigr)~.
\ee

A family of gauged models is obtained by gauging the currents~\cite{Martinec:2017ztd} 
\begin{align}
\begin{aligned}
\label{J gen ns}
\cJ &\;=\; l_1 \jj^3_\sl + l_2 \jj^3_\su + l_3 {\P}_{\!t,L} + l_4 {\P}_{\!y,L}  \;, \\
\bar{\cJ} &\;=\;  r_1 \bar{\jj}^3_\sl +r_2 \bar{\jj}^3_\su + r_3 {\P}_{\!t,R} + r_4 {\P}_{\!y,R} \;,
\end{aligned}
\end{align}
where 
\bea 
\jj^3_{\sl} &=& 
\nfive\bigl(\cosh^2\! \rho \,\partial\tau + \sinh^2 \!\rho \,\partial \sigma \bigr) \,,
\quad ~~
\bar{\jj}^3_{\sl} 
~=~ 
\nfive\bigl(\cosh^2\! \rho\, \bar\partial\tau - \sinh^2\! \rho \,\bar\partial\sigma \bigr) \,, 
\nonumber \\[0.5mm]
\label{eq:J3su}
\jj^3_{\su} &=& 
 \nfive\bigl( \cos^2\!\theta \, \partial \psi - \sin^2 \!\theta \,\partial \phi \bigr) \,, \phantom{\Big\{}
\quad ~~
\bar{\jj}^3_{\su} 
~=~ 
\nfive\bigl( \cos^2\!\theta \,\bar\partial \psi + \sin^2 \!\theta \, \bar\partial \phi \bigr) \,, \\[0.5mm]
&& \qquad {\P}_{\!t,L}\equiv \partial t
~~,~~~~
{\P}_{\!t,R}\equiv \bar\partial t
~~,~~~~
{\P}_{\!y,L}\equiv \partial y
~~,~~~~
{\P}_{\!y,R}\equiv \bar\partial y ~,
\nonumber
\eea
and where we impose that the currents be null by requiring 
\begin{align}
\begin{aligned}
\label{xi-norms}
0 &\;=\; \langle \boldsymbol{\ell},\boldsymbol{\ell} \rangle  \;\equiv\; \nfive (-l_1^2+l_2^2)-l_3^2+l_4^2  \,,\\
0 &\;=\; \langle \boldsymbol{r}, \boldsymbol{r}\rangle \;=\; \nfive (-r_1^2+r_2^2)-r_3^2+r_4^2\,.
\end{aligned}
\end{align}
From Eqs.\;\eq{eq:theta-xi} and \eqref{eq:J-def}, this corresponds to gauging the following Killing vector fields $\xi_a$,
\begin{align}
\begin{aligned}
\label{xione gen}
\xi_1 &\;=\;  l_1 \bigl( {\partial_\tau}-{\partial_\sigma} \bigr) - l_2 \bigl( {\partial_\psi}-{\partial_\phi} \bigr) 
+l_3 {\partial_t}-l_4{\partial_y} \,,\\
\xi_2 &\;=\;  r_1 \bigl( {\partial_\tau}+{\partial_\sigma} \bigr) -r_2 \bigl( {\partial_\psi}+{\partial_\phi} \bigr) 
+r_3 {\partial_t}-r_4{\partial_y} \,,
\end{aligned}
\end{align}
which are null in the metric \eqref{eq:metric} due to the conditions \eqref{xi-norms}.

\subsubsection{The Coulomb branch}
\label{sec:circ-CB}

To obtain the configuration of fivebranes on the Coulomb branch in a circular array given by Eqs.\;\eq{eq:CB1}, \eq{eq:CB2}, \eq{sphericalbipolar}, \eq{profile1}, we take
\be
\label{eq:CB-params}
l_1 =  l_2 = 1 \,, \quad~~  l_3= l_4 = 0 \,, \quad~~  r_1 =  1 \,,\quad~ r_2 = -1 \,, \quad~ r_3= r_4 = 0 \,,
\ee
so that the null Killing vectors $\xi_a$ to be gauged are as in Eq.\;\eq{eq:kvecs-ns} (here we reintroduce the subscript $\NS$ on $\phi_\NS$, $\psi_\NS$ to connect with the discussion around \eq{eq:metric-0}),
\begin{align}
\begin{aligned}
\label{xione coulomb}
\xi_1^{\NS} &\;=\;  \bigl( {\partial_\tau}+{\partial_{\phi_\NS}} \bigr) 
-\bigl({\partial_\sigma}+{\partial_{\psi_\NS}} \bigr)  \,, \\
\xi_2^{\NS} &\;=\;   \bigl( {\partial_\tau}+{\partial_{\phi_\NS}} \bigr) 
+\bigl({\partial_\sigma}+{\partial_{\psi_\NS}} \bigr) \,,
\end{aligned}
\end{align}
corresponding to the currents 
\be \label{currents-circ-CB}
\cJ = \jj^3_\sl+\jj^3_\su \,, \qquad \bar\cJ = \bar \jj^3_\sl - \bar \jj^3_\su \,.
\ee
The associated one-forms $\theta_a$ and quadratic form $\Sigma$ are
\bea
\label{theta coulomb}
\theta_1^{\NS} &=&  \nfive \left[ -\left(\cosh^2\rhoo \,  d\tau + \sinh^2\rhoo \,  d\sigma \right) - \left(\cos^2\theta \,d\psi- \sin^2\theta\, d\phi  \right) \right] \,,
\nn\\[4pt]
\theta_2^{\NS} &=&  \nfive\left[ +\left(  \cosh^2\rhoo \,  d\tau -\sinh^2\rhoo \,  d\sigma \right) - \left( \cos^2\theta \,d\psi+ \sin^2\theta \,d\phi  \right) \right] \,, 
\\[4pt]
\Sigma &=& \nfive \bigl( \sinh^2\rho + \cos^2\theta \bigr) \;=\; \nfive \Sigma_0 ~.\nn
\eea
The effective transverse geometry~\eqref{dsperp} resulting from null gauging has a fivebrane source singularity at $\rho=0,\theta=\pi/2$, which is the unit circle in the $x^1$-$x^2$ plane (see Eq.~\eqref{sphericalbipolar}). 
Choosing the gauge $\tau\tight=\sigma\tight=0$ and integrating out the gauge fields, one finds the effective geometry~\eqref{eq:CB1} where the harmonic function $\H$ is given in~\eqref{roundST}.

Equivalently, the preceding discussion can be translated into R-R coordinates, where the relevant quantities are the evaluation of the general expressions \eq{thetasigma to KAB}--\eq{dil-circ-array} on the solution data of this circular configuration.

\subsubsection{Circular supertubes}
\label{sec:circ-sup}

A generalization of the choice of null Killing vector leads to round supertube backgrounds~\cite{Martinec:2017ztd}.  
For the NS5-F1 supertube, one adds the following contributions
along $\bR_t^{~}\times \bS^1_y$ to the $AdS_3\times\bS^3$ Killing vectors for the Coulomb branch (c.f. Eq.~\eqref{tiltedKV})
\be
\delta \xi_1 = -k\Ry\bigl( \partial_t+\partial_y \bigr)
~~,~~~~
\delta \xi_2 = -k\Ry\bigl( \partial_t-\partial_y \bigr) ~,
\ee
or in terms of the $l_i$, $r_i$, one takes
\be
\label{eq:circ-sup-params}
l_1 =  l_2 = 1 \,, \quad  l_3=-k\Ry \,, \quad l_4 = k\Ry \,, \quad~~  r_1 =  1 \,,\quad~ r_2 = -1 \,, \quad~ r_3= r_4 = -k\Ry \,.
\ee
The parameter $k$ appears here through the Killing vectors that we gauge; in the downstairs solution, it plays precisely the same role as the parameter $k$ in section \ref{sec:fivebranes}, as we now describe.

The resulting background is T-dual to an NS5-P supertube with only the $k^{\rm th}$ harmonic of the transverse scalar with $++$ polarization excited on the fivebranes, so that the state of the fivebrane supertube is specified by
\be
\label{roundST state}
\Bigl(\ket{++}^{~}_k\Bigr)^{\none\nfive/k} ~.
\ee
Again choosing the gauge $\tau,\sigma=0$ and eliminating the gauge fields, the effective NS5-F1 geometry downstairs in 9+1d is given by combining Eqs.~\eqref{eq:Seff-suptube}, \eqref{dil-suptube} with the round supertube data \eqref{roundST}. Rewriting this in Lunin-Mathur form in which the metric is
\begin{align}
\label{STgeom-2}
ds^2 &= \K  ^{-1}\bigl[ -(dt+\A)^2 + (dy + \B)^2 \bigr] + \H \, d\xx\!\cdot\! d\xx + ds^2_\cM
\end{align}
we see that the $\K$ function now includes a constant term,
\begin{align} 
\K   = 1+ \frac{k^2\Ry^2}{\nfive\Sigma_0} \,,
\end{align}
see Eq.\;\eq{eq:Seff-suptube} and the discussion below Eq.\;\eqref{dil-suptube}. The (5+1)-dimensional part of the above metric naively reduces to a $\bZ_k$ orbifold of $AdS_3\times\bS^3$ in the limit $\Ry\to\infty$~-- though as discussed in~\cite{Martinec:2017ztd,Martinec:2018nco,Martinec:2019wzw}, the resolution of the orbifold singularity differs from the standard orbifold construction of worldsheet string theory.

\subsubsection{General gaugings of circular profiles}

More general choices of the parameters $l_i$, $r_i$ lead to three-charge spectral flowed supertube solutions, both supersymmetric and non-supersymmetric~\cite{Giusto:2004id,Giusto:2004ip,Lunin:2004uu,Jejjala:2005yu,Giusto:2012yz,Chakrabarty:2015foa}, as shown in \cite{Martinec:2017ztd} and analyzed in~\cite{Martinec:2018nco,Martinec:2019wzw}.
So far, the discussion in this section \ref{sec:circular} parallels the presentation in~\cite{Martinec:2017ztd,Martinec:2018nco,Martinec:2019wzw}. Let us now connect this to the discussion in section \ref{sec:null-gaug-gen}.  To ease the notation while keeping the discussion fairly general, let us assume for the rest of this subsection that the $l_i$, $r_i$ are restricted by 
\be
\label{eq:li-ri-R}
l_1 \,=\,  l_2  \,, \quad~~   r_1 \,=\,  -r_2  \qquad \Longrightarrow \qquad l_3^2 \,=\, l_4^2 \,,\quad~~ r_3^2\,=\,r_4^2 \,,
\ee
where the implication follows from \eq{xi-norms}. The discussion that follows can be generalized straightforwardly to general null $l_i$, $r_i$. 

Let us perform a spectral flow to R-R coordinates using the coordinate transformation \eqref{bkgd specflow}. The upstairs geometry is transformed appropriately, and the Killing vectors transform to
\begin{align}
\begin{aligned}
\label{xione gen R}
\xi_1^{\R} &\;=\;  l_1 \bigl( {\partial_\tau}-{\partial_\sigma} \bigr) 
+l_3 {\partial_t}-l_4{\partial_y} \,,\\
\xi_2^{\R} &\;=\;  r_1 \bigl( {\partial_\tau}+{\partial_\sigma} \bigr) 
+r_3 {\partial_t}-r_4{\partial_y} \,,
\end{aligned}
\end{align}
which are null due to the conditions on $l_3$, $r_3$ in Eq.\;\eqref{eq:li-ri-R}.

The null currents $\cJ$, $\bar\cJ$ in \eq{J gen ns} also transform using the spectral flow coordinate transformation \eqref{bkgd specflow}. Using Eq.~\eqref{eq:J-def}, this gives rise to the following one-forms $\theta_a$,
\begin{align}
\begin{aligned}
\label{theta gen R}
\theta_1^{\R} &\;=\; \nfive\Big[-l_1 \Sigma_0 d\hat{v} -l_1 (\cos^2\theta d\psi_\R-\sin^2\theta d\phi_\R) 
 - l_3 dt - l_4 dy \Big]\,, \\
\theta_2^{\R} &\;=\; \nfive\Big[r_1 \Sigma_0 d\hat{u} -r_1 (\cos^2\theta d\psi_\R+\sin^2\theta d\phi_\R) 
 + r_3 dt + r_4 dy \Big] \,.
\end{aligned}
\end{align}
On choosing parameters for the circular array on the Coulomb branch, \eqref{eq:CB-params}, the Killing vectors \eq{xione gen R} simply reduce to $\xi_1^{\R} =  2\partial_{\hat{u}}\,,$ $\xi_2^{\R} =2\partial_{\hat{v}}$ as in Eq.~\eqref{eq:kvecs-r}, 
and the one-forms $\theta_a$ agree precisely with the expressions in Eq.~\eqref{thetasigma to KAB} evaluated on the data in Eqs.\;\eq{b-o-A-B}, \eqref{roundST}. 

Similarly, for the circular supertube, the Killing vectors  \eq{xione gen R} evaluated on the data \eq{eq:circ-sup-params} agree precisely with the Killing vectors given in the general discussion in Eq.\;\eqref{tiltedKV}, and so we see that the example of the circular supertube in section \ref{sec:circ-sup} is a special case of the much more general structure described in section 
\ref{sec:STfromCoulomb}.

\subsection{Elliptical source profiles}
\label{sec:ellipse}

Having shown that null gauging can be applied to general Lunin-Mathur solutions, and reviewed examples based on circular profiles, we now derive the closed-form Lunin-Mathur solution for an elliptical profile, obtaining a new family of explicit examples. 

As reviewed in section \ref{sec:LMgeom}, the $\sltwo\times\sutwo$ geometry used to construct the circular array of fivebranes in section \ref{sec:circ-CB}, is itself a Lunin-Mathur geometry~\eqref{greensfn} for the circular source profile given in Eq.\;\eqref{profile1},
with $k=1$. In the previous subsection we saw how this fits within the general structure of null gauging laid out in section~\ref{sec:null-gaug-gen}.  Indeed, the averaged profiles in the source integrals~\eqref{greensfn} and the circular array are the same~-- a uniform fivebrane charge density along a circle.

\begin{figure}[ht]
\centering
  \begin{subfigure}[b]{0.4\textwidth}
  \hskip 0cm
    \includegraphics[width=\textwidth]{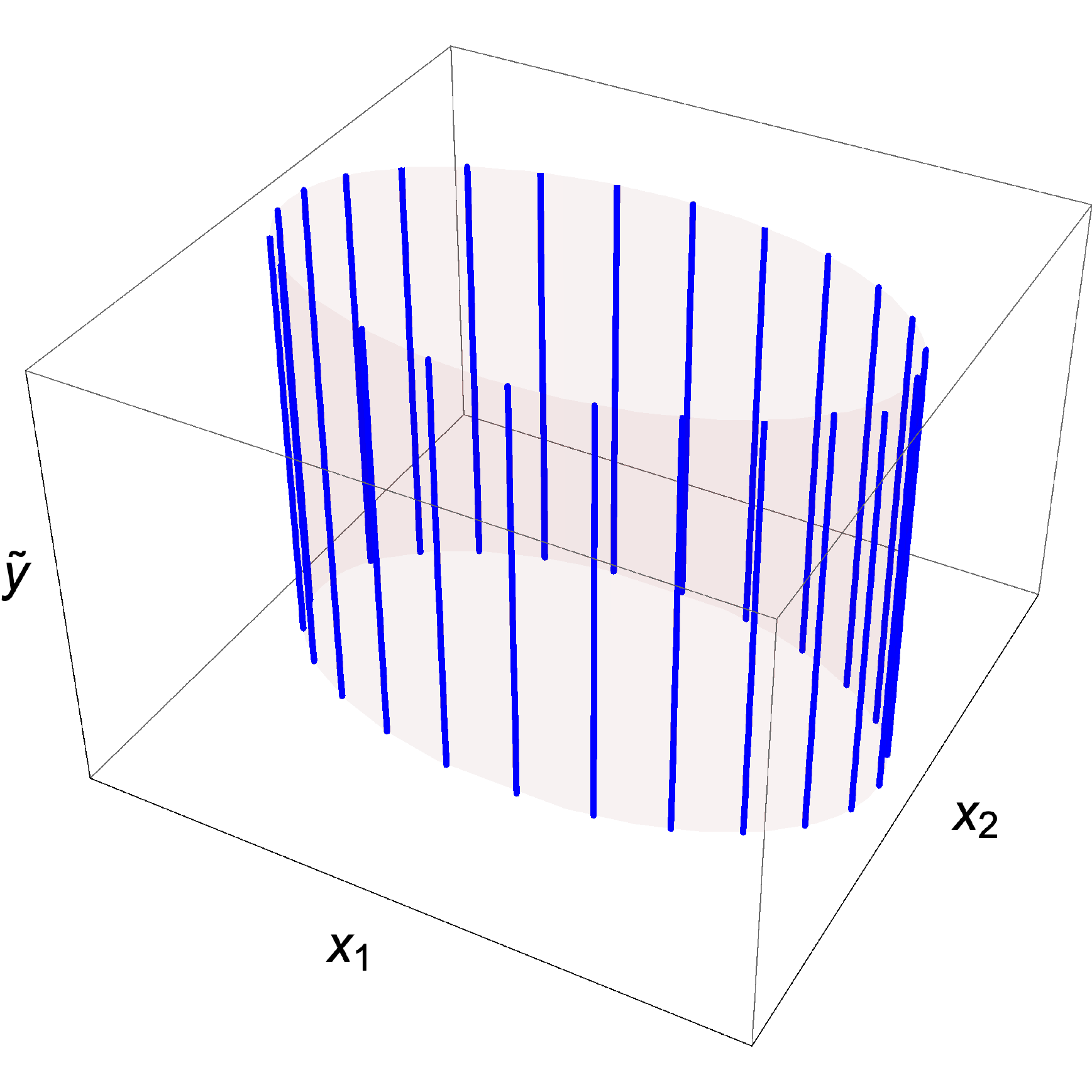}
    \caption{ }
    \label{fig:EllipseCoul}
  \end{subfigure}
\qquad\qquad
  \begin{subfigure}[b]{0.4\textwidth}
      \hskip 0cm
    \includegraphics[width=\textwidth]{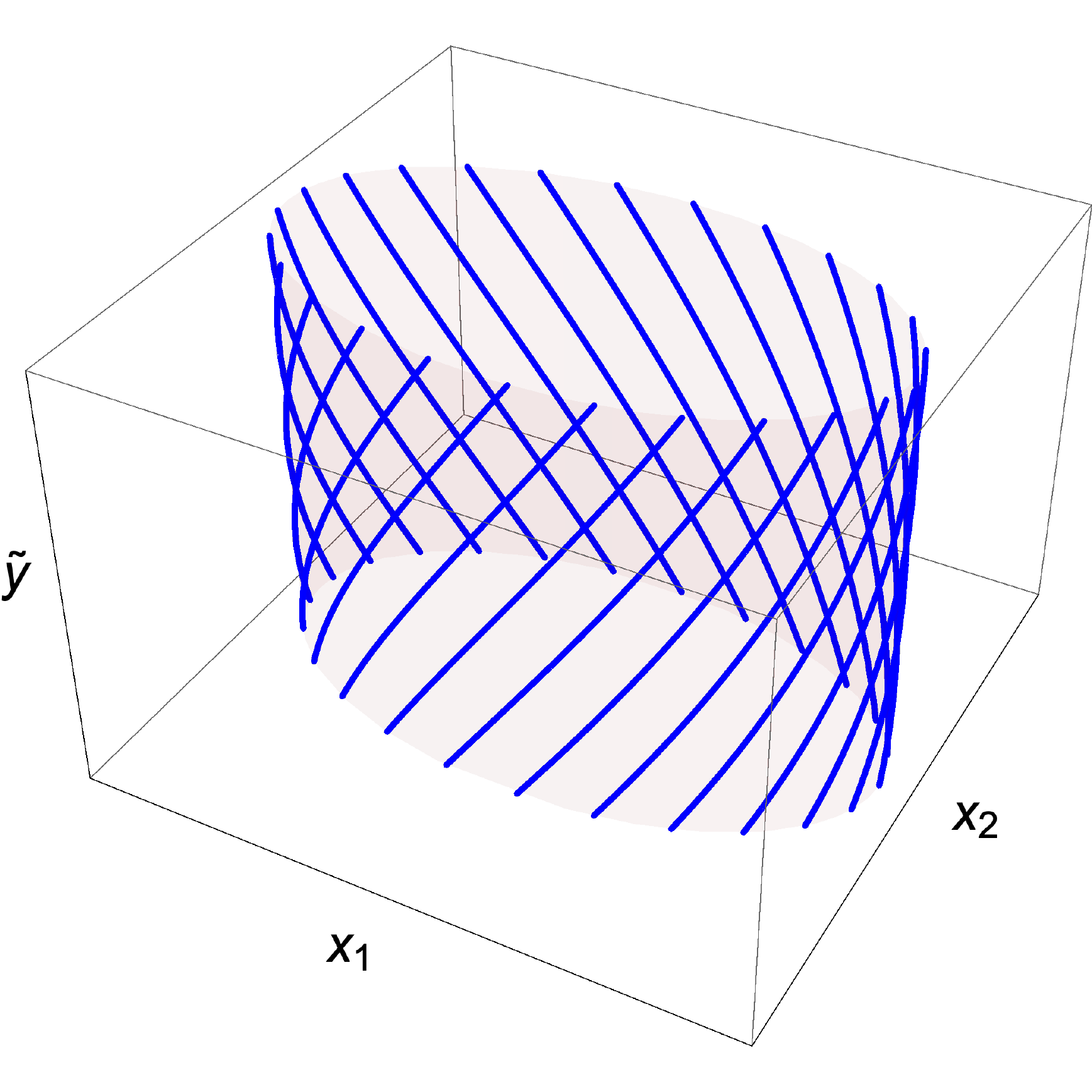}
    \caption{ }
    \label{fig:EllipseST}
  \end{subfigure}
\caption{\it 
On the left, a configuration of $\nfive=25$ NS5 branes on their Coulomb
branch uniformly distributed along an ellipse. On the right, the source profile for the associated Lunin-Mathur geometry with $k=4$.
In the supergravity limit, both configurations exhibit constant fivebrane charge density along the ellipse -- the separation between fivebrane strands is well below the string scale and so invisible in the classical sigma model background, which sees only a smeared source.
}
\label{fig:Ellipses}
\end{figure}

The (9+1)-dimensional geometry describing the elliptical array of static fivebranes on the Coulomb branch, Eqs.~\eq{eq:CB1}--\eq{eq:CB2} combined with \eq{Fellipse}, was worked out in~\cite{MariosPetropoulos:2005rtq}. An example of this configuration is depicted in Fig.\;\ref{fig:EllipseCoul}.
The source profile for the elliptical supertube is given by
\be
\label{Fellipse}
\F^1_{\!\supera} =  a_1 \cos \Bigl(\frac{k\:\! \tilde{v}}{\nfive} + \frac{2\pi \suplabel}{\nfive}\Bigr)\,, 
\qquad 
\F^2_{\!\supera} =a_2 \sin \Bigl(\frac{k\:\! \tilde{v}}{\nfive}  + \frac{2\pi \suplabel}{\nfive} \Bigr)\,,
\qquad \F^3_{\!\supera} \;\!=\;\! \F^4_{\!\supera} \;\!=\;\! 0 \,.
\ee
Once again, the sum over $m$ and integral over $v$ in the source profile $\F_{(m)}$ amounts to a uniform fivebrane charge density, now along an ellipse in the $x^1$-$x^2$ plane.  The Coulomb branch configuration corresponds to this source profile with $k=0$ (supergravity cannot resolve individual fivebrane sources, and so only sees the average of the sources along the ellipse).

We now exhibit the elliptical Lunin-Mathur supertube solution, an example of which is depicted in Fig.\;\ref{fig:EllipseST}. The harmonic function $\H$ and the two-form $\C$ are the same as those of the elliptical Coulomb branch configuration~\cite{MariosPetropoulos:2005rtq}.   In addition, we must solve for the harmonic one-form $\A$ and harmonic function $\K$ appearing in~\eqref{LMgeom}. It turns out that one can do so in closed form.\footnote{Multipole moments of ellipsoidal curves, as well as smeared elliptical profiles, were computed and studied holographically in~\cite{Kanitscheider:2006zf}. It would be interesting to further study elliptical profile solutions using recent developments in precision holography~\cite{Giusto:2019qig}.}

We shall work in the $AdS_3$ decoupling limit, however it is trivial to restore linear dilaton or flat asymptotics by adding appropriate constants to the harmonic functions $\K$ and/or $\H$, as described below equation~\eqref{greensfn}.

We define for future reference
\be
\label{eq:epsilon-def}
a^2 \;\!=\;\! \half(a_1^2+a_2^2)
~~,~~~~
\epsilon \;\!=\;\! \frac{a_1^2-a_2^2}{a_1^2+a_2^2}~.
\ee
The parameter $\epsilon$ characterizes the deformation from the circular array with $k=1$.
It is convenient to adopt the elliptical bipolar coordinates
\begin{align}
\label{ellbip}
x^1 = \sqrt{r^2\tight+a_1^2}\, \sin\theta\;\! \cos\phi
~~&,~~~~
x^3= r\cos\theta\;\!\cos\psi
\nn\\[.2cm]
x^2 = \sqrt{r^2\tight+a_2^2}\, \sin\theta\;\! \sin\phi
~~&,~~~~
x^4= r\cos\theta\;\!\sin\psi ~.
\end{align}
When $a_1=a_2$, these coordinates reduce to the spherical bipolars in Eq.\;~\eqref{sphericalbipolar}, with $r=a\sinh\rho$.   When $a_1\ne a_2$, the above coordinates are nicely adapted to the deformation of the source to an ellipse of semi-major (minor) axis $a_1$($a_2$), which continues to be the locus $r=0$, $\theta=\pi/2$. 
It is also convenient to define the functions (following~\cite{Bakas:1999ax} with some modifications, and with $a_3=a_4=0$)
\begin{align}
\label{AmBm}
\bA_m &= \sum_{i=1}^4 \frac{ x_i^2}{(r^2+a_i^2)^{m}}
\;,~~~~~
\bB_m = \sum_{i=1}^4 \frac{1}{(r^2+a_i^2)^{m}} 
\;,~~~~~
f = \prod_{i=1}^4(r^2+a_i^2) \;,
\nn\\[.3cm]
\Delta_i &=r^2+a_i^2
\;,~~~~~
\Delta = \Delta_1\Delta_2\cos^2\theta + r^2\sin^2\theta\bigl(\Delta_1\sin^2\phi+\Delta_2\cos^2\phi\bigr) \;.
\end{align}
The flat $\mR^4$ base metric is then given by
\begin{align}
\begin{aligned}
d\xx\!\cdot\! d\xx &\;=\; \frac{\Delta}{\Delta_1 \Delta_2} dr^2 + \Big[ r^2 \sin^2\theta + \cos^2\theta(\Delta_1 \cos^2\phi + \Delta_2 \sin^2 \phi) \Big] d\theta^2 + r^2 \cos^2\theta d\psi^2 \\
&+ \sin^2\theta \left( \Delta_1 \sin^2 \phi + \Delta_2 \cos^2 \phi \right) d\phi^2 + 2 (\Delta_2-\Delta_1) \sin\theta \cos\theta \sin \phi \cos \phi
\;\!  d\theta d\phi \,.
\end{aligned}
\end{align}
The functions $\H$ and $\K$, the one-forms $\A$ and $\B$, and the transverse $B$-field $\C$ are given by 
\begin{align}
\label{LMellipse}
\H  &= \frac{\nfive}{ f^{1/2}\bA_2}
= \frac{\nfive \sqrt{\Delta_1 \Delta_2}}{\Delta} \,,
\nn\\[.3cm]
\K   &=\frac{k^2}{\nfive}  \!\!\; \left[ a^2 \frac{\H }{\nfive} + \left( \frac{\H }{\nfive} \bigl( \hf \bB_{-1} - \bA_0 \bigr) -1 \right) \right]
\nn\\ 
&= \frac{k^2}{\nfive^2}\H  \Bigl[ a^2+\frac{1}{2}\Bigl(
(\Delta_1+\Delta_2)\cos^2\theta + \bigl(2r^2-(\Delta_1-\Delta_2)\cos2\phi\bigr)\sin^2\theta - \nfive \H ^{-1}\Bigr)\Bigr]\,,
\nn\\[.3cm]
\A &= a_1a_2 k\frac{\H}{\nfive}   \Bigl( \frac{x^2 dx^1}{r^2+a_2^2}- \frac{x^1dx^2}{r^2+a_1^2} \Bigr)
\\
&= -\frac{a_1a_2 k \sin\theta}{2\;\! \Delta}\Bigl[ 
\bigl((\Delta_1\tight+\Delta_2)-(\Delta_1\tight-\Delta_2)\cos2\phi\bigr)\sin\theta \,d\phi
-(\Delta_1\tight-\Delta_2)\cos\theta\sin2\phi \,d\theta\Bigr] \,,
\nn\\[.3cm]
\B &=  a_1a_2 k\frac{\H}{\nfive}  \Bigl( -\frac{x^4dx^3}{r^2}+ \frac{x^3dx^4}{r^2} \Bigr)
\nn\\
&= a_1a_2  k\frac{ \sqrt{\Delta_1\Delta_2}}{\Delta} \cos^2\theta \,d\psi  \,,
\nn\\[.25cm]
\C &= \frac{\nfive \Delta_1\Delta_2\cos^2\theta}{\Delta} d\phi\wedge d\psi
+ \frac{\nfive (\Delta_1-\Delta_2) r^2 \sin 2\theta \sin 2\phi}{4\Delta} d\theta\wedge d\psi \,. \nn
\end{align}
The expression for $\H$ was found in~\cite{Bakas:1999ax}, and the transverse $B$-field $\C$ in~\cite{MariosPetropoulos:2005rtq}, while the expressions for $\K$, $\A$ and $\B$ are new; in all cases, they arise from the source integrals~\eqref{greensfn} with the source profile~\eqref{Fellipse}. The analysis leading to these expressions is described in appendix~\ref{app:ellpert}.  
The geometry~\eqref{LMellipse} is a 1/2-BPS nonlinear gravitational wave on $AdS_3$ after reversing the spectral flow~\eqref{bkgd specflow} that relates NS and R boundary conditions.

This set of asymptotically $AdS_3\times \bS^3$ supergravity solutions for the elliptical profiles~\eqref{Fellipse} corresponds to a particular set of coherent states of the holographically dual CFT built out of the angular momentum eigenstates
\be
\label{ellipse state}
\ket{N_+,N_-}=
\Bigl(\ket{++}^{~}_k\Bigr)^{N_+} 
\Bigl(\ket{--}^{~}_k\Bigr)^{N_-} 
~~,~~~~
k(N_++N_-)=\none\nfive \equiv N~.
\ee
The ellipse solution in supergravity with parameters $a_1$, $a_2$ is dual to the state~\cite{Kanitscheider:2006zf}
\be
\big|\{a_1,a_2\} \big\rangle = \Big( \frac{N}{Q_1Q_5} \Big)^{\frac N{2k}}
\sum_{n=0}^{N/k} \Bigl[\frac{(N/k)!}{n!((N/k)-n)!}\Bigr]^{\frac 12} \Big(\frac{a_1+a_2}2\Big)^{\frac Nk-n} \Big(\frac{a_1-a_2}2\Big)^n \,\Big|{\frac Nk-n,n}\Big\rangle
\ee
with
\be
Q_5=n_5 ~,~~~ Q_1=\frac{n_1 g_s^2}{V_4} ~.
\ee
There is a natural generalization involving parameters $a_3,a_4$ and  describes a state which also includes $\ket{+-}^{~}_k$ and $\ket{-+}^{~}_k$ modes, in which case the source profile extends out into the $x^3$-$x^4$ plane.  We expect that this further generalization can also be solved using the methods described here; indeed the expressions for $\H,\A,\K$ in terms of $\bA_m,\bB_m,f$ hold for general $a_i$, and solve the Laplace equation away from sources.
%


\subsubsection{Elliptical Coulomb branch array}
\label{sec:ellip-CB}

We can now take the quantities $\H$, $\K$, $\A$, $\B$ of the elliptical Lunin-Mathur geometry, add a flat $\bR_t\times \bS^1_y\,$ to obtain a (10+2)-dimensional configuration, and perform null gauging as described in general in section \ref{sec:null-gaug-gen} to recover a 9+1d geometry with a linear dilaton throat and an $AdS_3\times\bS^3$ cap. As a first example, we now take the ellipse with $k=1$ and gauge $\xi_1 \,=\, 2\partial_{\hat u} \,,$~~$\xi_2 \,=\, 2\partial_{\hat v}\,$ as in Eq.~\eqref{eq:kv-1}.

The procedure parallels the discussion around Eqs.~\eq{thetasigma to KAB}--\eq{dil-circ-array}. Evaluating these general expressions on the elliptical solution data \eq{LMellipse}, we obtain the downstairs fields of the elliptical Coulomb branch configuration studied in~\cite{MariosPetropoulos:2005rtq},
\be
\label{dsperp-ell}
ds^2 = -dt^2 + dy^2 + \H\,  d\xx\tight\cdot d\xx + ds^2_\cM
~~,~~~~
B = \C_{ij} \, dx^i\wedge dx^j  \;, \quad e^{2\Phi} = \H \;,
\ee
where the smeared array of sources is now determined by the $\H$ given in Eq.~\eqref{LMellipse}.

\subsubsection{Elliptical supertubes from null gauging}
\label{sec:ellip-sup-ng}

Following the general procedure outlined in section~\ref{sec:STfromCoulomb}, the elliptical supertube is obtained by tilting the gauged null isometry into the $v,u=t\pm y$ directions.  One obtains the effective theory~\eqref{eq:Seff-suptube} with the specific harmonic forms/functions given in~\eqref{LMellipse}. As described below \eq{dil-suptube}, the downstairs solution is the elliptical supertube with linear dilaton asymptotics.

We will be particularly interested in the degeneration limit $a_2\to 0$ where the supertube strands collide with one another.  To this end let us consider the vicinity of the semi-minor axis by scaling 
$a_2\to \lam a_2,r\to \lam r, \phi\to \pm\frac\pi2+\lam\varphi$; in this limit the metric reduces to 
\begin{align}
ds^2 &= \frac{\lam\;\!\nfive}{a_1\sqrt{r^2+a_2^{\;\! 2}}} \bigg[\big(r^2+a_2^{\;\! 2}\cos^2\!\theta\big)\Big(\frac{-dt^2+dy^2}{(kR_y)^2} \Big) 
+ \big(r^2+a_2^{\;\! 2}\big) \Big(d\theta^2 + \cos^2\!\theta\, d\psi^2\Big) 
\\
&\hskip 3cm
+ dr^2 + a_1^{\;\! 2} d\chi^2 + 2a_1 a_2\sin\theta\,\frac{dt\,d\chi}{k R_y} + 2a_2\sqrt{r^2+a_2^{\;\! 2}}\,\cos^2\!\theta \,\frac{dy\,d\psi}{k R_y} \bigg]
\nn
\end{align}
where we have defined $\chi = \varphi\sin\theta$.  The geometry along $r,y,\theta,\psi$ can be written as
\be
\label{rythetapsi}
\frac{\lam\;\!\nfive}{a_1\sqrt{r^2+a_2^{\;\! 2}}} \bigg[
dr^2 + r^2\Big(\frac{dy}{kR_y}\Big)^2
+ (r^2+a_2^{\;\! 2}) \bigg(d\theta^2 + \cos^2\!\theta \Big(d\psi + \frac{a_2}{\sqrt{r^2+a_2^{\;\! 2}}} \frac{dy}{k R_y}  \Big)^2\bigg)
\bigg]
\ee
and shows that the degenerations are smooth at the supertube locus at $r=0,\theta=\frac\pi2$, apart from the usual $\bZ_k$ orbifold singularity that results from the supertube profile $\F(v)$ tracing the ellipse $k$ times.  
In string theory, this conical defect is not singular, as each winding is separated from its neighbors along the T-dual $\ytil$ circle by a stringy amount, as one sees in figure~\ref{fig:EllipseST}; indeed, the analysis of D-brane probes of the circular supertube in~\cite{Martinec:2019wzw} showed that the fractional branes that wrap the orbifold fixed point are much heavier than fundamental strings.  On the other hand, as the ellipse degenerates in the limit $\lambda\to0$, true vanishing cycles develop which are degenerate even in string theory~-- the D-branes wrapping them become massless in the limit.
The two-cycle in question is parametrized by the semi-minor axis of the ellipse parametrized by $\theta$ (with $\phi=\pm\frac\pi2$), which is the polar direction of a two-sphere whose azimuthal direction is the linear combination of $y,\psi$ in the last term in~\eqref{rythetapsi}.   The size of this $\bS^2$ vanishes as $\lam\to0$, and as we will show in section~\ref{subsec:DBI}, the vanishing of this cycle is not resolved by stringy $\alpha'$ effects; rather, D-branes wrapping this cycle become massless as $\lam\to 0$ and string perturbation theory breaks down.

\subsubsection{Linearizing elliptical configurations around circular configurations}

If we set $a_1=a_2$, the elliptical coulomb branch configuration and the elliptical supertubes reduce to their circular counterparts, discussed in sections \ref{sec:circ-CB} and  \ref{sec:circ-sup}  respectively.
Using the parameter $\epsilon$ defined in Eq.\;\eqref{eq:epsilon-def}, one can linearize elliptical configurations around their circular counterparts. 
This will connect with the next section where we will discuss vertex operators, including those that describe the deformation from circular configurations to the corresponding elliptical configurations.

Linearizing the upstairs elliptical supertube in $\epsilon$, $G = G^{(0)}+ \epsilon \;\! G^{(1)}+\cO(\epsilon^2)$, the first-order metric perturbation takes the form (anticipating that we will choose the gauge $\tau=\sigma=0$, we will suppress terms proportional to $d\tau$ and $d\sigma$ here and in the following few equations, to avoid unnecessary clutter)
\begin{align}
\label{deltaG}
G^{(1)} &\;\!=\; \nfive \frac{a^2}{r^2+a^2} \Bigl[ \cos 2\phi \,d\theta^2 + \frac14 \cos 2\phi\,(\sin2\theta)^2 \bigl(d\psi^2-d\phi^2\bigr)
-\sin2\phi\,\sin2\theta\,d\phi\,d\theta \Bigr] ~.
\end{align}
In this linearized expression, there is no distinction between the spherical bipolar coordinates \eq{sphericalbipolar} and the elliptical bipolars \eq{ellbip}. For concreteness let us take spherical bipolar coordinates \eq{sphericalbipolar}. Furthermore, since we are suppressing terms proportional to $d\tau$ and/or $d\sigma$, this expression is valid for $\phi,\psi$ being either R-R or NS-NS coordinates, due to the relation \eq{bkgd specflow}. For comparison with the following section, let us work with the NS-NS sector solution.

To obtain the linearized elliptical Coulomb branch configuration downstairs, we gauge the appropriate NS-NS Killing vectors, \eq{xione coulomb}. Expanding similarly $\theta_{1,2}=\theta_{1,2}^{(0)}+\epsilon\;\! \theta^{(1)}_{1,2}+\cO(\epsilon^2)$ and  $\Sigma=\Sigma^{(0)}+\epsilon\;\! \Sigma^{(1)}+\cO(\epsilon^2)$, we obtain (we return to suppressing $\NS$ subscripts/superscripts)\\
\begin{align}
\label{linear ellipse thetas}
\theta_1^{(1)} &\;=~ \xi_1^i G_{ij}^{(1)} d\varphi^j ~=\; -\nfive\frac{a^2\sin2\theta}{4(r^2+a^2)}\Bigl( 2\sin2\phi\,d\theta 
+\cos2\phi \,\sin2\theta(d\phi+d\psi) \Bigr),
\nn\\
\theta_2^{(1)} &\;=\; -\xi_2^i G_{ij}^{(1)} d\varphi^j \;=\; \nfive\frac{a^2\sin2\theta}{4(r^2+a^2)}\Bigl( 2\sin2\phi\,d\theta 
+\cos2\phi \,\sin2\theta(d\phi-d\psi) \Bigr),
\\
\Sigma^{(1)} &\;=\; -\frac12\xi_1^i G_{ij}^{(1)} \xi_2^j \;=\; \nfive\frac{a^2}{4(r^2+a^2)} \cos2\phi\,(\sin2\theta)^2 \;,
\nn
\end{align}
again suppressing terms proportional to $d\tau$, $d\sigma$.
We will discuss the corresponding vertex operators around Eq.~\eqref{ellipsevertex} below.
Upon performing null gauging, one obtains the expansion of the downstairs elliptical Coulomb branch configuration \eq{dsperp-ell} around the corresponding circular Coulomb branch configuration. Explicitly, choosing the gauge $\tau=\sigma=0$ and expanding the downstairs metric as $G_{\text{d}} = G_{\text{d}}^{(0)}+ \epsilon\:\!  G_{\text{d}}^{(1)}+\cO(\epsilon^2)$, we obtain
\begin{align}
\begin{aligned}
G_{\text{d}}^{(1)} &= \nfive \left[ \frac{a^2\cos 2\phi}{r^2+a^2}  d\theta^2 
- \frac{\sin 2\theta \sin 2\phi }{\Sigma_0}d\theta d\phi 
+\frac{(\sin 2\theta)^2\cos 2\phi}{4 a^2(r^2+a^2) \Sigma_0^{\;\! 2}} \Big( r^4 d\psi^2 - (r^2+a^2)^2 d\phi^2 \Big)\right].
\end{aligned}
\end{align}
Note that $G_{\text{d}}^{(1)}$ is singular as $\Sigma_0\to0$, with the last term going as $1/\Sigma_0^{\;\! 2}$. However this singularity is an artifact of the fact that we have made an expansion around the circular profile, and at the source the ansatz quantities \eq{roundST} are singular (even though the full metric \eq{eq:metric-0} is smooth).
The perturbation is trying to move the location of the source by changing the source profile function, and so successive orders in perturbation theory involve higher and higher powers of the unperturbed harmonic function.  However in the full ellipse configuration, the successively larger inverse powers of $\Sigma_0$ sum up to a smooth background where the singularity of the ansatz quantities \eq{LMellipse} is now at the elliptical source locus.  More generally, all the Lunin-Mathur geometries~\eqref{LMgeom}--\eqref{greensfn} are nonsingular, as long as the source profile $\F$ does not self-intersect.

Proceeding similarly, one obtains the linearized terms in the downstairs B-field. Furthermore, a similar analysis can be performed for the tilted null gauging for the elliptical supertube described above in section \ref{sec:ellip-sup-ng}.

\section{Linearized marginal perturbations}
\label{sec:CoulombfromLM}

We see that downstairs in 9+1d, the circular array of fivebranes on the Coulomb branch arises from gauging null isometries of a smooth geometry upstairs in 10+2d, namely the group manifold $\cG$ of~\eqref{Gupstairs2}.  A simple modification of the null isometries being gauged yields NS5-F1 and NS5-P supertubes.  Generalizing the background to an arbitrary Lunin-Mathur geometry, one can gauge its null isometries to obtain more general Coulomb branch configurations of fivebranes; tilting the gauging yields more general supertubes.  

BPS vertex operators that deform $AdS_3\times \bS^3$ represent the linearized marginal perturbations of the circular Coulomb branch array and the corresponding supertubes.  Upstairs, these add BPS supergravitational wave perturbations to $AdS_3\times \bS^3$ towards more general Lunin-Mathur geometries~\eqref{LMgeom}.  In what follows, we explore these connections.

\subsection{Linearized marginal deformations of the circular array}

Following \cite{Kutasov:1998zh,Argurio:2000tb} we can construct the BPS vertex operators in $AdS_3\times \bS^3$ that represent linearized deformations of the background that preserve the null Killing vectors being gauged.  By the state-operator correspondence of the worldsheet CFT, the vertex operators create on-shell string states.  The constraints imposed by the left and right BRST operators
\begin{align}
\label{BRST charges}
\cQ = \oint \big( cT \tight+ \gamma G \tight+ \tilde c\cJ \tight+ \tilde\gamma \lamb + \textit{ghosts} \big)
~~,~~~~
\bar\cQ = \oint \big( \bar c\bar T \tight+ \bar\gamma \bar G \tight+ \tilde {\bar c}\bar\cJ \tight+ \tilde{\bar\gamma} \bar\lamb + \textit{ghosts} \big) \;,
\end{align}
ensure that these excitation modes solve the appropriate linearized wave equation about the given background, and have physical polarizations. 
The null currents $\cJ,\bar\cJ$ being gauged have the form~\eqref{J gen ns}, and for their worldsheet superpartners one has
\be
\lamb = l_1 \;\! \psi^3_\sl + l_2 \;\! \psi^3_\su + l_3 \;\! \psi^t + l_4 \;\! \psi^y
~~,~~~~
\bar\lamb = r_1 \;\! \bar\psi^3_\sl + r_2\;\! \bar\psi^3_\su + r_3 \;\! \bar\psi^t + r_4 \;\! \bar\psi^y ~.
\ee
In particular, the $L_0+\bar L_0$ constraint arising from the zero modes of the stress tensor is the general linearized wave equation; and for the Ramond sector, its superpartner $G_0$ and $\bar G_0$ constraints act as generalized Dirac operators; the nonzero mode contributions restrict to physical polarization states.  
As described in~\cite{Martinec:2018nco}, the null gauging contributions to the BRST operator require the polarizations and momenta of the excitations to be transverse to the null isometries being gauged (polarizations and momenta along the null isometries are also transverse, but are trivial in the BRST cohomology; thus null gauging removes two dimensions from the target space). 
In the Ramond sector, these constraints act as null-projected gamma matrices ${\boldsymbol\ell}\cdot\Gamma$, $\,{\boldsymbol r}\cdot\Gamma$ (where $\Gamma$ are $O(10,2)$ gamma matrices) that restrict spinor polarizations on the left and right, respectively.

The generic form of a vertex operator in our background builds upon of center-of-mass wavefunctions
\be 
\cV_{\rm com} = \Phi_{\jsl\msl\bmsl}^{\sl} \Psi_{\jsu\msu\bmsu}^{\su}
e^{i w_y (t + \ytil )\Ry + n_y(t+ y)\Rytil}
\ee
which are eigenfunctions of the wave operator on $\sltwo\times\sutwo\times \bR_t\times\bS^1_y$.
Our group theory conventions for the $\sutwo$ and $\sltwo$ WZW models are summarized in Appendix~\ref{app:conventions}.  
Unitarity and normalizability restrict the range of allowed $j_{\sl}$ and $j_{\su}$ for affine $\sltwo$ and $\sutwo$ representations.  For the underlying bosonic current algebra vertex operators $\Phi$ and $\Psi$, one has the restrictions%
\footnote{Note: Our conventions for $j_\sl$ differ from~\cite{Kutasov:1998zh,Argurio:2000tb} and related works, where $j_\sl^{\rm there} = j_\sl^{\rm here}-1$.}
\begin{equation}
\label{jrange}
\frac12 < j_\sl < \frac{n_5+1}{2} 
~~,~~~~
0 \leq j_\su \le \frac{n_5}{2}-1 \;.
\end{equation}
The basic argument on the range of $\sltwo$ representations is that the $j_{\sl}=\half$ representation is at the bottom of a continuum $j_\sl = \half+i\lambda$ of radial wave states, and so the representation effectively has a ``constant'' wavefunction which is not normalizable.  The $\jsl=\half(\nfive+1)$ representation is obtained from $\jsl=\half$ by a combination of spectral flow and field identifications (to be discussed below in section~\ref{sec:FZZduality}), and so is also not allowed.  The absence of these representations is consistent with the fact that the two-point function either blows up or vanishes at these ends of the allowed range~\cite{Giveon:1999px}.

These zero mode operators are then decorated with polynomials $\cP,\bar\cP$ in the various worldsheet currents and their fermionic superpartners,
\be
\cP(\partial t, \partial y, \partial X^i, \jsl, \jsu;\psi^I)
\,\bar\cP(\bar \partial t, \bar\partial y, \bar\partial X^i, \bar\jsl, \bar\jsu; \bar\psi^I)
\,\cV_{\rm com}~,
\ee
(where $X^i$ are coordinates on $\bT^4$) to generate the polarization state of the corresponding string mode.
The analysis of the BRST constraints on the polarization state can be largely done separately in each worldsheet chirality, combining the results at the end to construct complete vertex operators (as usual, the center-of-mass wavefunction is common to both chiralities).

For the circular fivebrane array on the Coulomb branch, the currents being gauged are those in equation~\eq{currents-circ-CB}.
If we consider only the left-moving (holomorphic) half of the vertex operator, there are four possibilities in the BRST cohomology that preserve spacetime supersymmetry~\cite{Kutasov:1998zh,Chang:2014jta}:
\begin{align}
\label{vertexops}
\cV_j^{-} &= e^{-\varphi} \Phi^{\sl}_{j+1, m_{\sl}=j+1} ( \psi_{\su} \Psi_{j}^{\su})_{(j+1), m_\su= -(j+1)}  \nn\\
\cV_j^{+} &=e^{-\varphi} (\psi_{\sl} \Phi_{j+1}^{\sl})_{j, j} \Psi^{\su}_{j, - j}  \\
\cS_j^{\pm} & = e^{-(\varphi+\tilde\varphi)/2}\bigl(S^{\pm} \Phi_{j+1}^{\sl} \Psi_{j}^{\su}\bigr)_{j+1/2}  ~~, \nonumber
\end{align}
where the sign of $m_\su$ in the $\cV$'s is dictated by the relative sign between $\jj^3_\sl$ and $\jj^3_\su$ in the null current $\cJ$; also, $\varphi$ is the bosonized $\beta\gamma$ ghost of the worldsheet BRST formalism~-- the $\varphi$ charge indicates the ``picture'' of the vertex (here chosen to be $-1$ for NS, $-\half$ for R)~\cite{Friedan:1985ge} (similarly $\tilde\varphi$ arises from bosonizing the spinor ghosts for null gauging).  We have exhibited vertex operators involving the $\cD^+$ discrete series representations of $\sltwo$; there are similar charge conjugate vertex operators involving $\cD^-$ representations.  We defer the issue of identifications among these operators until section~\ref{sec:FZZduality}.  Some details about the construction of spacetime supersymmetry in the worldsheet formalism are presented in appendix~\ref{app:spacetimesusy}.

For the supergravitons $\cV$ of interest, in the $-1$ picture the polarization is carried by the choice of polarization of the fermion $\psi^a$ in $\sltwo$ or $\sutwo$, the superpartner of the bosonic currents $j^a$ of the WZW model on $\cG$; the BPS condition and the worldsheet BRST constraints allow the above possibilities, 
where the notation $(\psi_{\sl} \Phi_{j+1}^{\sl})_{j j}$ indicates the projection on the quantum numbers of the tensor product in $\sltwo$ of the spin one fermion $\psi$ and the spin $j\tight +1$ bosonic vertex operator $\Phi_{j+1}$ (similarly for $(\psi_{\su} \Psi_{j}^{\su})_{j+1, m= - (j+1)}$).
For supertubes, these are the only 1/2-BPS vertex operators; for the Coulomb branch there are twice as many supersymmetries (see Appendix~\ref{app:spacetimesusy}), and for instance supergraviton operators with fermion polarizations along $\bT^4$ and suitable quantum numbers in $\sltwo\tight\times\sutwo$ are also BPS~\cite{Kutasov:1998zh}.

For the R operators $\cS$, the $\pm$ superscript on the spin operator $S^\pm$ refers to the allowed polarizations on its $\bT^4$ component~-- the $\sltwo\times\sutwo$ part is fixed by the total spin being $j\tight+\hf$ in both factors (see~\cite{Kutasov:1998zh}, eq. 3.18).  Thus the polarization quantum number for the left chiral vertex operators are $\cV^\alpha$ and $\cS^A$, while for the right chiral vertex operators one has $\bar\cV^{\dot\beta}$ and $\bar\cS^{B}$.%
\footnote{For type IIB, recalling that the spinor labels refer to momentum modes on the T-dual IIA fivebrane; for type IIA, the right-moving part of the R vertex carries the spin polarization $\bar\cS^{\dot B}$.}

In the null gauging formalism, the vertex operators live in 10+2 dimensions; however, the worldsheet BRST constraints also impose the requirement that the vertex operators commute with an additional piece of the BRST charge that comes from gauging the null currents; this limits the polarizations and the center-of-mass momentum of the vertex operators to the 9+1d physical spacetime.  Satisfying this requirement is automatic for the above vertex operators~\eqref{vertexops} -- they don't change the asymptotic energy $E$ conjugate to the time $t$ and so keep the state on the BPS bound; since they don't involve $t$ or $y$, they are physical for both the circular Coulomb branch solution and the round supertube.
               
One can then construct the 1/2 BPS vertex operators by combining left- and right-moving contributions
\begin{align}
\label{BPSverts}
4~ {\mathrm{NS}\mhyphen \mathrm{NS}} &: \quad \cV_j^{\pm}\bar{\cV}_j^{\pm} \, \nn\\
4~ {\mathrm{R}\mhyphen \mathrm{R}} &: \quad \cS_j^{\pm}\bar{\cS}_j^{\pm} \,
\\
8~\text{fermions} &: \quad\cV_j^{\pm}\bar{\cS}_j^{\pm} \,
,\quad \cS_j^{\pm}\bar{\cV}_j^{\pm}
~. \nn
\end{align}
The resulting allowed vertex operators with $j=0,\half,\dots,\frac{\nfive}2-1$ yield $4(\nfive-1)$ marginal deformations of the Coulomb branch, corresponding to changing the relative positions of the fivebranes in the transverse $\bR^4$, as well as RR deformations which turn on gauge modes on the fivebranes.

\subsection{Marginal deformations of the round supertube}
\label{sec:ST defs}

For the circular supertube configuration described in section~\ref{sec:circ-sup}, the role of the vertex operators changes.  
Rather than describing the moduli space of fivebranes, the vertex operators change the distribution of quanta carried by the fivebrane. The background configuration corresponds to the state
\be
\Bigl(\ket{++}^{~}_k\Bigr)^{\none\nfive/k} ~.
\ee
The vertex operators that perturb this supertube to some other 1/2-BPS state~\eqref{halfBPS} must therefore transform some number of the background $\ket{++}^{~}_k$ modes into modes of other polarizations and mode numbers.

The null currents for the supertube, corresponding to gauging the tilted isometry~\eqref{tiltedKV}, are
\be
\label{gaugecurrents-sup}
\cJ = \jj^3_\sl + \jj^3_\su - \alphab ( \partial t - \partial y )
~~,~~~~
\bar\cJ = \bar \jj^3_\sl - \bar \jj^3_\su - \alphab ( \bar\partial t + \bar\partial y )\;,
\ee
with $\alphab=k\Ry$\,. Under the gauging of these currents, the operators~\eqref{vertexops} continue to be physical, since they are independent of $t,\ytil$.

The vertex operators~\eqref{vertexops} act in the following way on the excitation spectrum of the background, as one may deduce by matching quantum numbers:%
\footnote{One can also identify the corresponding 1/2-BPS vertex operators at the symmetric orbifold locus in the spacetime CFT moduli space, and show that they mediate the same transitions among 1/2-BPS ground states~\cite{Larsen:1999uk,Lunin:2001pw,Lunin:2001jy,Lunin:2002bj,Kanitscheider:2006zf,Taylor:2007hs} (see for instance the discussion in section 5 of~\cite{Bena:2016agb}).  These OPE coefficients are not renormalized across the moduli space~\cite{Gaberdiel:2007vu,Dabholkar:2007ey,Pakman:2007hn}.}
\begin{align}
\label{roundSTperts}
\cV_j^{++}\equiv\cV_j^{+}\bar \cV_j^{+}  &: ~~ (\lvert ++ \rangle_k)^{2j+1} ~\longrightarrow~ \lvert ++\rangle_{(2j+1)k}
\nn\\[.2cm]
\cV_j^{--}\equiv\cV_j^{-}\bar \cV_j^{-}  &:~~ (\lvert ++ \rangle_k)^{2j+1} ~\longrightarrow~ \lvert - -\rangle_{(2j+1)k}
\nn\\[.2cm]
\cV_j^{+-}\equiv\cV_j^{+}\bar \cV_j^{-}  &: ~~(\lvert ++ \rangle_k)^{2j+1} ~\longrightarrow~ \lvert +-\rangle_{(2j+1)k}
\\[.2cm]
\cV_j^{-+}\equiv\cV_j^{-}\bar \cV_j^{+}  &: ~~(\lvert ++ \rangle_k)^{2j+1} ~\longrightarrow~ \lvert - +\rangle_{(2j+1)k} ~~.
\nn
\end{align}
The RR operators build the bosonic fivebrane excitations $\ket{A B}_n$ with ``internal'' polarization; in particular
$(\cS^+\bar\cS^- - \cS^-\bar\cS^+)$ creates $|00\rangle_{2jk}$ modes, which play a prominent role in the superstratum literature~(see e.g.~\cite{Bena:2015bea,Bena:2016agb,Bena:2016ypk,Bena:2017geu,Bena:2017xbt,Bena:2017upb,Bena:2018bbd,Ceplak:2018pws,Heidmann:2019zws,Heidmann:2019xrd,Mayerson:2020tcl}).

These operators die exponentially at large radius, and so only deform the structure of the cap in the geometry; in particular they do not introduce or extract strings from the system, and so simply rearrange what's already there without changing the energy $E$ measured by asymptotic observers.

Note that the insertion of $\cV^{++}_0$ acts ``trivially".  This is due to the fact that this vertex operator is the zero mode of the dilaton, discussed in~\cite{Giveon:1998ns},
which indeed does not change the background. Rather, it just evaluates to a
c-number due to fact that the dilaton is a fixed scalar at the F1-NS5 source, effectively counting the number of background mode excitations (proportional to the central charge of the spacetime CFT in the $AdS_3$ decoupling limit).  

Similarly, according to the map~\eqref{roundSTperts} the operator $\cV_0^{--}$ changes a background $\lvert ++ \rangle_k$ mode to a $\lvert - - \rangle_k$ mode; this is the perturbation that deforms the circular supertube source profile toward an ellipse, analyzed in section \ref{sec:ellipse} at the fully nonlinear level.  
The ellipse perturbation of the circular profile corresponds to the vertex operator
\be 
\label{ellipsevertex}
\cV^{--}_{j=0} =  \Bigl(\lambda\,\Phi^\sl_{111} \bigl(\psi_\su^-\bar\psi_\su^+\Psi^\su_{000}\bigr) + \lambda^*\,\Phi^\sl_{1,-1,-1} \bigl(\psi_\su^+\bar\psi_\su^-\Psi^\su_{000}\bigr) \Bigr).  
\ee

Written in terms of Euler angles, the $\sltwo$ wavefunction $\Phi^\sl_{111}$ is 
\be
\Phi^\sl_{111} = e^{-2i\tau}\frac{a^2}{r^2+a^2} = e^{-2i\tau}\frac{1}{\cosh^2\rho} ~,
\ee
while the $\sutwo$ center-of-mass contribution is trivial.
We then transform the vertex operator from the $-1$ picture to the zero picture, which roughly speaking turns the fermions $\psi^\pm_\su$ carrying the polarization information into the corresponding currents $\jj^\pm_\su$ of~\eqref{eq:J3su-app}, again written in Euler angles; putting it all together, one finds the first-order metric perturbation to the upstairs geometry given in Eq.\;\eq{deltaG} above.
One can thus check that the first order deformation of the gauged WZW model~\eqref{thetasigma to KAB},~\eqref{roundST} agrees with the expansion of the ellipse geometry~\eqref{LMellipse} to first order in $\epsilon$.

One can also consider the action of spectral flow in the worldsheet $\sltwo$ and $\sutwo$ current algebras on these vertex operators.
These spectral flows generate the shifts%
\footnote{Left and right spectral flows for $\sltwo$ are identical because we work on the universal cover.}
\begin{align}
\label{specflow}
m_{\su} \rightarrow m_{\su} + \frac{n_5}{2}w_{\su} 
~~&,~~~~
m_{\sl} \rightarrow m_{\sl} + \frac{n_5}{2}w_{\sl} 
\nn\\[.3cm]
\bar m_{\su} \rightarrow \bar m_{\su} - \frac{n_5}{2}\bar w_{\su} 
~~&,~~~~
\bar m_{\sl} \rightarrow \bar m_{\sl} + \frac{n_5}{2} w_{\sl} ~.
\end{align}
The null constraints continue to be satisfied for $w_\su=\bar w_\su=w_\sl\equiv w$.
On the Coulomb branch, this shift amounts to a large gauge transformation, and so does not yield a physically distinct vertex operator; the moduli space of marginal deformations of the background is indeed $4(\nfive-1)$ dimensional, corresponding to shifting the relative positions of the fivebranes in the transverse $\bR^4$.
Thus while there are many, many Lunin-Mathur geometries, corresponding to all the deformations of the Fourier modes of the profile functions, there is only a $4(\nfive-1)$ parameter family of distinct Coulomb branch geometries.  The different choices of Lunin-Mathur geometry do however yield different expressions for the background geometry on the Coulomb branch; it is natural to suppose that these are related to one another by (perhaps stringy) diffeomorphisms, though this point is perhaps worth further investigation.  

For the round supertube, the shift~\eqref{specflow} is no longer a gauge transformation, and so yields a physically distinct vertex operator generating new marginal deformations.  We expect that the $\cV$ vertex operators implement the transformation of the mode excitations
\begin{equation}
\bigl(\lvert ++ \rangle_{k}\bigr)^{2j+1+ w n_5} ~\longrightarrow~ \lvert \alpha\dot\beta
\rangle_{(2j+1+w n_5)k}
\end{equation}
while the RR operators $\cS$ generate $\ket{A B}_{(2j+w n_5)k}$ modes, which are excitations of the NS5-brane tensor gauge field and internal scalar.

Ordinarily, the exactly marginal deformations of a string background are limited in number, however for the round supertube we are looking for an essentially arbitrary number of such perturbations when $\none$ is asymptotically large, which \naive ly seems rather peculiar.  It is here that the two-time nature of the geometry upstairs in the parent group manifold enters in a useful way; one can have vertex operators that cost zero energy in the asymptotic time $t$ while having nonzero energy in the cap time $\tau$ that allows a wide variety of cap deformations to be on-shell.  But because they cost zero energy in the asymptotic time $t$, they represent exact moduli of the system, at least classically.  As mentioned in section \ref{sec:fivebranes}, quantum mechanically the moduli space is compact and so rather than consisting of a continuum of deformations, quantization of the moduli space leads to a discretuum of ${ \exp}[2\pi\sqrt{2\none\nfive}\,]$ states~\cite{Rychkov:2005ji}.  The vertex operators~\eqref{BPSverts} are moving us around in this discretuum.

We can unwind $w_{\sl}$ back to the zero winding sector using gauge spectral flow~\cite{Martinec:2018nco}, which shifts the winding to the $y$ circle.  Consider the vertex operators~\eqref{BPSverts} multiplied by massless exponentials of $t,\ytil$
\be
\exp\bigl[- iw_y(t - \ytil )\bigr]  ~~;
\ee
then $q$ units of this spectral flow shift a string's zero mode quantum numbers by
\begin{align}
\label{LargeGaugeTransf}
\wsl \to \wsl + q 
~,~~~
\wsu  \to \wsu  + q
~,~~~
\bwsu \to \bwsu - q 
~,~~~
w_y\to w_y + kq ~~.
\end{align}
If we start with the vertex operators~\eqref{BPSverts} and perform a spectral flow with $q=w_\sl\equiv w$, we shift the state to an equivalent representative in the zero winding sector in $\sltwo$.  The winding is then shifted to the $y$ circle, with $\delta w_y= kq$; at the same time one unwinds the $w_\su$ winding of BPS vertex operators generated by~\eqref{specflow} (which satisfy the null constraints when $\bar w_\su= - w_\su=w$).%
\footnote{Note that the above shifts are for the NS5-F1 supertube; in the NS5-P supertube obtained by T-duality on the $y$ circle, the spectral flow now acts to shift the $\ytil$ circle momentum quantum $n_\ytil=w_y$, as one expects from the fact that the vertex operators act in this duality frame to change the moding and polarization of momentum excitations on the fivebranes.}  
In a sense, these two ``pictures'' of the vertex operators~-- one with winding along the $\sltwo$ angular direction, and one with winding along the $y$ circle~-- represent the view of the string as seen by observers in the $AdS_3\times\bS^3$ cap, and by asymptotic observers.

How does the accounting of $\sutwo$ spin work now, since these different pictures seem to differ in the spin carried on the vertex operator by $w \nfive/2$?  The physical conserved charges carried by F1 string vertex operators are the charges associated to all the currents that commute with the gauge currents (modulo gauge transformations of course), see~\cite{Martinec:2019wzw}.  
These charges are independent of the gauge spectral flow ``picture'' we use to describe a given string state.  
For instance the currents%
\footnote{Our T-duality conventions are $(\partial y,\bar\partial y) = (-\partial\ytil,\bar\partial\ytil)$.} 
\be
\cL = - k \jj^3_\su + \nfive \partial(y/\Ry)
~,~~~~~
\bar\cL = k \bar \jj^3_\su + \nfive \bar\partial(y/\Ry)
\ee
measure the {\it physical} spin of the string; when we do gauge spectral flow, we change the left and right $\sutwo$ spins by $\nfive/2$ and the winding on the $y$ circle by $k$ units, and the two effects cancel.

The fact that $y$ circle winding comes in multiples of $k$ under gauge spectral flow suggests that there are additional perturbations corresponding to the remaining winding numbers along~$y$.  
We can also allow $\omega_y \neq p k$, \ie\ consider vertex
operators such as
\begin{equation}
\label{general winding vertex}
\Phi_{j+1}^{\sl} (\bar\psi_{\su}\psi_{\su} \Psi_{j}^{\su})_{j+1} \, e^{i w_y (t +  \ytil )}
~ ,
\end{equation}
which implement
\begin{equation}
\label{generalshift}
\bigl(\lvert + + \rangle_k\bigr)^{2j+1} ~\longrightarrow~ \lvert - - \rangle_{(2j+1)k +
  w_y n_5} ~ .
\end{equation}
The $y$ circle winding of such operators cannot be entirely transferred to the cap picture unless $w_y$ is a multiple of $k\nfive$.  The subscript on the fivebrane modes effectively is an accounting of little string winding number, which in the round supertube background comes in such multiples.  More generally, each fundamental string winding corresponds to $\nfive$ units of little string winding, so we can only add multiples of $\nfive$ compared to the $w_y=0$ sector when creating F1 strings winding the $y$ circle.  We have assumed that $k$ and $\nfive$ are relatively prime so that the supertube binds all the fivebranes together;%
\footnote{Otherwise one has $gcd(k,\nfive)$ distinct fivebrane strands, and a moduli space of the strands which has coincidence limits where perturbation theory breaks down.  Actually, since the fivebranes are compact, one must quantize this moduli space, and the wavefunction has support in this ill-understood region.}
then one can check that one can generate excitations $\lvert\alpha\dot\beta\rangle_n$ of any mode number $n$ through a suitable choice of $j,k,w_y$, together with a suitable polarization choice for the vertex (though as noted in~\cite{Argurio:2000tb} the range $j=0,\frac12,...,\frac \nfive2-1$ means that states with $(2j\tight+1)k\tight+w_y\nfive=\nfive\tight-1$ mod $\nfive$ are absent from the spectrum).

One might think that, because the winding cannot be shifted away via a gauge transformation, these vertex operators do not correspond to marginal deformations.  But as discussed in~\cite{Martinec:2018nco}, background string winding energy has been subtracted from the energy budget seen by the worldsheet description, while the energy cost of perturbative strings is explicit.  A consistent accounting procedure shows that all the operators~\eqref{general winding vertex} cost zero energy at fixed total string winding charge $n_1$, and implement the transitions~\eqref{generalshift} among the 1/2-BPS states.

\subsection{1/4-BPS excitations}
\label{sec:quarterBPS}

We can also consider vertex operators with non-zero momentum $n_y$ in addition to winding $w_y$,
i.e. take
\begin{equation}
\label{Vpw}
\cP(\partial t, \partial y, \partial X^i, \jsl, \jsu;\psi)\, \Phi_{j+1,m_\sl,j+1}^{\sl} (\bar\psi_{\su} \Psi_{j}^{\su})_{j+1,m_\su,j+1} 
e^{i w_y (t + \ytil )\Ry + n_y(t+ y)\Rytil}
 ~,
\end{equation}
where again $\cP$ is a polynomial in the indicated fields and their derivatives.
The right-moving structure is kept the same as the vertex operators~\eqref{vertexops}.
Therefore, one can solve the right null and Virasoro zero-mode constraints
\begin{align}
0 &= 2\bmsl -2-2\bmsu - \bwsu\nfive - k\Ry\big( E+P_{y,R} \big)  ~,
\nn\\[.2cm]
0 &= -E^2+P_{y,R}^2 + 4\bmsu\bwsu+\nfive\bwsu^2 
\end{align}
in the same way as for 1/2-BPS operators, by setting 
$E\tight=-P_{y,R} \tight= -\frac{n_y}{\Ry}\tight+w_y\Ry$ and 
$\bmsl\tight=j\tight+1,\bmsu\tight=j,\bwsu\tight=0$.
The left Virasoro constraint then in general requires
left-moving oscillator excitations to compensate the cross term $n_y w_y$,
\begin{equation}
-w_y n_y = \msu\wsu+\frac\nfive4 \wsu^2 + N_L - \frac12 ~.
\end{equation}
The left null constraint then imposes~\cite{Martinec:2017ztd,Martinec:2018nco}
\be
\label{axialnull}
-kn_y = \msl + \msu + \frac\nfive2\wsu + J^3(\cP) ~.
\ee
We thus construct three-charge states of type $\lvert - - \rangle_{2jk + w_y n_5}$ 
(or similarly other polarizations) carrying momentum charge in addition to F1 and NS5 charge, perturbatively about the round supertube. 
In particular, the massless RR operators with $w_y\tight=w_\su\tight=0$ are supergravity excitations that just add cap momentum $\msl-\bmsl$ and angular momentum $\msu$ in multiples of $k$ times the basic unit of momentum quantization in the cap,%
\footnote{As discussed in~\cite{Martinec:2018nco}, momentum quantization in the $\sltwo$ factor of the sigma model upstairs is related to momentum quantization on the $y$ circle by a factor $k$, due to the null constraints of the round supertube built on a background of $\ket{++}_k$ modes.  This is how the worldsheet theory sees momentum fractionation of the spacetime CFT excitations.  This feature explains the factor of $k$ between the quantization of $\msl$ and $n_y$ in the constraint~\eqref{axialnull}.}
while turning background $\ket{++}$ modes into an $\ket{AB}$ mode.  It is natural to identify a subset of these modes with those that deform the round supertube into a {\it superstratum} built on excitations~\cite{Bena:2015bea,Bena:2016ypk,Bena:2017xbt} 
\be
\big({\mathfrak J}_{-1}^+\big)^m\big({\mathfrak L}^{~}_{-1} - {\mathfrak J}_{-1}^3\big)^{n} \big|AB\big\rangle_{2jk}  
\ee
where ${\mathfrak L}_{-1} , {\mathfrak J}_{-1}^3, {\mathfrak J}_{-1}^+$ are generators of the superconformal algebra of the spacetime CFT.  We note that there are more operators in supergravity than just these, because $\msl$ and $\msu$ need not separately be multiples of $k$, only their sum; these are ``fractionated''  modes of the sort discussed in~\cite{Bena:2016agb}.

While some of the 1/4-BPS operators such as those just described can be identified with supergravity modes, it seems clear that there are many more modes to be had by using oscillator excitations of the strings.
Naively one might anticipate that the elliptic genus of such strings counts a large BPS entropy of three-charge states along the lines of~\cite{Giveon:2015raa}, which one might compare to the BPS entropy of the spacetime CFT.  Indeed, an approach to estimating this entropy~\cite{Bena:2008nh,Bena:2008dw} investigated supertube probes placed in a region of deep redshift, and found a result $S\sim Q^{5/4}$, less than the BTZ entropy $S\sim Q^{3/2}$ but more than the entropy $S\sim Q$ of BPS fundamental strings in global $AdS_3$.
However, many of these states are expected to lift away from the BPS bound due to interactions; an example of this phenomenon was analyzed in conformal perturbation theory about the symmetric orbifold point in the onebrane-fivebrane moduli space in~\cite{Hampton:2018ygz}.  
It would be useful to estimate the entropy of the states generated by vertex operators of the sort~\eqref{Vpw} and compare with~\cite{Bena:2008nh,Bena:2008dw} as well as a recent analysis of superstratum configurations~\cite{Shigemori:2019orj} that arrived at a similar entropy count.  Note that there are eight rather than four physical bosonic oscillator polarizations of string excitation, because the strings are not absorbed into the fivebranes where the polarizations would be restricted to $\bT^4$.  If the polarizations transverse to the fivebranes are highly excited then the strings are puffing out in the transverse direction and this takes us further from the Higgs branch, apart from the occasional self-intersections of the supertube source profile discussed below; if only the polarizations along the $\bT^4$ are excited, then we stay close to the Higgs branch -- the string's wavefunction in the transverse space is the same as the unexcited BPS state.

\section{Generalized FZZ duality}
\label{sec:FZZduality}


The naive sigma model action in a background Lunin-Mathur geometry describes a supergravity limit in which the fivebrane source has been smeared over a contour as in equation~\eqref{greensfn}.  However, string theory can probe the background in ways that extend beyond supergravity.  One of the stringy features of the background is known as FZZ duality~\cite{FZZref,Kazakov:2000pm}, an example of the Calabi-Yau/Landau-Ginsburg correspondence~\cite{Martinec:1988zu,Vafa:1988uu}.   In its original form, the duality relates the $\sltwo/\uone$ gauged WZW model to $\cN=2$ Liouville theory;
in its non-supersymmetric incarnation, it relates the bosonic coset model to Sine-Liouville theory.
Each perturbation of the coset geometry has an FZZ dual description, and while the geometrical perturbations might not resolve individual fivebrane sources in the background, the same perturbations in the dual description do contain such information about the fivebranes' locations.  We review the structure of FZZ duality in this section, and apply it in the following section to the localization of fivebrane sources in the Lunin-Mathur backgrounds.


\subsection{Bosonic FZZ duality}

The origin of FZZ duality lies in the structure of $\sltwo$ current algebra representation theory (see for instance~\cite{Maldacena:2000hw}).  The discrete series representations of the bosonic $\sltwo$ theory correspond to highest weight operators $\Phi_{jm\bar m}^{w,\epsilon}$, where the $\epsilon=\pm$ superscript denotes the choice of representation $\cD^\pm$ and $w$ is a spectral flow quantum number.  The compact $U(1)$ subalgebra can be gauged, yielding a set of coset theory (also known as parafermion) operators $V_{jm}^{\epsilon}$; one has the relation
\begin{align}
\label{bosonpf}
\Phi_{jm}^{w,\epsilon} &= V_{jm}^{\epsilon} \exp\Bigl[i\frac{2}{\sqrt{\nfivehat}} \Bigl(m+\frac{\nfivehat}2 w\Bigr) Y\Bigr] \;,
\end{align}
with the conformal dimensions 
\be
h[\Phi_{jm}^{w,\epsilon}] = -\frac{j(j-1)}{\nfive} -mw-\frac{\nfivehat w^2}{4} \;.
\ee
Here, $\nfivehat=\nfive+2$, and $Y$ bosonizes the current $j^3$ of the bosonic affine algebra (that is gauged to generate the coset), and we have suppressed the antiholomorphic structure.  Where possible, we will also suppress this $\pm$ superscript with the understanding that it is implicit in the range of $m$.

\begin{figure}[ht]
\centering
  \begin{subfigure}[b]{0.35\textwidth}
  \hskip .5cm
    \includegraphics[width=\textwidth]{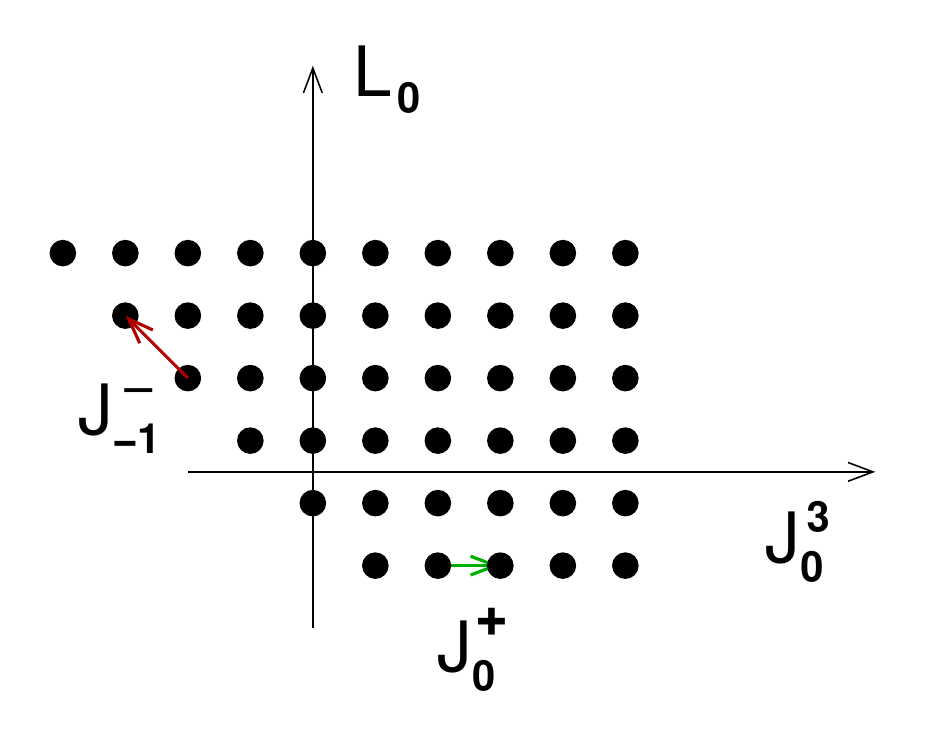}
    \caption{ }
    \label{fig:geodesics}
  \end{subfigure}
\qquad\qquad
  \begin{subfigure}[b]{0.35\textwidth}
      \hskip -.5cm
    \includegraphics[width=\textwidth]{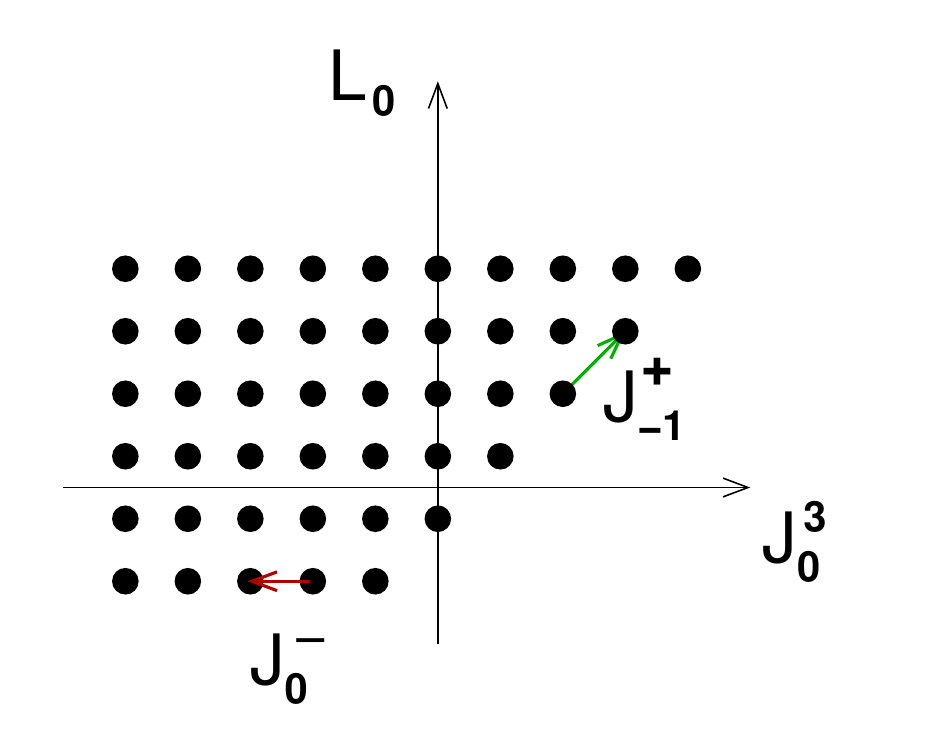}
    \caption{ }
    \label{fig:wound_string}
  \end{subfigure}
\caption{\it Affine $\sltwo$ representations:  (a) $\cD^+$, and (b) $\cD^-$.  A unit of spectral flow tips one into the other.}
\label{fig:Dpm}
\end{figure}

FZZ duality in bosonic $\sltwo$~\cite{Giveon:1999px,Maldacena:2000hw,Giveon:2015cma,Giveon:2016dxe} results from the spectral flow
\be
j_{n}^{3} \longrightarrow j_n^3+\frac\nfivehat2 w\delta_{n,0} ~~,~~~~ j_{n}^{\pm} \longrightarrow j_{n\mp w}^\pm~.
\ee
The affine weight diagrams of $\cD^\pm$ (see figure~\ref{fig:Dpm}) rotate into one another under a unit of spectral flow.  One has the correspondence of highest weights
\be
\label{affine-pf hw}
\Phi_{jj}^{w=0,\epsilon=+} = \Phi_{{\frac{\nfivehat}2}-j,j-{\frac{\nfivehat}2}}^{w=1,\epsilon=-}
~~\Longleftrightarrow~~
V_{jj}^{\epsilon=+} = V_{{\frac{\nfivehat}2}-j,{j-\frac{\nfivehat}2}}^{\epsilon=-,w=1} ~,
\ee
indeed the conformal dimensions and $j^3$ charges match.  On the left side of the correspondence, one can move away from the highest weight by the action of the zero mode operator $j^+_0$, which flows to the action of $j^+_{-1}$ on the other side; similarly, $j^-_{-1}$ flows to $j^-_0$.  One thus has the descendant relation
\begin{align}
\label{affine-pf}
& \Phi_{jm}^{w=0,\epsilon=+} = (j_0^{+})^{m-j}\, \Phi_{jj}^{w=0,\epsilon=+} = (j_{-1}^+)^{m-j}\,\Phi_{{\frac{\nfivehat}2}-j,j-{\frac{\nfivehat}2}}^{w=1,\epsilon=-}~,
\nn\\[.3cm]
&\hskip .5cm 
(j_{-1}^-)^{m-j}\, \Phi_{jm}^{w=0,\epsilon=+} =  (j_{0}^-)^{m-j}\,\Phi_{{\frac{\nfivehat}2}-j,j-{\frac{\nfivehat}2}}^{w=1,\epsilon=-} ~.
\end{align}
One of these dual states is not living in the zero mode representation only, but has some oscillator excitations above it.  One particular example of the duality 
\be
j_{-1}^-\Phi_{j,j}^{w=0,\epsilon=+} = \Phi_{{\frac{\nfivehat}2}-j,{j-1-\frac{\nfivehat}2}}^{w=1,\epsilon=-}
\ee
relates a graviton vertex on the left-hand side to a tachyon vertex on the right-hand side.  For $j=1$, both sides have zero $j^3$ charge and thus descend to operators in the coset theory.

\begin{figure}[ht]
\centering
  \begin{subfigure}[b]{0.25\textwidth}
    \includegraphics[width=\textwidth]{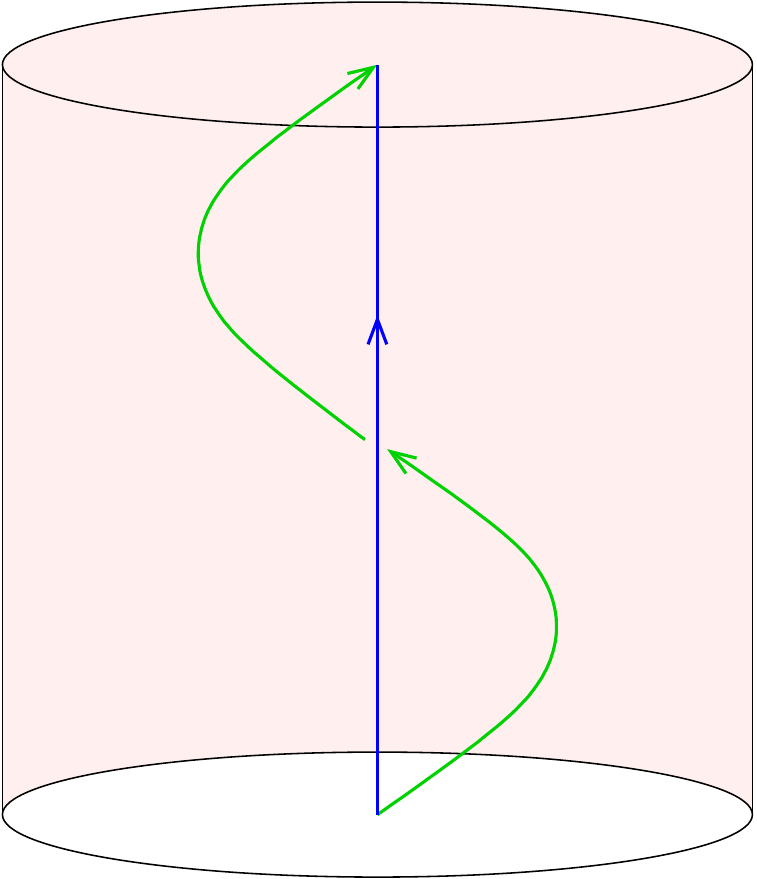}
    \caption{ }
    \label{fig:geodesics-2}
  \end{subfigure}
\qquad\qquad
  \begin{subfigure}[b]{0.25\textwidth}
      \hskip .5cm
    \includegraphics[width=\textwidth]{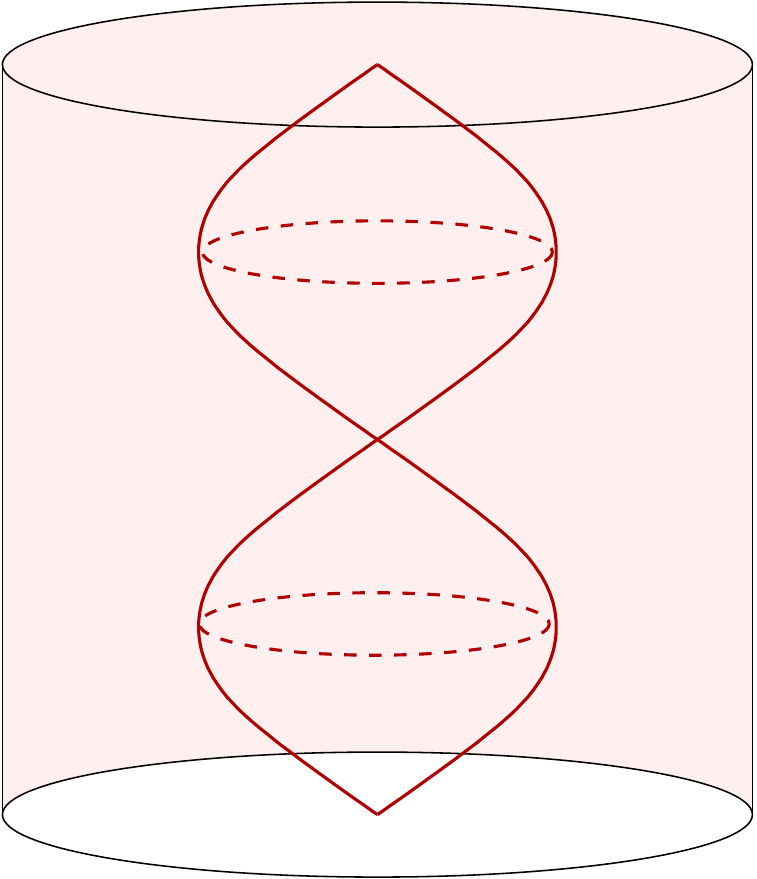}
    \caption{ }
    \label{fig:wound_string-2}
  \end{subfigure}
\caption{\it 
Classical string solutions in $AdS_3$: (a) highest weight states describe unexcited strings sitting at the origin (blue trajectory), with zero mode descendants describing strings oscillating about the origin having no oscillator excitations (green trajectory); (b) winding sectors describe strings that ``wind'' the origin, with zero mode descendants exciting a breathing mode oscillation (red trajectory).  FZZ duality relates such classical solutions to unwound strings with a coherent excitation of the lowest oscillator mode.
}
\label{fig:AdS3boundstrings}
\end{figure}

The duality applies to arbitrary $\sltwo$ descendant states.  Indeed, an analysis of the semiclassical limit provides an understanding of how it arises.  For strings on a group manifold, classical solutions factorize into a product of left- and right-moving group elements 
\be
\g(\xi) = \g_\ell(\xi_-) \g_r(\xi_+) 
\ee
where $\xi_\pm = \xi_0\pm\xi_1$ are the worldsheet coordinates.  These states should be thought of as coherent states of current algebra descendants such as~\eqref{affine-pf}.
For geodesics in $\sltwo$ oscillating about $\rho=0$, one has
\begin{align}
\label{gengeodesic0}
\g^{w=0}_{\alpha,\rho_+,\rho_-}(\xi) = \bigl(e^{\half\rho_-\,\sigma_1} \, e^{\frac i2\alpha \xi_- \sigma_3}\bigr)\bigl(e^{\frac i2\alpha \xi_+ \sigma_3} \, e^{-\half\rho_+\sigma_1}\bigr)
\end{align}
with $\rho_\pm = \rho^\smallfrown\pm\rho_\smallsmile$.
Recalling the Euler angle parametrization of $SU(1,1)$
\be
g = e^{\frac i2(\tau-\sigma)\sigma_3} \, e^{\rho\sigma_1} \, e^{\frac i2(\tau+\sigma)\sigma_3}
\ee
one sees that this classical solution of the WZW equations of motion describes unexcited strings whose center of mass travels an elliptical trajectory oscillating between inner radius $\rho_\smallsmile$ and outer radius $\rho^\smallfrown$
\be
\sinh^2\rho = \cos^2(\alpha\xi_0)\sinh^2\rho_\smallsmile+\sin^2(\alpha\xi_0)\sinh^2\rho^\smallfrown
\ee
(see~\cite{Maldacena:2000hw}; note we are working in $SU(1,1)$ rather than $\sltwo$ in order to diagonalize the compact generator).  
The quantum numbers of this motion are
\begin{align}
\cE &= \frac{\nfive}4\tr\bigl[\partial_\xi g \partial_\xi g^{-1}\bigr] = \frac{\nfive}2 \alpha^2
\\[.2cm]
\jj^3 &= -\frac{i\nfive}2 \tr\bigl[(\partial_\xi g)g^{-1}\sigma_3\bigr] = \nfive\alpha\cosh(\rho_-)
\nn\\[.2cm]
\bar \jj^3 &= -\frac{i\nfive}2 \tr\bigl[g^{-1}(\partial_\xi g)\sigma_3\bigr] = \nfive\alpha\cosh(\rho_+)  ~~;
\end{align}
thus one has $j\sim \frac\nfive2\alpha$, while $\rho_\pm$ code $m,\bar m$ (or rather coherent states thereof).  The spectral flow of this geodesic motion
\be
\g^w_{\alpha,\rho_+,\rho_-}(\xi) = e^{ \frac i2w\xi_-\sigma_3} \, \g^{w=0}_{\alpha,\rho_+,\rho_-}(\xi) \, e^{\frac i2w\xi_+\sigma_3}
\ee
describes a circular string that gyrates around the origin between the same two limits; see figure~\ref{fig:AdS3boundstrings}.

The FZZ duality~\eqref{affine-pf} in this semiclassical context relates this gyrating round string to a coherent oscillator excitation (of the first Fourier mode) on top of the geodesic motion~\eqref{gengeodesic0}.  The oscillating classical string solution of interest is
\begin{align}
\gosc^{\rm osc}_{\alpha,\rho_+,\rho_-}(\xi) &= \gosc^{\rm osc}_\ell(\xi_-) \gosc^{\rm osc}_r(\xi_+)
\nn\\[.3cm]
\gosc^{\rm osc}_\ell (\xi_-) &=  \exp\Bigl[\frac{\rho_-}2\bigl( e^{-i\xi_-} \sigma_+ + e^{i\xi_-} \sigma_-\bigr)\Bigr]
\cdot\exp\Bigl[\frac i2\alpha\,\xi_- \, \sigma_3 \Bigr]
\\[.3cm]
\gosc^{\rm osc}_r (\xi_+) &=  
\exp\Bigl[\frac i2\alpha\,\xi_+ \, \sigma_3 \Bigr]
\cdot\exp\Bigl[-\frac{\rho_+}2\bigl( e^{i\xi_+} \sigma_+ + e^{-i\xi_+} \sigma_-\bigr)\Bigr]
\nn
\end{align}
One finds that if one takes the conjugate state of $g_{\rm osc}$ by sending $\alpha\to-\alpha$ (the classical equivalent of sending $\cD^+$ to $\cD^-$), one arrives at a trajectory in the loop group of $\sltwo$ that is precisely the same as $\g^{w=-1}_{1-\alpha,\rho_+,\rho_-}$ :
\be
\label{semiclassicalFZZ}
\gosc^{\rm osc}_{-\alpha,\rho_+,\rho_-}(\xi)
= \g^{w=-1}_{1-\alpha,\rho_+,\rho_-}(\xi) ~~.
\ee
This equivalence is precisely the realization of FZZ duality on semiclassical coherent states!
Note that we can also spectral flow this relation.  The limit $\rho_\pm\to0$ recovers the primary highest weight states which sit at $\rho=0$ and travel up the $\tau$ direction.  Thus we see that at the classical level, FZZ duality is simply a global identification of coordinates in the loop group.

One can understand qualitatively the reason why the duality maps $j$ to $\half\nfive-j$ in the following way.  Consider the $\cD^-_j$ highest weight state with $m=-j$.  Classically, this is a string that sits at $\rho=0$ and simply moves up the time direction with momentum $j$~\cite{Maldacena:2000hw}.  Quantum mechanically, the wavefunctions for small $j$ are concentrated near $\rho=0$ out to a distance of order $\rads/j$.  Now consider the $\cD^+_j$ states with $w=1$.  Their behavior is affected by the ``Lorentz force'' 
\be
\label{lorentzforce}
F_\rho = H_{\rho\tau\sigma} \,\partial_0\tau \,\partial_1\sigma = H_{\rho\tau\sigma} \, j\, w
\ee
due to the background $B$-field.  The larger $j$ is, the larger the radial force trying to stretch the string.  At some point ({\ie\ when $j\sim \nfive/2$) the radial force trying to stretch the string compensates the tension trying to shrink it, and due to its reduced effective tension the string wavefunction spreads out radially on the scale $\rads/(\frac\nfive2-j)$.%
\footnote{This is also the mechanism that restricts the principal quantum number of the allowed bound states to $j\lesssim \frac\nfive2$; when the momentum in the $\tau$ direction is too large, the Lorentz force exceeds the tension, the circular string ceases to oscillate and simply expands radially to infinity.}
Thus the states have wavefunctions peaked around the same classical solution, with the same extent in space in addition to having the same quantum numbers and representation structure, providing overwhelming evidence for the proposal that they are in fact two dual descriptions of the same state.

\begin{figure}[h!]
\centering
\includegraphics[width=.6\textwidth]{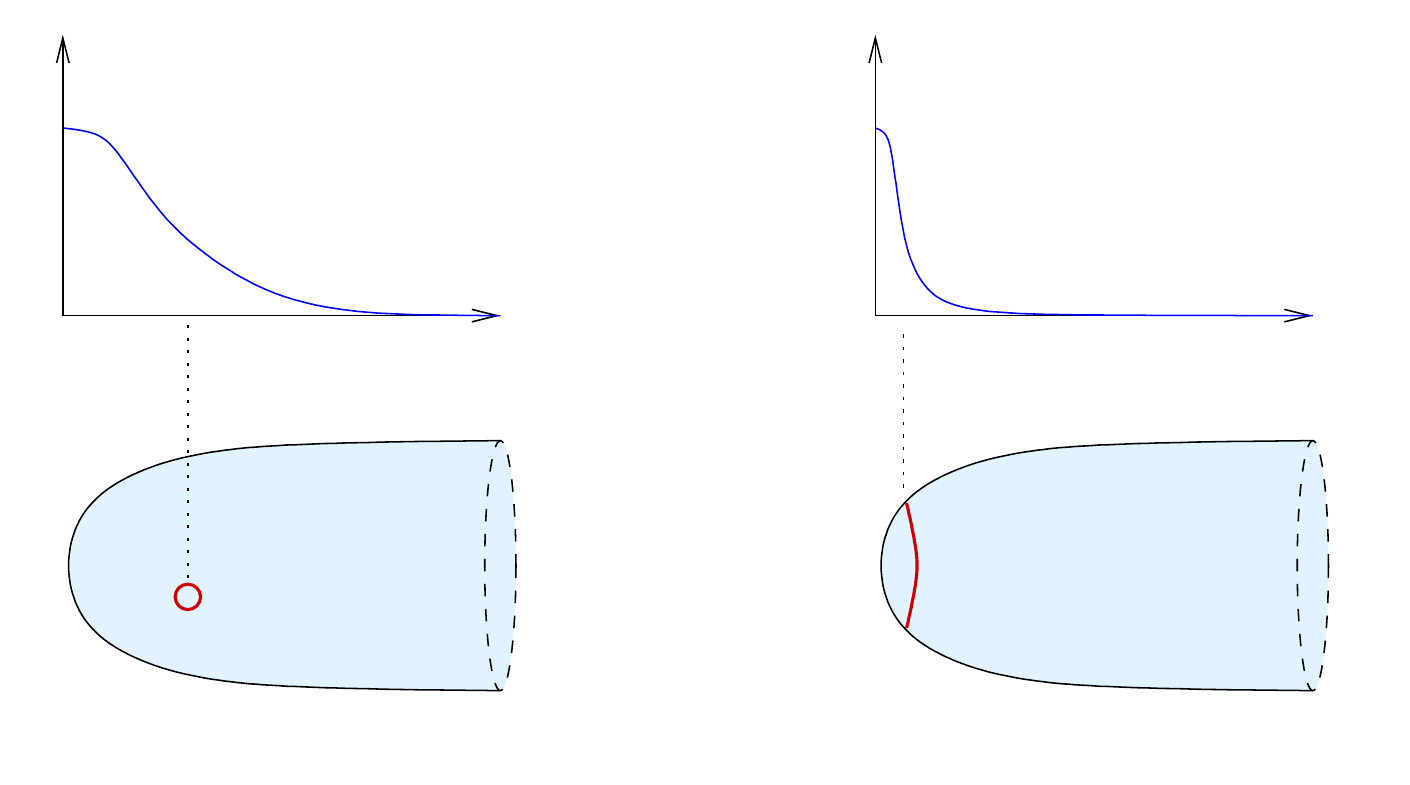}
\caption{\it Different regimes of the string wavefunction are made manifest by the FZZ dual descriptions.  On the left, center of mass motion of a localized string with oscillator excitation; on the right, the extent of the breathing excitation of the winding regime.  These are the tails of the wavefunction in two different directions in configuration space around a single peak -- the classical solution, a pointlike string at the tip of the geometry, which is identical for the two FZZ dual descriptions.}
\label{fig:FZZduals}
\end{figure}

The classical solution for these two FZZ dual descriptions of the state is identical.  However, the form of the corresponding vertex operators emphasizes different regimes of loop space configurations~-- different ways that the string can fluctuate away from this classical solution.  For the string with $\alpha=2j/\nfive$ and oscillator excitations, the quantum number $j$ describes the way the center of mass wavefunction of such a string falls off as $e^{-j\rho}$ away from the origin.  For the string with $\alpha=1-2j/\nfive$ and a unit of winding but without oscillator excitations, the wavefunction behavior $e^{-(\half\nfive-j)\rho}$ describes a steeper falloff (at small $j$) of the breathing mode excitation of the wound string away from the classical solution; see figure~\ref{fig:FZZduals}.


\subsection{Supersymmetric FZZ duality}
\label{sec:superFZZ}

In the supersymmetric theory we have a similar story.  The supersymmetric WZW model consists of the bosonic theory together with a set of free fermions transforming in the adjoint representation.  For $\sltwo$, we bosonize the fermions $\psi^\pm$ in terms of a boson $H$, and write primary fields
\be
\label{superaffine}
\medhat\Phi_{jm}^{\eta,w} = \Phi_{jm}^w \exp\bigl[i\sqrt2\,\eta H \bigr] ~.
\ee
The super-parafermion theory has $\cN=2$ supersymmetry, and in particular a $U(1)$ $\cR$-symmetry current which is bosonized in terms of a scalar $\cH$.  The superconformal parafermion primary representations are then given (in the sector with $w=0$) by a bosonic parafermion operator $V$ times an exponential of~$\cH$
\be
\label{superpf}
\medhat V_{jm}^\eta = V_{jm}^{~} \exp\Bigl[ i\sqrt{\frac{8}{\nfive\nfivehat}} \,\Bigr(m+\frac{\nfivehat}{2}\eta\Bigr)\cH\Bigr]
\ee
where $\eta\in\bZ$ specifies the spectral flow in the $\cR$-charge.  
We then have two ways to represent super-affine primaries $\medhat\Phi$ -- either as bosonic affine primaries $\Phi$ times a highest weight fermion operator as in~\eqref{affine-pf}, or alternatively as a super-parafermion $\medhat V$ times an exponential of the boson $\cY$ that bosonizes the total $J^3$ current, which adds the fermionic contribution $\psi^+\psi^-$ to the bosonic current $j^3$. 

The relation between the bosons $H,Y$ and $\cH,\cY$ is given for instance in~\cite{Martinec:2001cf} (see also~\cite{Chang:2014jta}):
\begin{align}
Y &= \sqrt{\frac{\nfivehat}\nfive}\,\cY + \sqrt{\frac{2}{\nfive}} \,\cH
\nn\\
H &= \sqrt{\frac{\nfivehat}\nfive}\,\cH + \sqrt{\frac{2}{\nfive}} \,\cY  ~.
\end{align}
We then have
\begin{align}
\label{superaffine-2}
\medhat\Phi_{jm}^{\eta,w} &\equiv \Phi_{jm}^w \exp\Bigl[i \sqrt2\, \eta H\Bigr] =  \medhat V_{jm}^{\eta+w} \exp\Bigl[ i\frac{2}{\sqrt\nfive} \Bigl(m+(\eta+w)+\frac\nfive2 w\Bigr)\cY \Bigr] ~;
\end{align}
We see that at large $\nfive$, fermion number measured by $H$ and $\cR$-charge measured by $\cH$ are the same up to small corrections, so we can identify the $\cR$-charge spectral flow quantum number $\eta$ with fermion number up to small corrections, at least when $j$ is small.  As a consequence, we can think of vertex operators $\medhat V^{\pm 1}_{jm}$ with $j\ll\nfive$ as being $-1$ picture supergravitons, while $\medhat V^0_{jm}$ vertex operators are ``tachyons''.  This interpretation breaks down for $j\sim\nfive/2$.

The supersymmetric version of FZZ duality is then
\be
\label{superFZZ}
\medhat V_{jj}^\eta =  \medhat V_{\frac{\nfivehat}2-j,j-\frac{\nfivehat}2}^{\eta+1}  ~.
\ee
This relation and its descendants lift to the parent super-affine theory as
\begin{align}
\label{upstairs superFZZ}
\medhat\Phi_{j,j+n}^{\eta,w=0,\epsilon=+} &= \bigl( J_{-1}^+\bigr)^{n} \medhat \Phi_{\frac\nfivehat2-j,j-\frac\nfivehat2}^{\eta,w=1,\epsilon=-} 
~~ ,~~~
 \bigl( J_{-1}^-\bigr)^{n}\medhat\Phi_{jj}^{\eta,w=0,\epsilon=+} = \medhat \Phi_{\frac\nfivehat2-j,j-n-\frac\nfivehat2}^{\eta,w=1,\epsilon=-} 
\nn\\[.3cm]
\medhat\Phi_{j,-j-n}^{\eta,w=0,\epsilon=-} &= \bigl( J_{-1}^-\bigr)^{n} \medhat \Phi_{\frac\nfivehat2-j,j+\frac\nfivehat2}^{\eta,w=-1,\epsilon=+} 
~,~~~
\bigl( J_{-1}^+\bigr)^{n} \medhat\Phi_{jj}^{\eta,w=0,\epsilon=-} =  \medhat \Phi_{\frac\nfivehat2-j,-j+n+\frac\nfivehat2}^{\eta,w=-1,\epsilon=+} 
\end{align}
as one can see by comparing the conformal dimensions, $\cR$ and $J^3$ charges of the operators (note that for our application, we are only interested in highest weight operators with $|m|=j$).  Thus we see that FZZ duality is related to the unit spectral flow in $\cY$ that relates the $\cD^\pm$ representations in the super-affine theory.

The quantum numbers $\eta,w$ refer back to the underlying bosonic WZW times fermion structure related to the bosons $H,Y$, while for our application it is more useful to adopt the parametrization in terms of the $\cR$-charge boson $\cH$ and the supercurrent boson $\cY$, since it is a linear combination of $\cY_\sl$ and $\cY_\su$ that is being gauged.  We see that in this presentation, a flow of $w$ shifts the $\cR$-charge, which depends on the combination $\hat\eta=\eta+w$, as well as the $\cY$ charge.  Thus if we start off with a state with $\eta=1$ and small $m$, so that the $\cR$-charge is of order $\nfive$ in units of the discretization of $m$, a flow by $w=1$ gets us to a state whose underlying super-parafermion has zero $\hat\eta$ contribution to the $\cR$-charge; the latter comes then entirely from a $j^3$ value of order $\nfive$.

In particular, the background metric downstairs has the same asymptotic as $\medhat V_{jm}^\eta$ with $j=-m=1$, $\eta=1$ (or equivalently its conjugate $j=m=1$, $\eta=-1$), and is FZZ dual via~\eqref{superFZZ} to the $\cN=2$ Liouville interaction which is the downstairs ``tachyon'' $\medhat V_{jm}^\eta$ with $j=m=\frac k2$ and $\eta=0$, and its conjugate.  
We can lift either of these vertices upstairs to a supercurrent algebra primary~\eqref{superaffine-2}.  For the graviton, this operator is essentially
\be
\label{Vgrav}
\medhat V_{1,-1}^{\eta+w=-1} = \medhat\Phi_{1,1}^{\eta=1,w=0}  = \psi^+\Phi_{1,1}^{w=0} 
\ee
and its conjugate,
which carry no $\cY$ charge (since the total $J^3$ charge vanishes, cancelling between the bosonic and fermionic contributions, and so survives the coset projection (note that we always use $m$ to refer to the eigenvalue of the bosonic $j^3$).  Similarly, the $\cN=2$ Liouville perturbation lifts to
\be
\label{Vtach}
 \medhat V_{\frac\nfivehat2-1,1-\frac\nfivehat2}^{\eta+w=0} = \medhat\Phi_{\frac\nfivehat2-1,1-\frac\nfivehat2}^{\eta=1,w=1} 
\ee
which also is $\cY$-independent because the spectral flow shift $w=1$ cancels against the $m$ value, $m+\frac\nfive2 w=0$.

There is also a similar structure for $\sutwo$ primaries $\Psi$, parafermions $\Lambda$, and their super-versions $\medhat\Psi$ and $\medhat\Lambda$
\begin{align}
\Psi_{jm}^{w} &= \Lambda_{jm} \, \exp\Bigl[i\frac{2}{\sqrt{\nfivetil}} \Bigl(m+\frac{\nfivetil}2 w\Bigr) Y_\su\Bigr]
\nn\\[.3cm]
\medhat \Lambda_{jm}^{\tilde\eta} &= \Lambda_{jm}^{~} \exp\Bigl[ i\sqrt{\frac{8}{\nfive\nfivetil}} \,\Bigr(-m+\frac{\nfivetil}{2}\tilde\eta\Bigr)\cH_\su\Bigr]
\nn\\[.4cm]
\medhat\Psi_{jm}^{\eta.w} &= \Psi_{jm}^{w} \exp\Bigl[i \sqrt2\,\eta H\Bigr] =  
\medhat \Lambda_{jm}^{\tilde\eta} \exp\Bigl[ i\frac{2}{\sqrt\nfive} \Bigl(m+\tilde\eta+\frac\nfive2 w\Bigr)\cY_\su \Bigr] ~.
\end{align}
where $\tilde\eta=\eta-w$.  The spectral flow relation for $\sutwo$ is
\be
\medhat \Lambda_{j,j}^{\eta-w=0}=\medhat \Lambda_{\frac\nfivetil2-j , j-\frac\nfivetil2}^{\eta-w=-1}
~~,~~~~
\medhat \Psi_{j,j}^{\eta=0,w=0}=\medhat \Psi_{\frac\nfivetil2-j , j -\frac\nfivetil2}^{\eta=0,w=+1}
\ee
where $\nfivetil=\nfive-2$.

One can then make an entire $(a,a)$ ring of antichiral primary ``tachyons'' (and their FZZ dual supergravitons) by considering the $\frac{\sltwo}{\uone}$ vertex operators downstairs with $\cR$-charge $n/\nfive-1$ and $\eta=0$, and tensoring them with antichiral primaries from the $\frac{\sutwo}{\uone}$ theory with $\cR$-charge $-n/\nfive$ to make operators with total $\cR$ charge equal to minus one   
\be
\label{chiring}
\cV_j^+ = \medhat V_{j+1,j+1}^{\hat\eta=-1} \, \medhat \Lambda_{j,j}^{\tilde\eta=0}
=  \medhat V_{\frac\nfive2-j,j-\frac\nfive2}^{\hat\eta=0} \, \medhat \Lambda_{j,j}^{\tilde\eta=0} ~~.
\ee
Similarly, the $(c,c)$ ring is built upon chiral primary vertex operators in the $\cD^-$ representation of $\sltwo$.
Here $\hat\eta=\eta+w$, $\tilde\eta=\eta+w$; the first expression is again a graviton vertex, while the second is a tachyon.
These will again lift to operators upstairs which have zero charge under the gauge current, but this time by virtue of a nontrivial cancellation between the $\sltwo$ and $\sutwo$ contributions:
\be
\label{FZZduals}
\cV_j^+ = \medhat \Phi_{j+1,j+1}^{\eta=-1,w=0}\,\medhat \Psi_{j,j}^{\eta=0,w=0}
=  \medhat \Phi_{\frac\nfive2-j,j-\frac\nfive2}^{\eta=-1,w=1}\, \medhat \Psi_{j,j}^{\eta=0,w=0}
\ee
Note that typically both factors on the RHS have nonzero $\cY_\sl$ and $\cY_\su$ contributions, but they form a linear combination that is orthogonal to (\ie\ proportional to) the null current being gauged; in other words, this form of the vertex operators differs from~\eqref{chiring} by a gauge transformation.

So let's take stock of the physical vertex operators at our disposal.  All vertex operators upstairs consist of a fermion in the $-1$ picture to satisfy the GSO projection, together with a bosonic center of mass wavefunction $\Phi_{jm}$ and $\Psi_{jm}$ for $\sltwo$ and $\sutwo$, respectively.  The super-Virasoro constraints require $j_\sl=j_\su+1$ for the bosonic operators $\Phi_{jm}$, $\Psi_{jm}$; spacetime supersymmetry requires $|m|=j$ for both, and that the fermion polarization change the spin of that $\sutwo$ contribution to match that of $\sltwo$, or vice versa.  The null gauging constraints determine the relative sign of $m_\sl$ and $m_\su$.  We have the binary choice as to whether the $\sltwo$ center of mass contribution is from $\cD^+$ or $\cD^-$.  We also have equivalence relations generated by gauge spectral flow, and FZZ duality in both $\sltwo$ and $\sutwo$.

Putting everything together, one has the following 1/2-BPS vertex operators and their equivalences, restricting attention to the right-moving component for simplicity:
%
%
\begin{centermath}
\arraycolsep=1.4pt\def\arraystretch{1.1}
\begin{array}{|c|c|c|c||c|c|c|c||c||c|c|c|c|}
\hline
j_\sl & \bar m_\sl & \bar\eta_\sl & w_\sl & j_\su & \bar m_\su & \bar \eta_\su & \bar w_\su & w_y & \bar\cH_\sl & \bar\cH_\su & \bar\cY_\sl & \bar\cY_\su \\[2pt]
\hline
\hline
j\tight+1 & j\tight+1 & -1 & 0 & j & j & 0 & 0 & w & \frac{2j}\nfive \tight-1 & -\frac{2j}\nfive & \frac{2j}\nfive & \frac{2j}\nfive \\[2pt]
\hline
\frac\nfivehat2\tight-j\tight-1 & j\tight+1\tight-\frac\nfivehat2& -1 & 1 & j & j & 0 & 0 & w & \frac{2j}\nfive \tight-1 & -\frac{2j}\nfive & \frac{2j}\nfive & \frac{2j}\nfive \\[2pt]
\hline
\frac\nfivehat2\tight-j\tight-1 & j\tight+1\tight-\frac\nfivehat2 & 0 & 0 & j & j & -1 & -1 & w\tight-k & \frac{2j}\nfive \tight-1& -\frac{2j}\nfive & \frac{2j}\nfive \tight-1 & \frac{2j}\nfive \tight-1 \\[2pt]
\hline
\frac\nfivehat2\tight-j\tight-1 & j\tight+1\tight-\frac\nfivehat2 & 0 & 0 & \frac\nfivetil2\tight-j & j\tight-\frac\nfivetil2 & -1 & 0 & w\tight-k & \frac{2j}\nfive \tight-1 & -\frac{2j}\nfive & \frac{2j}\nfive \tight-1 & \frac{2j}\nfive \tight-1 \\[2pt]
\hline
\hline
j\tight+1 & j\tight+1 & 0 & 0 & j & j & 1 & 0 & w & \frac{2(j\tight+1)}\nfive & 1\tight-\frac{2(j\tight+1)}\nfive & \frac{2(j\tight+1)}\nfive & \frac{2(j\tight+1)}\nfive \\[2pt]
\hline
j\tight+1 & j\tight+1 & 0 & 0 & \frac\nfivetil2\tight-j & j\tight-\frac\nfivetil2 & 1 & 1 & w & \frac{2(j\tight+1)}\nfive & 1\tight-\frac{2(j\tight+1)}\nfive & \frac{2(j\tight+1)}\nfive & \frac{2(j\tight+1)}\nfive \\[2pt]
\hline
j\tight+1 & j\tight+1 & 1 & -1 & \frac\nfivetil2\tight-j & j\tight-\frac\nfivetil2 & 0 & 0 & w\tight-k & \frac{2(j\tight+1)}\nfive & 1\tight-\frac{2(j\tight+1)}\nfive & \frac{2(j\tight+1)}\nfive\tight-1 & \frac{2(j\tight+1)}\nfive\tight-1 \\[2pt]
\hline
\frac\nfivehat2\tight-j\tight-1 & j\tight+1\tight-\frac\nfivehat2 & 1 & 0 & \frac\nfivetil2\tight-j & j\tight-\frac\nfivetil2 & 0 & 0 & w\tight-k & \frac{2(j\tight+1)}\nfive & 1\tight-\frac{2(j\tight+1)}\nfive & \frac{2(j\tight+1)}\nfive\tight-1 & \frac{2(j\tight+1)}\nfive\tight-1 \\[2pt]
\hline
\end{array}
\end{centermath}
{{Table 1. }\textit{Equivalences among vertex operators.  The first block of four are all equivalent representatives of the same state; similarly, the entries in the second block of four are all equivalent to one another.}}
%
%

\vskip .2cm
\noindent
with the analogous structure for left-moving vertex operators (for which one should flip the signs of $m_\su,\eta_\su,w_\su$).  
The first row of the first block is a graviton operator with polarization in $\sltwo$, while the second row is its FZZ dual tachyon; then the third row is the spectral flow of this tachyon, which switches the fermion number and winding of the tachyon over to the $\sutwo$ factor; finally, FZZ duality on the $\sutwo$ factor switches us back to a graviton operator but now with polarization in the $\sutwo$ factor.  In the process, the $\sltwo$ representation has been flipped from a $\cD^+$ representation to a $\cD^-$ representation.  The same procedure relates the entries in the second block, which starts off as a $\cD^+$ graviton state polarized in $\sutwo$, and ends up with a $\cD^-$ graviton operator polarized in $\sltwo$.

Note that after the map $j\to\frac\nfivetil2-j$ and $w\to-w+k$, the second block of vertex operators are the charge conjugates of the first set (in reverse order, so that the first row of the first block is conjugate to the last row of the second block, \etc).  In particular, the vertex operator for the first set with $j=w=0$ has vacuum quantum numbers for the graviton operator with $\sltwo$ polarization, as does the second set with $j=\nfivetil=\nfive-2$, $w=-k$.

In worldsheet $\cN=2$ terms, these vertex operators comprise the chiral and twisted chiral ring of the worldsheet CFT.  The (c,c) chiral ring takes both left- and right-moving contributions from the second block, while its conjugate (a,a) ring takes both from the first block; and the twisted chiral (c,a) and (a,c) rings take one chirality from each block.  Note, though, that one must take the representative in each block with the same value of the principal quantum number $j$ and winding numbers $w_\sl,w_\su,w_y$ on each side.

\subsection{General backgrounds}

We expect FZZ duality to be a generic feature of Lunin-Mathur geometries.  The existence of two null Killing vectors with compact spatial orbits provides perturbative strings with energy, momentum, and winding quantum numbers.  Because the spatial orbits degenerate, the winding will not be conserved, and a winding string is not topologically distinguishable from one with a coherent excitation of oscillator modes.  The Lorentz force argument given above suggests that when a wound string carries too much center-of-mass energy, it starts to expand out toward the boundary; as a result, there will be an upper bound on the energy of strings that are bound to the cap.  States slightly below the bound will have wavefunctions that spread out on the same scale as the lowest center-of-mass modes in the cap, and a picture similar to that of the semiclassical coherent states~\eqref{semiclassicalFZZ} begins to emerge.

In conformal perturbation theory by 1/2-BPS deformations about the circular supertube backgrounds described by null gauged WZW models, the operator identifications of FZZ duality will deform smoothly.  Thus one expects that there will be FZZ dual pictures of the background across the 1/2-BPS configuration space.  We now turn to a discussion of this generalized FZZ duality.

\section{Locating fivebranes}
\label{sec:locatingbranes}

The harmonic forms and functions~\eqref{greensfn} of the supertube background~\eqref{LMgeom} average over the fivebrane source locations.  Thus the round supertube background~\eqref{roundST} naively respects a $\uone\times\uone$ isometry, whereas the exact source distribution breaks one of the $\uone$'s to $\bZ_\nfive$.  
The vertex operator $\cV_0^{++}$ of~\eqref{Vgrav} represents the zero mode of the dilaton, and has the same asymptotic behavior as the background metric and $B$-field; its trivial $\bS^3$ zero mode wavefunction also respects  $\uone\times\uone$.  However its FZZ dual~\eqref{Vtach} has a zero mode wavefunction that only respects a discrete $\bZ_\nfive$ rotations in the angular coordinate $\phi$ instead of arbitrary shifts, and so the complete string background knows about the locations of the underlying fivebranes.  

In this section we explore this nonperturbative structure (in $\alpha'$) of the closed string background, and how it codes the underlying locations of the fivebranes which are hidden from supergravity.  We will in addition be interested in the structure of D-branes ending on the fivebranes, since they also precisely probe the fivebrane locations.  These two aspects fit together into a coherent picture of the exact source structure.  We begin with the FZZ dual description of fivebranes on the Coulomb branch and its relation to a Landau-Ginsburg orbifold, and discuss how D-branes that stretch between the fivebranes are realized in this presentation.  We then relate these branes to their geometrical picture upstairs in the Lunin-Mathur geometry, generalizing the analysis of~\cite{Martinec:2019wzw}.
Finally, we extend the discussion to exhibit the structure underlying general supertube backgrounds; we will explore the implications for black hole formation and the nature of fuzzballs in the next section.

\subsection{The Coulomb branch}

The downstairs coset geometry, which we have presented in terms of null gauging, can also be viewed as a $\bigl(\frac\sltwo\uone\times\frac\sutwo\uone\bigr)/\bZ_\nfive$ coset orbifold~\cite{Sfetsos:1998xd,Giveon:1999px}.  
The FZZ dual background is a ``tachyon'' condensate, consisting of $\cN\tight=2$ Liouville theory for the first factor, and a Landau-Ginsburg model with superpotential $\cW=\sfZ^\nfive$ for the second factor.  The $\bZ_\nfive$ orbifold projects onto states with integer $\cR$ charge of the worldsheet $\cN=2$, which allows a chiral GSO projection and spacetime supersymmetry.  Thus one has a background $\cN=2$ superpotential of the form
\be
\label{W0}
\cW_0 = \sfZ^\nfive - \lambda_0 \, e^{\nfive\sfX} = \prod_{\ell=1}^\nfive \Bigl( \sfZ - \mu_\ell\, e^{\sfX} \Bigr) 
\ee
where $\mu_\ell= \lambda_0^{1/\nfive} \exp[2\pi i \ell/\nfive]$.  

The duality between the Landau-Ginsburg orbifold and the $\bigl(\frac\sltwo\uone\times\frac\sutwo\uone\bigr)/\bZ_\nfive$ coset orbifold is a non-compact version of the Calabi-Yau/Landau-Ginsburg correspondence~\cite{Ooguri:1995wj}.  In the compact CY/LG correspondence~\cite{Martinec:1988zu,Vafa:1988uu}, the target space is specified by the locus $\cW=0$ in a weighted projective space.  In this non-compact version, the superpotential zeroes $\mu_\ell$ encode the locations of the fivebranes in $\sfZ$ which parametrizes the $x^1\tight+ix^2$ plane (though the map between the two parametrizations is somewhat nontrivial).  These $\nfive-1$ complex parameters are moduli of the theory.  The $\bZ_\nfive$ orbifold theory also has marginal deformations in twisted sectors, which are elements of the twisted-chiral ring; these yield another $\nfive-1$ complex parameters $\tilde\mu_\ell$ which parametrize the locations of the fivebranes in the $x^3\tight+ix^4$ plane.  Together, the $\mu_\ell$ and $\tilde\mu_\ell$ specify the relative locations of the fivebranes in $\bR^4$.  Together, these deformations constitute an $\cN=2$ description of the full $\cN=4$ structure of the moduli space of worldsheet CFTs for supertubes.

The zeros of the superpotential code the locations of the fivebranes.  The moduli space of fivebranes is explored by varying the locations of the zeros.
One can organize the general Landau-Ginsburg superpotential as
\be
\label{spotl def}
\cW =  \cW_0+\sum_{j=1/2}^{\nfive/2-1} \lambda_j \,\sfZ^{2j} e^{(\nfive-2j)\sfX}
\ee
where the chiral operators are the tachyon version of~\eqref{chiring}, \ie\ the FZZ duals of the 1/2-BPS graviton vertex operators. 
Note that $\cV^{--}$ is chiral in the $\cD^+$ representation, and antichiral in the $\cD^-$ representation, while $\cV^{++}$ is antichiral in the $\cD^+$ representation and chiral in the $\cD^-$ representation.  By considering all the tachyon representatives of the chiral ring for all the allowed spins, and allowing the $\lambda_j$ to be complex, we cover all the marginal deformations coming from $\cV^{++}$ and $\cV^{--}$.   Similarly, $\cV^{+-}$ and $\cV^{-+}$ form the twisted chiral ring of marginal deformations.

The $\lambda_j$ have fixed $\bZ_{\nfive}$ momentum charges; their discrete Fourier transform 
\be
\mu_\ell = \sum_j  \lambda_j \exp\bigl[ 4\pi i \, \ell j/\nfive\bigr]
\ee
codes the deformations of the positions of the individual fivebranes in the $x^1\tight+ix^2$ plane.%
\footnote{At nonlinear order, there is a nontrivial map between the fivebrane locations on $\bR^4$ and the $\lambda_n$; see below.}
For example, the linearized deformation from the circular array to the ellipse is governed by the coefficient $\delta\lambda_{\nfive/2-1}$, with higher orders in the perturbation introducing other superpotential coefficients scaling as $\delta\lambda_{\nfive/2-p}\sim (\delta\lambda_{\nfive/2-1})^p$, as one can see by carrying out the discrete Fourier transform of the elliptical array.  Indeed, one sees from Table~1 that the winding tachyon representative of $\cV_0^{++}$ has $\sutwo$ spin zero and corresponds to a deformation of $\lambda_0$ in~\eqref{W0}, while $\cV_0^{--}$ has $\sutwo$ spin $\hf\nfive\tight-1$ and corresponds to a deformation of $\lambda_{\nfive/2-1}$ in~\eqref{spotl def}.

The Liouville theory has a linear dilaton $\Phi = Q\sfX$, but the exponential potential wall $\cV=|\nabla\cW|^2$ keeps fundamental strings from exploring the strong coupling region at large $\sfX$ near the fivebrane sources.  In the effective theory after the gauge projection, this is one way to understand why there is no strong coupling singularity, and a sensible perturbative expansion.  However, at points along the moduli space where zeros of the superpotential degenerate, say $\mu_\ell=\mu_{\ell'}$, the bosonic potential $|\nabla\cW|^2$ vanishes along the direction
\be
\sfZ=\mu_\ell \, e^{\sfX}
\ee
and so strings {\it can} explore the strong coupling region.  This feature is related to the fact that there is a perturbative string description for the linear dilaton throat of coincident fivebranes, but not single fivebranes in isolation~\cite{Callan:1991at}.  The difference $\mu_\ell-\mu_{\ell'}$ controls a local effective Liouville cosmological constant for a ``little throat'' involving just the two nearly coincident fivebranes; the limit $\mu_\ell\to\mu_{\ell'}$ removes the local Liouville wall and allows strings to explore a strong coupling region near this pair of fivebranes.

To see this developing little throat, consider the superpotential
\be
\label{littlethroat}
\cW = \bigl( \sfZ - (\mu_0+\epsilon)e^\sfX\bigr)\bigl( \sfZ - (\mu_0-\epsilon)e^\sfX\bigr) \,\mfw(\sfX,\sfZ)  
\ee
where $\mfw$ is a homogeneous polynomial of degree $\nfive-2$ in $\sfZ$ and $e^\sfX$.
Along the direction $\sfZ=\mu_0 \, e^{\sfX}$, the bosonic potential at large $\sfX$ is
\be
\label{littlethroatpotl}
\cV\big({\sfZ\tight=\mu_0 e^{\sfX}}\big) \sim \epsilon^4 e^{2(\sfX+\bar\sfX)}\Bigl( \bigl|\partial_\sfZ\cU\bigr|^2 + \bigl|\partial_\sfX\cU+2\,\cU \bigr|^2\Bigr)
\sim {\textit{const.}} \,\epsilon^4 e^{\nfive(\sfX+\bar\sfX)}
\ee
and so there is an effective Liouville ``cosmological constant'' along this direction in field space that vanishes like $\epsilon^4$, which controls the degree to which perturbative strings can access strong coupling effects at a given energy scale.

In the group sigma model describing the circular array of fivebranes, the Landau-Ginsburg picture of the source in terms of a winding tachyon condensate lifts from the coset downstairs to the group sigma model upstairs, using the vertex operators that deform the theory along the moduli space which have representatives such as~\eqref{FZZduals} upstairs.  Upon shifting from the $-1$ picture to the $0$ picture, FZZ duality becomes~\eqref{upstairs superFZZ}, and once again the FZZ dual to a graviton vertex is a tachyon winding the angular direction of $AdS_3$ 
\be
\label{lifted chiral ring}
\Bigl( J_{-1}^- \medhat\Phi_{j+1,j+1}^{\eta=0,w=0,\epsilon=+}\bigr) \medhat\Psi_{jj}^{\eta=0,w=0} = \medhat\Phi_{{\frac{\nfivehat}2}-j-1,{j-\frac{\nfivehat}2}}^{\eta=0,w=1,\epsilon=-} \, \medhat\Psi_{jj}^{\eta=0,w=0}
\ee
where for simplicity we have suppressed the antiholomorphic structure.  
The $\cN=2$ structure also lifts.%
\footnote{At least, a Lorentzian version of it; the standard presentation of $\cN=2$ supersymmetry arises if we continue $\sltwo$ to the Euclidean hyperboloid $\sfH_3^+$.  The conditions for $\cN=2$ supersymmetry on group manifolds were explored in~\cite{Rastelli:2005ph,Spindel:1988sr}.}  
The tachyon operators on the RHS of~\eqref{lifted chiral ring} form the lifted chiral ring, and thus a set of $\cN=2$ superpotential deformations upstairs.

The winding tachyon picture of the moduli space is FZZ dual to a geometrical picture where the Lunin-Mathur geometry upstairs deforms the $\sltwo\times\sutwo$ group manifold by a nonlinear gravitational wave.   
As discussed in section~\ref{sec:ST defs}, for the gauging that leads to the Coulomb branch of fivebranes, BPS deformations of the Lunin-Mathur geometries with higher spin $j\ge\frac\nfive2$ are gauge equivalent to those built from the $\sltwo$ primaries~\eqref{BPSverts}-\eqref{jrange}.  This leads to a $4(\nfive\tight-1)$ parameter family of Lunin-Mathur geometries built on the harmonic objects~\eqref{greensfn} with 
\be
\label{CoulombProfile}
\F = \sum_{j=0}^{\nfivetil/2} {\sfa_{2j+1}} \, \exp\Bigl[ i \frac{(2j+1) k\sfv}{\nfive}\Bigr]
\ee
whose null gauging yields the smeared geometry of Coulomb branch fivebranes seen in the supergravity approximation.  Here we have implemented the twisted boundary condition~\eqref{twistedbc} by going to an $\nfive$-fold cover that wraps all the fivebranes together, and set $\sfv=(\ytil+t)/\Rytil$.
It is natural to suppose a one-to-one map between the profile parameters $\sfa_{2j+1}$ and the superpotential parameters $\lambda_j$ of~\eqref{spotl def}.  We now work out this map in the context of supertube backgrounds.

\subsection{Supertubes}
\label{subsec:supertubes}
 
Null gauging is compatible with a superpotential that depends on the coordinate $\sfv$, and after Wick rotation we can allow the $\cN=2$ superpotential to depend on the corresponding holomorphic coordinate and in particular impose the appropriate dependence of the zeros of the superpotential $\cW$ on this coordinate that yields the spiraling supertube.%
\footnote{It is not absolutely necessary to work in the Euclidean continuation, which is only mentioned here as it allows us to use the tools of holomorphy and various non-renormalization theorems of $\cN=2$ superspace to characterize the moduli space.  We strongly suspect that one can work directly in Lorentz signature, where spacetime supersymmetry and the associated null Killing vectors play a very similar role; indeed, worldsheet $\cN=2$ and spacetime supersymmetry are closely related~\cite{Banks:1987cy}. }
This reasoning suggests that the superpotential for the round supertube takes the form
\be
\label{ST spotl}
\cW_0 = \sfZ^\nfive e^{ik\sfv} - \lambda_0 \, e^{\nfive\sfX} = \prod_{\ell=1}^\nfive \Bigl( \sfZ \,e^{ik\sfv/\nfive} - \mu_\ell\, e^{\sfX} \Bigr) 
\ee 
with $\mu_\ell = \lambda_0^{1/\nfive} \,e^{2\pi i\ell/\nfive}$, generalizing the Coulomb branch superpotential which takes the same form but sets $k=0$. 

The deformed superpotential then allows the zeros $\mu_\ell$ to be displaced from the roots of unity.  Expanding the factorized form of $\cW$, the coefficients of the various powers of $\sfZ$ are symmetric polynomials in the $\mu_\ell$ in which no root appears more than once.
One can then rewrite the superpotential~\eqref{ST spotl} in a different basis of symmetric polynomials as
\be
\label{WSchur}
\cW = \biggl(\sfZ^\nfive e^{ik\sfv} \, \exp\Bigl[ -\sum_{n=1}^\infty \frac1n M^n \Bigl(\sum_\ell \mu_\ell^n\Bigr) \Bigr] \biggr)_+
~~,~~~~
M = \frac{e^\sfX\, e^{-ik\sfv/\nfive} }{{\sfZ}}
\ee
where the subscript $+$ indicates that one should expand the exponential in power series, and keep only the terms with non-negative total powers of $\sfZ$.  Remarkably, all the terms with higher powers of individual roots cancel among themselves.  We then allow the $\mu_\ell$ a more general $\sfv$ dependence, and write the zeroes in a discrete Fourier transform
\be
\label{DiscreteFT}
\mu_\ell = \lambda_0 \, e^{2\pi i \ell/\nfive} + \sum_{n=0}^{\nfive-1} \lambda_n \, e^{2\pi i (n+1)\ell/\nfive} ~.
\ee
Note that $\lambda_n$ need not be small; however if we do work only to linearized order in these Fourier amplitudes, substituting this expression in~\eqref{WSchur} one finds
\be
\cW = \cW_0 + \sum_{n=0}^{\nfive-1}  \dlam_n \, \lambda_0^{\nfive-n-1} \, \sfZ^n e^{(\nfive-n)\sfX} e^{ink\sfv/\nfive} ~.
\ee
Now expand $\lambda_n$ in an integer Fourier series
\be
\label{IntegerFT}
\lambda_n = \sum_{w_y} \lambda_{n,w_y} \, e^{iw_y \sfv}
\ee
and substitute in the superpotential; again if we only expand to leading order in $\dlam_n$ we find
\be
\cW = \cW_0 + \sum_{n=0}^{\nfive-1} \sum_{w_y} \dlam_{n,w_y} \, \lambda_0^{\nfive-n-1} \sfZ^n \exp\Bigl[{\bigl(\nfive-n\bigr)\sfX + i\Bigl(\frac{nk+ w_y\nfive}{\nfive}\Bigr)\sfv}\Bigr]  ~,
\ee
though we can expand equation~\eqref{WSchur} (with the substitutions~\eqref{DiscreteFT}, \eqref{IntegerFT}) to any desired order in the coefficients $\lambda_{n,w_y}$ to find the nonlinear relation between the $\sfv$ dependence of the locations of the zeroes and the coefficients of the scaling operators in the perturbed superpotential. 
For instance, consider once again the ellipse deformation of the circular array.  The fivebrane locations are
\be
\mu_\ell = a_+ e^{i\phi_\ell} + a_- e^{-i\phi_\ell}
~~,~~~~
a_\pm = \frac{a_1\pm a_2}{2}
~~,~~~~
\phi_\ell = 2\pi \ell/\nfive ~;
\ee
The elementary symmetric polynomials are
\be
\sum_{\ell=1}^\nfive \mu_\ell^n = \bigl( a_+^\nfive + a_-^\nfive \bigr) \delta_{n,\nfive} + 
\begin{cases}
\binom{n}{n/2}(a_+a_-)^{n/2} & n~\text{even} \\
0 & n~\text{odd}
\end{cases}
\ee
and the expression~\eqref{WSchur} sums to
\be
\cW = \Bigl( Z^\nfive e^{ik\sfv}\,\half\Bigl[1 + \sqrt{1-a^4 M^4\epsilon^2}\,\Bigr]\Bigr)_+
- \bigl( a_+^\nfive + a_-^\nfive \bigr) e^{\nfive X}
\ee
where again one is instructed to perform the series expansion in $\epsilon$ and keep only the terms with positive powers of $\sfZ$.

On the other hand, consider the tachyon vertex operators~\eqref{FZZduals} with a general winding as in~\eqref{general winding vertex}
\be
\label{vtach}
\cV_{j,w_y}^{\textit{tach}} = \medhat \Phi_{\frac\nfive2-j,j-\frac\nfive2}^{\eta=-1,w=1}\, \medhat \Psi_{j,j}^{\eta=0,w=0}
\,\exp\bigl[ i w_y \Ry v \bigr]
\ee
(and their Hermitian conjugates), which characterize the linearized perturbations of the chiral ring away from a Lunin Mathur geometry with a round source profile.  These vertex operators represent perturbations away from the supertube background given by $\cW_0$.  Bosonizing the gauge current as 
\be
\cJ= \sqrt\nfive\bigl(\partial\cY_\sl+\partial\cY_\su\bigr) - k\Ry\bigl(\partial t + \partial \ytil\bigr) \equiv \partial\Upsilon
\ee 
one can write the dependence of these vertex operators on the bosons $\cY_\sl,\cY_\su,t,\ytil$ as
\be
\exp\Bigl[i\frac{2j}{\sqrt\nfive}\bigl(\cY_\sl+\cY_\su\bigr) + iw_y\Ry\bigl(t+\ytil\bigr)\Bigr]
= \exp\Bigl[i \frac{2j}{\nfive}\Upsilon + i\Bigl(\frac{2jk+ w_y\nfive}{\nfive}\Bigr)\Ry v\Bigr] ~.
\ee
Thus the vertex operators~\eqref{vtach} represent perturbations of the background superpotential $\cW_0$ which sew together strings in the background to create ones with winding $(2j+1)k\tight+\nfive w_y$, as indicated in~\eqref{generalshift}, and we identify them with perturbations of the background superpotantial
\be
\cV_{j,w_y}^{\,\eff} =  \sfZ^{2j} \exp\Bigl[{\bigl(\nfive-2j\bigr)\sfX + i\Bigl(\frac{2jk+ w_y\nfive}{\nfive}\Bigr)\sfv}\Bigr] 
\ee
after stripping off the $\Upsilon$ dependence (which amounts to a gauge transformation).

It is natural to suppose a one-to-one map between the parameters $\sfa_{2j+1}$ deforming the profile shape in the $x^1\tight+ix^2$ plane via
\be
\label{STProfile}
\F = \sum_{p\ge k} {\sfa_{p}} \, \exp\Bigl[  \frac{ip\sfv}{\nfive}\Bigr] 
\ee
and the superpotential deformation parameters $\dlam_{2j,w_y}$.  The relation between these parameters follows from comparing the Fourier coefficients of $\F$ with the Fourier expansion of the zeroes $\mu_\ell$ of the superpotential.  To zeroth order, one finds $\sfa_k=\lambda_0$; to next order, one has
\be
\delta\sfa_{(n+1)k+w_y\nfive} = \dlam_{n,w_y}~.
\ee
Thus, the deformations of the source profile in the $x^1$-$x^2$ plane directly determine the deformation parameters of the superpotential, which are the coefficients of chiral ring operators.  The presentation~\eqref{WSchur} of the superpotential allows one to work out the full nonlinear map between the superpotential parameters $\lambda_{n,w_y}$ and the profile parameters $\sfa_p$.
This map is reminiscent of a similar relation between the superpotential deformation parameters and the natural flat coordinates on the moduli space for compact $\cN=2$ Landau-Ginsburg models (see for instance~\cite{Eguchi:1993xx,Eguchi:1993ty}).  Note that the
 $\sfa_{p}$ are natural flat coordinates on the moduli space -- they are the natural variables for the geometric quantization of the Lunin-Mathur geometries~\cite{Rychkov:2005ji}), in that their action reduces to that of a collection of harmonic oscillators.

Similarly, the deformations of the source profile in the $x^3$-$x^4$ plane are associated to the couplings to operators in the twisted chiral ring.  FZZ duality is simply the statement that the shape of the source profile is inextricably linked to the geometry it sources; the Landau-Ginsburg superpotential encodes the former, while the dual sigma model geometry encodes the latter.  In string theory, there is no sharp boundary between matter and geometry -- both are string condensates, and the background seamlessly transitions from one to the other, depending on what part of the wavefunction is being probed.  Vertex operators that deform the background have both a Landau-Ginsburg aspect and a geometrical aspect, as we see from Table 1.

The above considerations allow us to construct the map between the parameters of the superpotential and the shape of the profile.   The zeros of the superpotential code the locations of the fivebranes, which for the supertube evolve in $\sfv$ and are specified by the source profile $\F(\sfv)$.  In particular, collisions of the fivebrane strands correspond to colliding zeros of $\cW$, opening up a ``little throat'' of the sort~\eqref{littlethroatpotl} where strings can probe strong coupling and the nonabelian fivebrane physics that lurks there.  

\begin{figure}[h!]
\centering
\includegraphics[width=.5\textwidth]{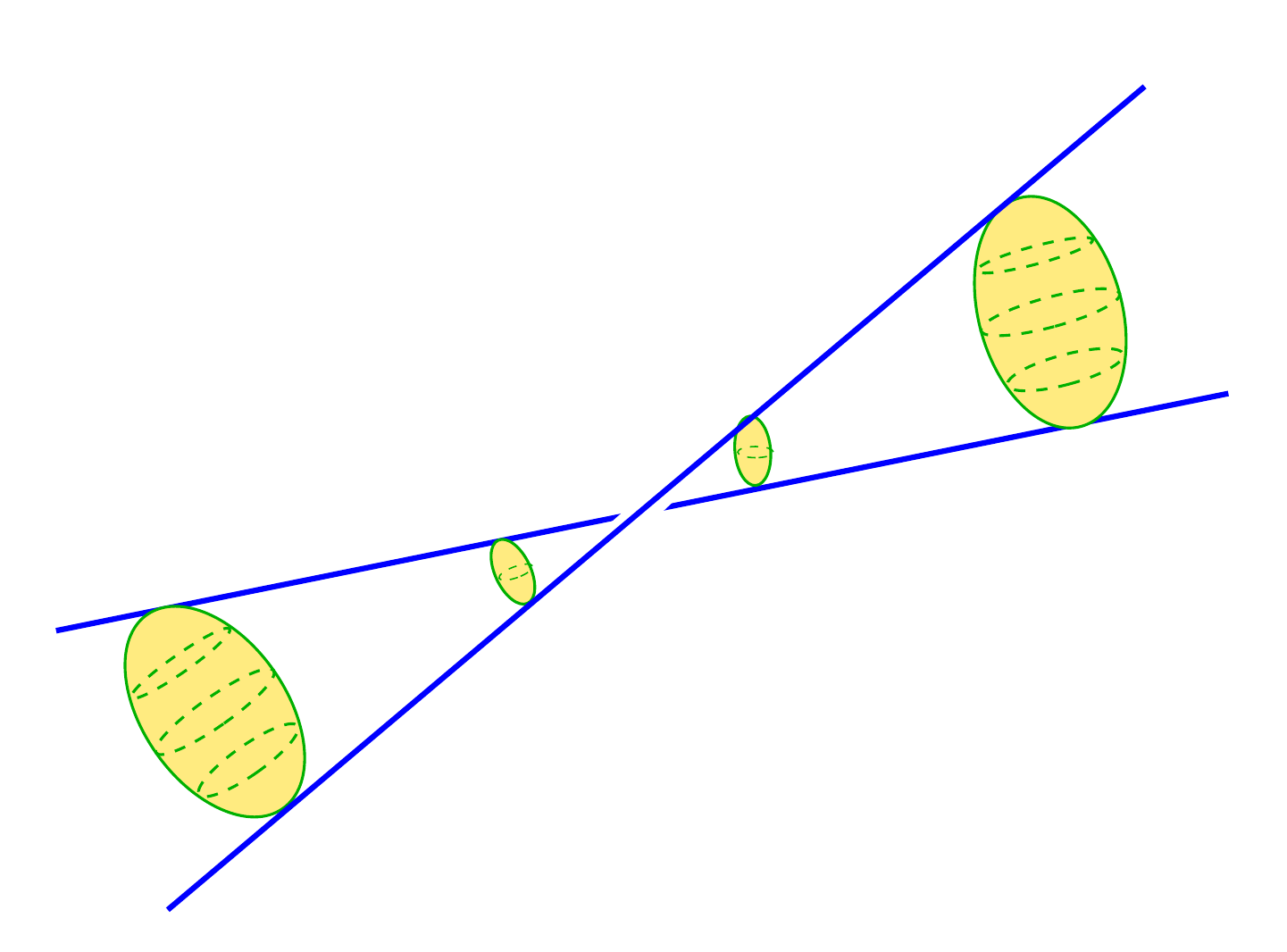}
\caption{\it Two supertube strands nearly intersect.  The local geometry has a pair of lines of Gibbons-Hawking points (blue), which specify the north and south poles of a line of two-spheres which shrink to a small size at the point of closest approach of the strands.  When the strands intersect, an $A_1$ singularity develops locally; the W-branes are now D-branes wrapping the local vanishing cycle and are massless at the intersection.}
\label{fig:STcrossing}
\end{figure}

The geometrical dual picture of such a near-intersection is depicted in figure~\ref{fig:STcrossing}.
The Lunin-Mathur geometry upstairs has been argued to be smooth up to orbifold singularities when the profile $\F$ self-intersects%
~\cite{Lunin:2001jy,Lunin:2002iz}.
The smearing of the Coulomb branch fivebrane source in the Green's function integrals~\eqref{greensfn} reduces the order of the singularity at the source from isolated double poles in the harmonic functions to a line source with a simple pole singularity along a closed contour.  For isolated singularities arising from a non-intersecting contour, this apparent singularity is a coordinate artifact~-- locally one has Gibbons-Hawking coordinates for the transverse space; the smeared singularity is an isolated Gibbons-Hawking point in the 4d space transverse to the contour, and the geometry is smooth.  That is, locally the transverse space is realized as a circle fibered over $\bR^3$ which degenerates smoothly at the Gibbons-Hawking point.  When there are pairs of strands approaching one another closely, then the path between the strands forms the polar direction of a homological two-sphere, with the fibered circle being the azimuthal direction.  When the contour self-intersects, one has coinciding Gibbons-Hawking points; the two-sphere vanishes at the intersection and the geometry develops an orbifold singularity.
This orbifold singularity is only resolved by strong-coupling effects, namely the condensation of W-branes wrapping the vanishing cycle.  In the remainder of this section, we develop this picture using D-brane probes -- in the Landau-Ginsburg picture generally, and in the geometric picture for the specific case of the elliptical profile.

\subsection{D-branes in Landau-Ginsburg theory}

Supergravity sees the supertube fivebrane source as having constant density smeared along a contour.  D-branes are able to localize the fivebranes along this contour; for instance, D-strings end on type IIB fivebranes, and those endpoints can be determined with Planckian precision.  In the null gauged sigma model, D-strings lift upstairs by extending them along the gauge orbits, becoming spherical spatial two-branes whose north/south poles are pinned to the degeneration locus of the gauge orbits, which is the supertube contour.  The type IIA picture is similar, adding one more dimension to these D-branes both downstairs and upstairs.

In~\cite{Martinec:2019wzw}, it was shown for the circular source profile that the brane upstairs carries a worldvolume magnetic flux, whose quantization discretizes the possible brane endpoints along the contour.  Thus in the geometrical picture dual to the FZZ tachyon superpotential, one can also see the localization of the fivebrane sources, even though in the effective sigma model geometry downstairs these locations are smeared.
The same structure will hold for the general Lunin-Mathur geometry with source profile~\eqref{CoulombProfile}, and again quantization effects serve to localize the fivebranes along the smeared source locus.  

D-branes also fill out the Landau-Ginsburg picture of the near-source structure.  We can use these D-branes to probe the nature of the singularities arising from source profile self-intersections.  We thus turn to a discussion of D-branes in general fivebrane supertube backgrounds.

There are two general classes of BPS D-branes: (1) A-branes that wrap special Lagrangian cycles; and (2) B-branes that wrap holomorphic cycles.  These two classes of branes are related by mirror symmetry.
The Landau-Ginsburg superpotential $\cW$ describes a surface $\cW(\sfX,\sfZ)\tight+\sfY^2=0$ in $\bC^3$ which is mirror to the geometry transverse to the NS5-branes%
~\cite{Witten:1993yc,Ooguri:1995wj,Hori:2001ax}.
Thus for instance a D-string stretching between NS5-branes is mirror to a D2-brane wrapping a holomorphic cycle of the surface, and is thus a B-brane in the Landau-Ginsburg picture%
~\cite{Ribault:2003ss,Israel:2004jt,Fotopoulos:2004ut}.%

Such branes can be described via open string tachyon condensation on stacks of branes and antibranes (for reviews, see for instance~\cite{Hori:2004zd,Jockers:2007ng}).  This structure is an example of a general story about D-brane categories~\cite{Witten:1998cd}.  Brane-antibrane pairs annihilate via condensation of the open strings that stretch between them, which become tachyonic when the branes are nearly coincident; however, the annihilation can leave behind lower branes when the open string tachyon profile has topological defects.  For instance, a vortex texture in the complex tachyon field on a $D_p$--$\bar D_p$  pair leaves behind a $D_{p-2}$ brane at the location of the vortex after tachyon condensation, because the topology requires the tachyon to to vanish and thus remain uncondensed in the vortex core.  Thus a boundary tachyon profile with vortex cores at points $\mu_\ell$ will yield an effective theory with branes localized at those points.
In the Landau-Ginsburg realization, the open string tachyon profiles are given by solutions to the matrix factorization of the bulk superpotential
\be
\cQ=\left( \begin{matrix} 0& \cJ(\sfX,\sfZ)\\ \cE(\sfX,\sfZ)&0 \end{matrix} \right)
~~,~~~~
\cQ^2 = \cW(\sfX,\sfZ)\, \One
\ee
where $\cE,\cJ$ are matrices whose entries are (holomorphic) polynomials in $\sfZ$ and $e^\sfX$ describing open string tachyon condensates on the brane-antibrane stack.  
The boundary tachyon condensate is large in the asymptotic regions of $\sfZ$ and annihilates brane-antibrane pairs in that region, leaving behind branes localized near zeros of the boundary superpotential $\cJ$ where the tachyon condensate is forced to remain small for topological reasons.  
A proposal to fit Liouville theory into the Landau-Ginsburg framework rewrites its superpotential as the negative power of a Landau-Ginsburg field $\sfU$, via $e^{\nfive\sfX} = \sfU^{-\nfive}$~\cite{Ooguri:1995wj}.  The localized branes called ZZ branes in Liouville theory are pointlike B-branes localized at $\sfU=0$ (and thus $\sfX=\infty$)%
~\cite{Eguchi:2003ik,Eguchi:2004ik,Ahn:2004qb,Hosomichi:2004ph,Israel:2004jt}.

In the Landau-Ginsburg orbifold, D-branes that are stretched between fivebranes of the circular array are products of B-branes~\cite{Elitzur:2000pq,Eguchi:2004ik,Israel:2005fn} -- a point ZZ brane in $\cN=2$ Liouville times a B-brane in the $\cW=\sfZ^\nfive$ minimal model which is localized near vanishings of the superpotential.  
The overall structure appears to be quite similar to constructions of branes using matrix factorization studied in~\cite{Ashok:2004zb,Brunner:2005fv,Enger:2005jk,Behr:2010ug}.  Given a factorization of the superpotential as in~\eqref{ST spotl}, a rank one factorization partitions the roots according to
\be
\cJ = \prod_{\ell\in \cI} \big(\sfZ - \mu_\ell \, e^\sfX \big)
~~,~~~~
\cE = \prod_{\ell\in \cI^c} \big(\sfZ - \mu_\ell \, e^\sfX \big)
\ee
where $\cI\cup\cI^c=\{\mu_\ell\}$, $\cI\cap\cI^c=\emptyset$.  The resulting D-brane worldvolume is naively the superposition of lines $\sfZ=\mu_\ell e^\sfX$ for $\mu_\ell\in\cI$, with further open string tachyon condensation binding these together into the composite D-brane of least action.  For the degenerating superpotential~\eqref{littlethroat}, it seems natural to choose $\cI$ to consist of the pair of fivebranes that are approaching one another.  The resulting D-brane will then be localized in the strong coupling region where the Liouville wall is receding, and become light in the limit that the zeros coalesce.

It would be useful to have a calculation of the disk partition function for these B-branes, since this yields the D-brane mass.  The expectation is that the mass of D-branes corresponding to colliding zeros of the worldsheet bulk superpotential will vanish when the superpotential degenerates, since these D-branes will be concentrated near the colliding zeros, a region of strong coupling.   Indeed, the boundary superpotential is degenerating along with the bulk superpotential.  We leave such a calculation to future work, and instead turn to the corresponding calculation of D-branes stretching between fivebranes in the particular case of the elliptical array, using the DBI action in the geometrical picture where a quantitative understanding can be achieved.

\subsection{DBI analysis}
\label{subsec:DBI}

The geometrical sigma model picture is associated to smooth Lunin-Mathur geometries upstairs which are sourced by a fivebrane density that is constant along the curve $\F(v)$, leading to a downstairs geometry~\eqref{dsperp} after the gauge projection.  This geometry has a fivebrane singularity smeared along a curve.  Upstairs this curve is not the locus of a singularity but rather is the degeneration locus of a spacelike combination of the null Killing vectors of the background.
The D-branes that stretch between fivebranes in the geometrical picture upstairs are spheroidal D2-branes in the Lunin-Mathur geometry.  The N/S poles of these probe branes lie along the curve, and the allowed pole locations are discretized by flux quantization on the D-brane~\cite{Martinec:2019wzw}.
The fivebranes thus sit at discrete points along this curve, diagnosed by nonperturbative effects in $\alpha'$ and D-brane probes, and this structure imparts a definite ordering to the fivebranes.  We can label the D-branes by pairs of fivebranes $(M,M+L)$, sequentially ordered along the contour $\F(v)$; then $L$ is the number of units of magnetic flux on the brane.

\subsubsection{Downstairs picture}

The DBI analysis for D-strings in the geometry \eqref{eq:CB1} is
particularly simple. The effective action on D-branes at leading order in the derivative expansion is 
\be
\cS_{DBI} = \int e^{-\Phi}\sqrt{\det(G+B)}~.
\ee
In the elliptical array (and therefore in its circular limit), the B-field~\eqref{LMellipse} vanishes in the interior of the ellipse in the $x^1$-$x^2$ plane, at the origin in $x^3$-$x^4$.
The dilaton and transverse metric~\eqref{LMellipse} are both proportional to the harmonic function $\sfH$, and for D1-branes these factors cancel in the DBI action.  Thus D-strings just see a flat background, and consequently run along straight lines across the interior of the ellipse. 

It is useful to derive the effective action for D-strings stretching between fivebranes in the elliptical source distribution, using the bipolar coordinates \eqref{ellbip}.  For simplicity, we will focus on D-strings inside the ellipse (corresponding to the $r=0$ section of the geometry) and located at constant $x_1$ (\ie\ parallel to the semi-minor axis):
\begin{equation}
  \label{dstringembeddown}
a_1 \sin\theta \cos\phi = c \, .
\end{equation}
 If we parametrize the brane worldvolume as follows
\begin{equation}
t = \xi_0 \, ,\quad 
\phi =\xi_1 \, ,\quad
\theta = \theta(\xi_1) 
\end{equation}
the brane action is
\begin{equation}
  \label{Dstringactiondown}
\mathcal{L} = \cos \theta \Big[(a_2^2 \cos^2\xi_1
  +a_1^2\sin^2\xi_1)\tan^2 \theta - (a_1^2 - a_2^2) \sin
              2\xi_1 \dot{\theta} \tan \theta + (a_1^2\cos^2\xi_1+
  a_2^2\sin^2\xi_1) \dot{\theta}^2 \Big]^{1/2} \, .
  \end{equation}
  Note that when $a_1 = a_2 = a$ this reduces to
  \begin{equation}
\mathcal{L} = a \sqrt{\sin^2\theta + \dot{\theta}^2\cos^2\theta } \, ,
\end{equation}
as we reviewed for branes in the parafermion disk in~\cite{Martinec:2019wzw}. 
It is straightforward to check that the straight line embedding $\eqref{dstringembeddown}$ solves the
equations of motion following from the action \eqref{Dstringactiondown}. We thus have on-shell
\begin{equation}
 \label{onshellactiondown}
\mathcal{L} = \frac{a_2 c}{a_1 \cos^2 \xi_1} \, .
  \end{equation}
By evaluating the action, and noticing that the location $\phi = \mu$ of the 
endpoints of the D-brane at $\theta = \pi/2$ is
defined by $a_1 \cos \mu = c$, one obtains the mass of the D-brane
\begin{equation}\label{D1mass}
M \propto a_2 \sin \mu \, ,
\end{equation}
which is, as expected, proportional to the length of the D-brane
segment. We will now describe these D-strings in the geometry
upstairs, and discuss how flux quantization discretizes the values of
$\mu$, reproducing the non-perturbative information about the
fivebrane locations discussed in the previous sections.

\subsubsection{Upstairs picture}

We now want to understand how the D-strings inside the ellipse lift in
the 10+2 geometry upstairs. We are interested in the $r=0$ section,
corresponding to the interior of the ellipse. The metric in this limit
can be written in the following way
\begin{align}
\begin{aligned}
\label{deltaG-2}
\Big(e^{-2\Phi}G\Bigr)_{r=0} &\;\!\propto\;  \sum_{i=1}^4 d\tilde x_i^2 +  (\textit{terms~proportional~to~}d\tau) \,,
\end{aligned}
\end{align}
where the factor of proportionality depends only on $n_1$ and $V_4$.
\begin{equation}
\tilde x_1 = \frac{a_1 \sqrt{\gamma}}{\sqrt{a_1 a_2}}\sin\theta \cos\phi \, ,\quad \tilde
  x_2 = \frac{a_2 \sqrt{\gamma}}{\sqrt{a_1 a_2}} \sin\theta \sin\phi\, , \quad \tilde x_3
    +i  \tilde x_4 = \sqrt{a_1 a_2} \cos \theta e^{i (\psi+\sigma)} \, 
  \end{equation}
  and $\gamma  = (a_1^2- a_1 a_2 +a_2^2)$.
  We omit the details of the terms proportional to $d\tau$ as 
  we are choosing the gauge $\tau = \sigma = 0$. Note that the
  coordinates $\tilde x_i$ satisfy the relation
\begin{equation}\label{3dellipsoid}
\frac{a_2}{a_1\gamma }\tilde x_1^2 +
\frac{a_1}{a_2\gamma }\tilde x_2^2 +\frac{1}{a_1 a_2}(\tilde x_3^2 +\tilde x_4^2 )= 1 \, .
\end{equation}
This equation describes a
three-dimensional ellipsoid with two equal axes.

A detailed analysis of D-branes in the
null gauged formalism has been performed in
\cite{Martinec:2019wzw}. As shown there, in the upstairs geometry the
determinant of the 
matrix $M=G +B +\cF$ evaluated on the brane worldvolume vanishes. The
prescription of \cite{Martinec:2019wzw} for extracting the physical
DBI action is to define
\be
\cS_{DBI} = \mub_0\int \sqrt{\cE} ~,
\ee
where
\be
\label{minorkey}
\cE^2\equiv\frac{d}{d\lambda} \det\bigl(M^\dagger M-\lambda\One \bigr)\Bigl|_{\lambda=0}
~~,~~~~
M=G+B+\cF ~~.
\ee
and $\mub_0$ is an overall constant.
We are interested in studying 2+2d branes in the above geometry that
reduces to the D-strings inside the ellipse upon gauging. Moreover,
when $a_1 = a_2$ these ``upstairs'' branes should reduce to the branes
considered in \cite{Martinec:2019wzw}. We thus consider the following embedding
\begin{equation}
  t  = \xi_0  \, ,\quad  \tau = -\xi_3 \, ,\quad \phi =
  \xi_1 \, ,\quad
\psi +\sigma  = \xi_2 \, ,
\end{equation}
and we make an ansatz for the worldvolume flux inspired from the
result of the circular distribution, namely we only turn on a
constant component $\mathcal{F}_{21} = 1$. 
The explicit expression for the matrix $M$ is not illuminating;
one can check that the kernel and cokernel of $M$ are spanned by the left
and right gauge directions on the brane, $\xi_2-\xi_3$ and
$\xi_2+\xi_3$, respectively. Additionally, by extracting the
effective action from the prescription~\eqref{minorkey} we find that
$\sqrt{\cE}$ reproduces the effective
action \eqref{Dstringactiondown}. 
Thus we find that the embedding 
\begin{equation}\label{embeddingup2}
a_1 \sin\theta \cos (\xi_1) = c 
\end{equation}
extremizes the DBI action, and on-shell we reproduce the result \eqref{D1mass}.
Note that the above embedding describes a section of
the ellipsoid \eqref{3dellipsoid}. Let us consider for example the lift of a D-string downstairs stretching along the
semi-minor axis, with endpoints at $\phi = \pm \pi/2$. The corresponding section $\tilde x_1 = 0$ is a two-dimensional oblate spheroid. Note that when $a_1 =
a_2$ this becomes an $\bS^2$ sitting inside a three-sphere, recovering the
description of D-branes in the parafermion disk. We are particularly
interested in the limit $a_2 \rightarrow 0$. In this case the $\tilde
x_1 = 0$ section is again a round $\bS^2$ whose size vanishes,
as one can also see from the $r= 0$ limit of the metric
\eqref{rythetapsi}. 

As mentioned in the previous section, the lift of the D-strings
downstairs to 2+2d branes upstairs provides a
semi-classical description of the locations of the fivebranes along
the contour $\F(v)$ through flux quantization. We can see this
explicitly in the ellipse example. In fact the discussion precisely
parallels the one in \cite{Martinec:2019wzw} for a circular
arrays of fivebranes. The brane worldvolume flux $\mathcal{F}$ can be
written in bipolar coordinates, by using the embedding
equation \eqref{embeddingup2}, as 
\begin{equation}
\mathcal{F} = \pm\frac{n_5 c \cot \theta }{\sqrt{a_1^2 \sin^2\theta - c^2}} \,d\theta
\wedge d\psi_{\NS} ~,
\end{equation}
where the two signs refer to the two branches of~\eqref{embeddingup2} (the north and south hemispheres of the brane).
This is the same result one obtains for the circular distribution of
fivebranes. Defining $\cos \mu = c/a_1$, flux quantization imposes
(see equation (6.16) in \cite{Martinec:2019wzw})
\begin{equation}
\mu = 2\pi j /n_5 \, ,\quad j = 0, \frac12, 1, \dots , \frac12 n_5 \, .
 \end{equation}
This relation discretizes the location of the N/S poles of the spheroidal branes, and thus of the endpoints of the D-strings along the fivebrane locus downstairs.
 
 D-branes in the elliptical supertube will have a similar structure.  One distinction from the Coulomb branch configuration is the loss of a moduli space for the D-string.  On the Coulomb branch, the D-brane can freely slide up and down along the fivebrane worldvolume as one sees in figure~\ref{fig:EllipseCoul}.  However, for the supertube of figure~\ref{fig:EllipseST}, trying to slide the D-string stretching between fivebrane strands will change its length; there will be some equilibrium position $v^*$, which moves up the $\ytil$ axis as $t$ increases so as to hold $v^*=t+\ytil$ fixed.  This behavior contrasts with the round supertube analyzed in~\cite{Martinec:2019wzw}, where the supertube spiral has an isometry along $v$, and thus the D-string in that example did have a moduli space corresponding to sliding it along the fivebrane strands.

\section{Discussion}
\label{sec:discussion}

In this work, we have pieced together a rather comprehensive picture of the 1/2-BPS configuration space of the onebrane-fivebrane system at the fully stringy level.  The ingredients of this portrait include the Lunin-Mathur supergravity solutions~\eqref{LMgeom} written in terms of a source profile $\F$ in~\eqref{greensfn}, and the duality to a Landau-Ginsburg description whose superpotential~\eqref{ST spotl} is specified in terms of this same source profile.  At special points in this configuration space corresponding to circular source profiles, the worldsheet theory can be solved exactly as a null gauged Wess-Zumino-Witten model.  Linear deformations away from these special points are in one-to-one correspondence with BPS vertex operators on the worldsheet; we exhibited the spectrum of these vertex operators and showed how they map to linearized deformations of the profile functions.  Furthermore, a prescription for the map at the fully nonlinear level was given.  We analyzed one particular deformation at the fully nonlinear level, wherein the circular profile is deformed to an ellipse.

The advantage of the null-gauging formalism is that it allows us to see certain effects that are non-perturbative in $\alpha'$ via a semiclassical analysis, such as how D-brane probes can be used to localize the fivebrane source, information that is invisible in the supergravity approximation.  So for instance, the round supertube with tilt quantum $k$ has a local $\bZ_k$ orbifold structure, but the stringy description~\cite{Martinec:2017ztd,Martinec:2018nco,Martinec:2019wzw} shows that the cycles are non-vanishing, and the D-branes that wrap them have nonzero mass.  On the other hand, the $\bZ_2$ singularity that results from a self-intersecting profile $\F$ {\it will} have vanishing cycle sizes and a strong coupling singularity, as we see for instance when the elliptical profile degenerates.

We also developed aspects of FZZ duality, which is an instance of the Calabi-Yau/Landau-Ginsburg correspondence.  For supertubes, this duality implies that the background geometry comes supplied with a string condensate near the fivebrane source, and perturbations of the geometry are inextricably linked to perturbations of this condensate.  For the circular source profile, the duality is tied to operator identifications in the worldsheet WZW model that relate graviton vertex operators to ``tachyon'' vertex operators for string winding modes; we showed that these identifications are a consequence of coordinate identifications in the loop group.  This string condensate is encoded in the worldsheet superpotential of the FZZ dual description.  The consequences of FZZ duality for the string spectrum are a generic feature of the 1/2-BPS configuration space, and the appearance of singularities on the geometry side due to self-intersections of the source profile are mirrored by degenerations of the Landau-Ginsburg superpotential and twisted superpotential.

Indeed, what we see going on in the cap is an iteration of the decoupling limit that produced the fivebrane throat to begin with.  Recall that there are two routes to little string theory, related by mirror symmetry.  The first brings together $\nfive$ fivebranes (in type IIA/IIB string theory), while the second shrinks the vanishing cycles of an $A_{\nfive-1}$ singularity (in type IIB/IIA string theory); in the process, one keeps fixed the mass of W-branes -- D-branes stretched between the NS5's or wrapped on the vanishing cycles of the ALE space.  The ``capped throat'' of the fivebrane description is the decoupling limit of the geometry sourced by the fivebranes, while the Landau-Ginsburg description is naturally related to the algebraic presentation of the ALE space as the hypersurface $\cW=0$~\cite{Giveon:1999px,Giveon:1999tq}.  These two decoupling limits do not lead to different theories, but rather to two different descriptions of the same theory~-- little string theory.  

When the source profile of the Lunin-Mathur geometry self-intersects, locally an $A_1$ singularity develops in the geometry~\cite{Lunin:2002iz}.  In the dual Landau-Ginsburg description, two of the zeroes of the superpotential coalesce, signalling the development of a ``little throat'' of two fivebranes coming together.  Thus, even after having taken the decoupling limit of $\nfive$ fivebranes, there are further limits where a subset of the fivebranes come together, and these further limits are an iteration of the Calabi-Yau/Landau-Ginsburg duality for this subsector.  But this further limit is taking place far down the throat of the full system of $\nfive$ fivebranes, which is already described by little string theory. We are seeing the dynamics of little string theory through its dual description in perturbative string theory and the dualities of the worldsheet dynamics, and these worldsheet dualities give us a window into the processes near threshold that initiate near-extremal black hole formation.

This picture of the 1/2-BPS configuration space reinforces the notion that the process of black hole formation in the onebrane-fivebrane system is a Coulomb-Higgs phase transition, in which\footnote{More precisely, the Coulomb branch states can be described on the Higgs branch (in this case little string theory), so the distinction is between Coulomb and ``pure-Higgs'' states~\cite{Bena:2012hf} which cannot be described in the Coulomb branch effective field theory; for more details see the comments in~\cite{Martinec:2019wzw} and references within.} 
\begin{itemize}
\vspace{-0.5mm}
\item
Smooth capped (though perhaps stringy) geometries are realized when fivebranes are slightly separated on their Coulomb branch; 
\vspace{-0.5mm}
\item
Strong coupling and/or infinite redshifts develop when fivebranes manage to come together; and 
\vspace{-0.5mm}
\item
W-branes which become light in that limit condense to push the system slightly onto the Higgs branch, which is the black hole phase.
\vspace{-0.5mm}
\end{itemize}
The formation of stretched fundamental strings in D-brane scattering was studied in~\cite{Douglas:1996yp}, where it was shown that in the perturbative regime, stretched strings are likely to form when the impact parameter of the D-branes is less than $O(v^{1/2}\lstr)$, where $v$ is the relative velocity of the D-branes.  The formation of W-branes between approaching strands of NS5-branes is the S-dual of this process.

We can now present a refinement of the picture of evolution of perturbed BPS states given in~\cite{Eperon:2016cdd,Marolf:2016nwu}, which analyzed such perturbations at the level of supergravity (see also~\cite{Keir:2016azt,Keir:2018hnv,Chakrabarty:2019ujg}).  A feature of 1/2-BPS Lunin-Mathur geometries (and the related 3-charge geometries obtained by fractional spectral flow in the spacetime CFT~\cite{Giusto:2004id,Giusto:2004ip,Giusto:2012yz}) is that the supertube source profile is the locus of an {\it evanescent ergosurface}, where the globally null Killing vector field $\partial_u$ becomes orthogonal to the Killing vector field $\partial_y$.
In~\cite{Eperon:2016cdd}, it was noted that perturbations localized along the evanescent ergosurface cost very little energy, and null geodesics are stably trapped there; it was then argued that such perturbations could serve as the source of a non-linear classical instability of the geometry.
(Of course, the geometry cannot be linearly unstable, since it lies on the BPS bound.) 

The analytic analysis in~\cite{Eperon:2016cdd} that led to these conclusions was performed in an ``eikonal limit'' of large angular momentum $j$ for the classical perturbations (and further supported by numerics for general $j$).  However in light of our results, one can simply work in the fivebrane decoupling limit and at finite $j$, as we have done above; the zero-energy quantum perturbations are then none other than our 1/2-BPS vertex operators (the corresponding solutions to the linearized wave equation were obtained in~\cite{Martinec:2018nco,Chakrabarty:2019ujg}%
\footnote{The modes considered in detail in these papers were gravitons with polarization  along the $\bT^4$ compactification, which are simpler to analyze but are not BPS.  The zero-energy BPS modes considered here are gravitons with polarization along the transverse space to the fivebranes, together with associated RR fields and fermions in the same supermultiplet.}).
The evanescent ergosurface is none other than the locus of the source profile $\F(v)$, and the tendency of perturbations to be trapped there is simply telling us that the source wants to move in response to the perturbation.

The relevant physics employs the large space of marginal deformations of the supertube profile that can cost little or no energy to excite, and so a perturbed solution can wander off in any of these directions, to a faraway region in configuration space.  However just because the perturbed configuration starts to explore the moduli space does not mean that a singularity develops on a short time scale via some sort of runaway.  Rather, near the BPS bound, the system wanders around in a slow scrambling dynamics along the moduli space of 1/2-BPS configurations, which may eventually lead to strong coupling physics when fivebranes intersect.

A subsequent analysis~\cite{Marolf:2016nwu} suggested that, if the system is allowed to shed angular momentum, it should evolve toward an entropically favored ``typical'' supertube~-- the density of states is peaked within spacetime angular momentum ${\mathfrak J}^3_\su\lesssim \sqrt{\none\nfive}$.  The geometry in this regime is stringy, and so an understanding of the physics of this regime cannot be achieved by studying the supergravity limit. 
This suggested evolution assumes that the system is coupled to some external bath that allows it to shed $\bS^3$ angular momentum with very little cost in energy.  The more general story is that the fivebrane profile starts to change adiabatically once the system is perturbed.  Giving the BPS geometry a little kick by adding a small amount of energy, the system starts to explore the nearby BPS configuration space of supertubes.  If the system can shed angular momentum, then Fermi's Golden Rule suggests that it will do so and indeed evolve toward more compact source profiles due to their larger density of states, but if the system is kept at fixed angular momentum it will gyrate within the available phase space until it reaches a self-intersection and strong-coupling dynamics ensues.
The ensemble of states in this regime is dominated by small black holes (localized in six dimensions)~\cite{Banks:1998dd}. 
If the angular momentum is large, self-intersections are rare because the radius of gyration $a$ of the supertube is large; but as the angular momentum decreases, the supertube is more tightly coiled, and the time decreases for the supertube to come close enough to a self-intersection that strong coupling effects arise.  
The evolution toward self-intersection of a profile with only the first two modes excited is shown in figure~\ref{fig:TwoMode}. These self-intersections allow fivebrane joining/splitting processes and other non-perturbative effects.

\begin{figure}[ht]
\centering
  \begin{subfigure}[b]{0.35\textwidth}
  \hskip 0cm
    \includegraphics[width=\textwidth]{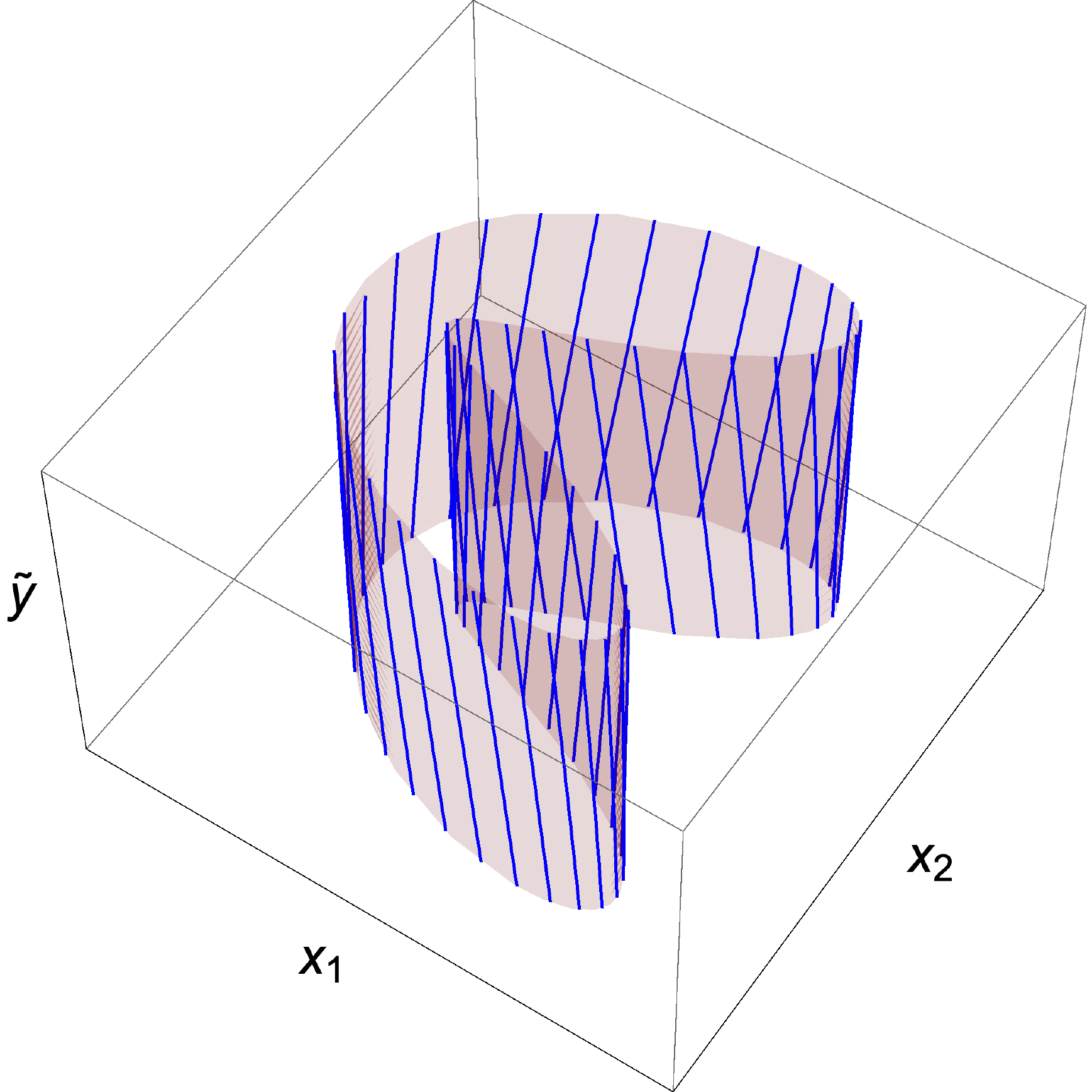}
    \caption{ }
    \label{fig:TwoModeSTn50k1_4}
  \end{subfigure}
\qquad\qquad
  \begin{subfigure}[b]{0.35\textwidth}
      \hskip 0cm
    \includegraphics[width=\textwidth]{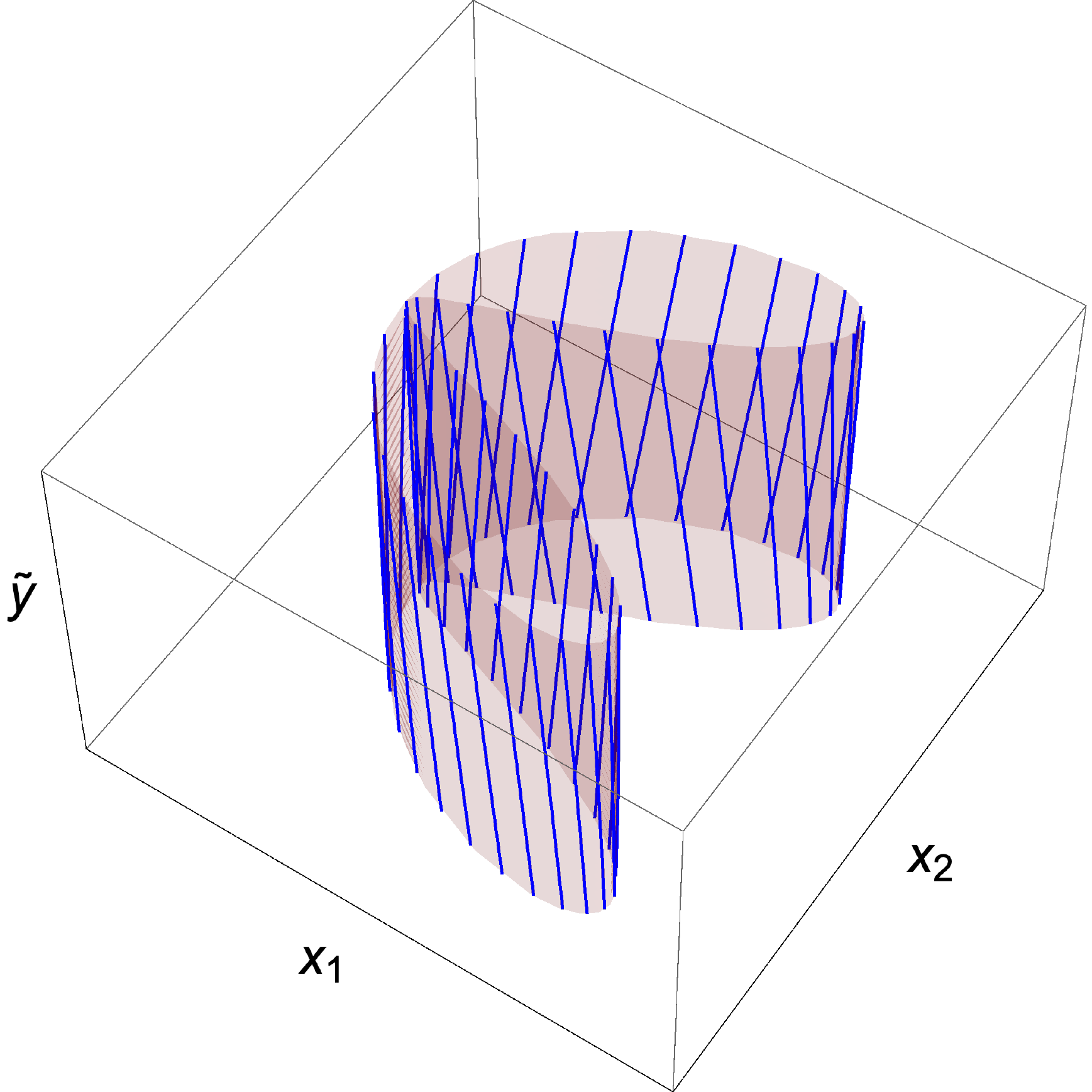}
    \caption{ }
    \label{fig:TwoModeSTn50k1_6}
  \end{subfigure}
\caption{\it 
The evolution of the source profile eventually generates fivebrane self-intersections.  Depicted here are profiles exciting the two lowest modes.  Of course, when an intersection appears in the $x^1$-$x^2$ plane, the source is generically separated in the $x^3$-$x^4$ plane; self-intersections are relatively rare events, but do occur with finite probability.  
}
\label{fig:TwoMode}
\end{figure}

If there is sufficient energy above the BPS bound, another channel is available.  Rather than moving with greater velocity along the moduli space, the supertube can offload some of its angular momentum onto a gas of strings and gravitons higher up the throat, leaving behind a more compact fivebrane supertube profile at smaller radius with a greater likelihood of colliding with itself.  

For three-charge NS5-F1-P systems,  it was argued in~\cite{Bena:2011zw} that the entropically favored three-charge BPS configurations for spacetime angular momenta ${\mathfrak J}^3_{\su} < \half \none\nfive$ involve a black hole plus a supertube (see figure~\ref{fig:MoultingSpectrum})~-- the system sheds its angular momentum onto a supertube at the 1/2-BPS threshold costing no energy, and the energy above the 1/2-BPS bound is piled into a zero angular momentum black hole (which might then relax back to extremality).  This can happen when fivebrane profiles self-intersect as in figure~\ref{fig:TwoMode}, since they may reconnect in such a way that the worldvolume splits into two strands which then exchange energy and angular momentum until reaching the entropically favored configuration.

\begin{figure}[h!]
\centering
\includegraphics[width=.6\textwidth]{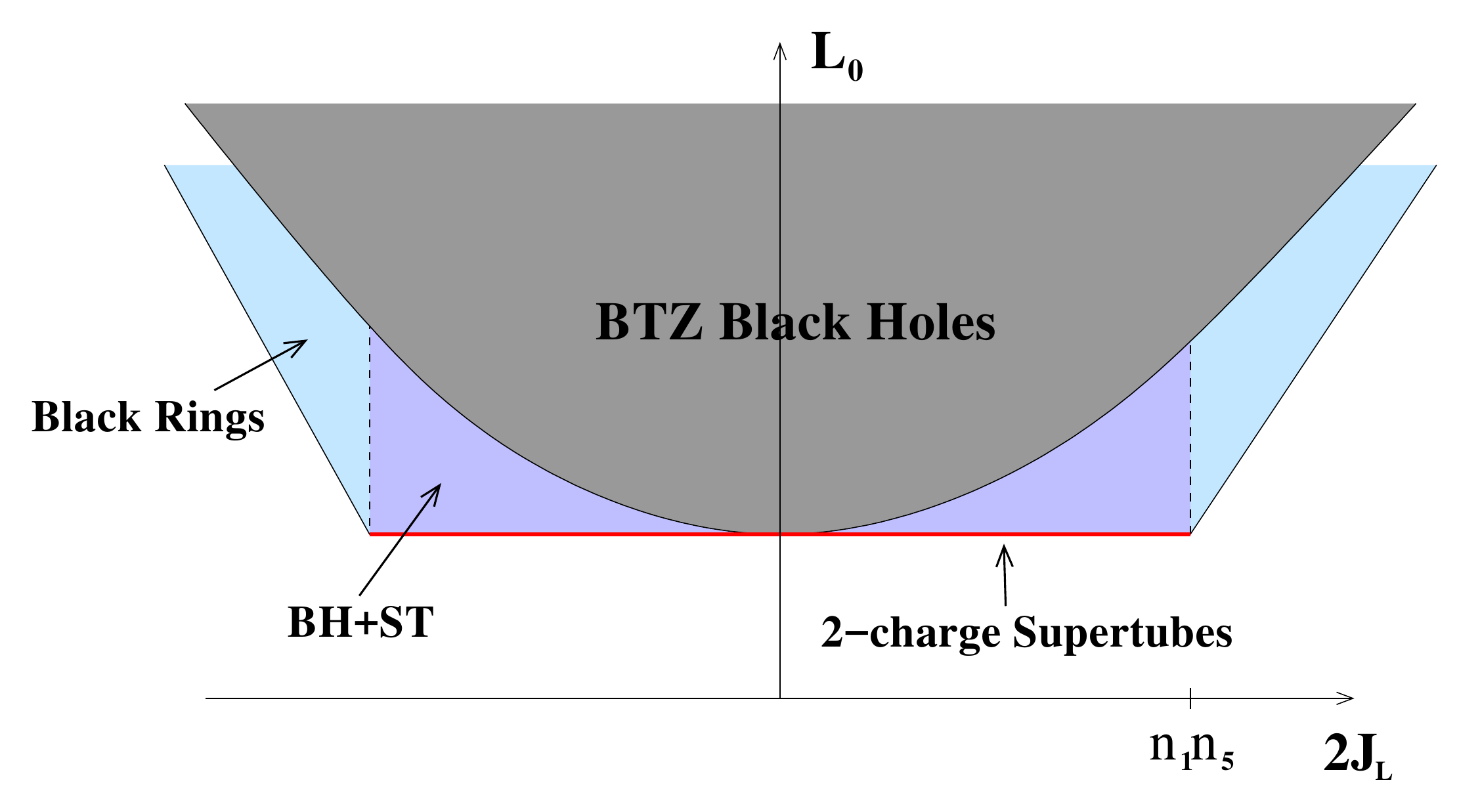}
\caption{\it Phase diagram of the spacetime CFT, where $J_L={\mathfrak J}^3_{\mathrm{su}}$, arising in the $AdS_3$ decoupling limit.}
\label{fig:MoultingSpectrum}
\end{figure}

Of course, the supertube profile need not {\it exactly} self-intersect in order for strong-coupling effects to arise.  The strong coupling effects in question involve W-brane creation; for type IIB, this process is S-dual to the fundamental string creation process analyzed in~\cite{Douglas:1996yp}.  To the extent that their analysis of the effective super Yang-Mills theory applies also to NS5-brane dynamics, the range of impact parameters over which W-brane creation occurs increases with the relative velocity of the supertube segments that are approaching one another. 
Once created, a gas of W-branes will bind together the cluster of fivebranes in question.  Thus in a moduli space approximation, low-energy excitations of the system will correspond to motion along the moduli space.  When segments of the supertube undergo a near-collision, they will have some relative velocity; the more the energy above extremality, the larger this relative velocity will be, and the larger the region over which brane creation processes are likely.  The W-brane creation process appears to be the typical route toward black hole formation in the onebrane-fivebrane system near the 1/2-BPS threshold.

The same sort of analysis may also apply to three-charge superstratum backgrounds, assuming that the known superstratum construction built upon round supertubes~\cite{Bena:2015bea} can be extended to more general 1/2-BPS source profiles.  Given that superstrata can be constructed that have the same quantum numbers as black hole geometries, one expects that the evolution of perturbed superstrata will teach us about the black hole formation process in holographic settings in a framework where we have an unusual degree of control over the bulk description.  

It would be interesting if our explicit elliptical supertube profiles could serve as the seed for new superstratum constructions.
Analogues of the vertex operators~\eqref{BPSverts} which deform the base supertube profile are also expected in superstrata, and there is no energetic barrier to the profile evolving toward a self-intersection as discussed above, leading once again to W-brane creation and the formation of a near-extremal black hole.
The deformations that take one along the generalized superstratum configuration space preserve (0,4) supersymmetry on the worldsheet.  Such vertex operators were briefly discussed in section~\ref{sec:quarterBPS}.  The Landau-Ginsburg side will correspond to (0,4) deformations of the superpotential(s), whose geometrical duals are at present unknown.  We hope that the tools developed here may enable further progress on understanding the 1/4-BPS configuration space, including the microstates of large supersymmetric three-charge black holes.


\vspace{1mm}

\section*{Acknowledgements}

We thank 
Andrea Galliani
for useful discussions, and early collaboration on the construction of the spacetime supercharges.
The work of EJM and SM is supported in part by DOE grant DE-SC0009924 and a UChicago FACCTS collaboration grant. 
EJM also thanks the Simons Foundation for support as a Fellow in Theoretical Physics for 2018-19.
The work of DT is supported by a Royal Society Tata University Research Fellowship. For hospitality during the course of this work, we all thank the CEA Saclay; EJM and SM also thank the Centro de Ciencias de Benasque.

\vspace{1mm}


\appendix


\section{Conventions}
\label{app:conventions}


In this appendix we record several conventions.
\vspace{-1mm}

\refstepcounter{subsection}
\subsection*{\thesubsection \quad SU(2)} \label{sec:sutwoapp}

We parametrize SU(2) via Euler angles as
\be \label{eq:su2-param}
g_{\su} ~=~ e^{\frac{i}{2}(\psi-\phi)\sigma_3}e^{i \theta \sigma_1}e^{\frac{i}{2}(\psi + \phi)\sigma_3} 
~= ~
\begin{pmatrix} \cos\theta \;\! e^{i\psi}  &~~  i \:\!  \sin\theta\:\!  e^{-i\phi} \\ 
i \:\!  \sin\theta\:\!  e^{i\phi}  &~~ \cos\theta \;\! e^{-i\psi} \end{pmatrix}\,
\,.
\ee
These are the conventions used in~\cite{Martinec:2018nco} (note that~\cite{Martinec:2017ztd} had conventions related to these by $\phi \to - \phi$).
Here the $\sigma_a$ are the usual Pauli matrices, explicitly
\be
\sigma_1 = \left(\begin{array}{cc} 0 &1 \\ 1 &
    0 \end{array}\right), \qquad \sigma_2 = \left(\begin{array}{cc} 0 &-i \\ i &
    ~~0  \end{array}\right), \qquad \sigma_3 = \left(\begin{array}{cc} 1 &~~0 \\ 0 &
    -1  \end{array}\right) \,.
\ee
The generators and structure constants of the Lie algebra $\su(2)$ are as usual
\be
T^a_\su = \frac{1}{2}\sigma_a \,, \qquad \quad \T{(f_\su)}{ab}{\:\! c} = i\:\! \epsilon_{abc} \,, \qquad  \epsilon_{123}=1\,.
\ee
We define the following worldsheet current one-forms,
\bea 
\label{eq:J3su-app}
{\jj}^1_{\su} &=& \nfive \tr \left[ (- i T^1_\su )\partial g_{\su} \;\! g_{\su}^{-1}  \right] ~=~  
\nfive\,\big[ \cos(\psi\tight-\phi)\, \partial\theta + \hf \sin2\theta \sin(\psi\tight-\phi) \big(\partial\psi+\partial\phi\big)\big]
\nn\\[3pt]
{\jj}^2_{\su} &=& \nfive \tr \left[ (-i T^2_\su )\partial g_{\su} \;\! g_{\su}^{-1}  \right] ~=~  
\nfive\,\big[ -\sin(\psi\tight-\phi)\, \partial\theta + \hf \sin2\theta \cos(\psi\tight-\phi) \big(\partial\psi+\partial\phi\big)\big]
\nn\\[3pt]
{\jj}^3_{\su} &=& \nfive \tr \left[ (-i T^3_\su )\partial g_{\su} \;\! g_{\su}^{-1}  \right] ~=~  \nfive\bigl[ \cos^2\!\theta \, \partial \psi - \sin^2 \!\theta \,\partial \phi \bigr]
\\[3pt]
\bar{\jj}^1_{\su} &=&\nfive\tr \big[ (-i T^1_\su) g_{\su}^{-1} \bar\partial g_{\su} \;\!   \big] ~=~ 
\nfive\,\big[ \cos(\psi\tight+\phi)\, \bar\partial\theta + \hf \sin2\theta \sin(\psi\tight+\phi) \big(\bar\partial\psi-\bar\partial\phi\big)\big]
\nn\\[3pt]
\bar{\jj}^2_{\su} &=&\nfive\tr \big[ (-i T^2_\su) g_{\su}^{-1} \bar\partial g_{\su} \;\!   \big] ~=~ 
\nfive\,\big[ \sin(\psi\tight+\phi)\, \bar\partial\theta - \hf \sin2\theta \cos(\psi\tight+\phi) \big(\bar\partial\psi-\bar\partial\phi\big)\big]
\nn\\[3pt]
\bar{\jj}^3_{\su} &=&\nfive\tr \big[ (-i T^3_\su) g_{\su}^{-1} \bar\partial g_{\su} \;\!   \big] ~=~ \nfive\bigl[ \cos^2\!\theta \,\bar\partial \psi + \sin^2 \!\theta \, \bar\partial \phi \bigr] ~.
\nn
\eea
The relation between the one-forms $\jj_\su^a$ and the current operators $J_\su^a$ is
\be
\label{one-forms to ops}
J_\su^a = i\:\! \jj_\su^a  ~,\quad~~  \bar J_\su^a = -i \:\!  C \:\! \bar \jj_\su^a C^{-1} ~,
\ee 
where $C$ is the charge conjugation matrix 
($C\tight=i\sigma_2$ in the fundamental representation, and thus one has $\bar J_\su^1=i\bar \jj_\su^1, \bar J_\su^2=-i\bar \jj_\su^2, \bar J_\su^3=i\bar \jj_\su^3$; note in particular that $J_\su^\pm\tight= J_\su^1\tight\pm iJ_\su^2\tight=i\jj_\su^\pm$, but $\bar J_\su^\pm = i\bar \jj_\su^\mp$).
Note that $J_\su^a$ is related to $\bar J_\su^a$ by flipping the sign of $\phi$, as it should be.

Corresponding to the current operators ${J}^a_{\su}$ are vector fields $\sfV_\su^a$ which are right-invariant in SU(2) (and which generate left translations), and corresponding to the $\bar{J}^a_{\su}$ are vector fields $\bar{\sfV}_\su^a$ which are left-invariant in SU(2). In our conventions these vector fields satisfy
\be
[\sfV_\su^a,\sfV_\su^b] = i\epsilon_{abc} \sfV_\su^c \;, \qquad [\bar{\sfV}_\su^a,\bar{\sfV}_\su^b] = i\epsilon_{abc} \bar{\sfV}_\su^c \;.
\ee
(explicitly, the components of $J_\su^a$ and $\sfV_\su^a$ are related by 
$(J_\su^{a})_i=2 G_{ij}(\sfV_\su^a)^{j} $, similarly $(\bar J_\su^{a})_i=2 G_{ij}(\bar\sfV_\su^a)^{j} $).
The action of $J_\su^\pm,\bar J_\su^\pm$ then fills out the center-of-mass wavefunctions, in particular one has
\be
\label{su2rep}
\Psi_{j,-j,j} = e^{-i 2j\phi} \sin^{2j}\theta
~~,~~~~
\Psi_{j,j,-j} = e^{i 2j\phi} \sin^{2j}\theta
\ee
(up to an overall normalization).
From~\eqref{eq:J3su-app}--\eqref{su2rep} one can write explicit expressions for the vertex operators~\eqref{ellipsevertex}, as well as the linearized deformations~\eqref{linear ellipse thetas} that lead to the elliptical source profile; and more generally all the other 1/2-BPS vertex operators~\eqref{BPSverts} that deform the background.  

The supersymmetric $\sutwo$ level $\nfive$ current algebra consists of current operators $ J_\su^a$ and their fermionic superpartners $\psi_\su^a$ having the OPE structure
\begin{align}
J_\su^a(z)\, J_\su^b(0) &\sim \frac{\frac12 \nfive \, \delta^{ab}}{z^2} + \frac{i\epsilon_{abc} J_\su^c(0)}{z}
\nn\\
J_\su^a(z)\, \psi_\su^b(0) &\sim i\epsilon_{abc}\frac{\psi^c_\su(0)}{z}
\\
\psi_\su^a(z)\, \psi_\su^b(0) &\sim \frac{\delta^{ab}}{z}
\nn
\end{align}
with the Killing metric $\delta^{ab} = {\rm diag}(+1,+1,+1)$ (recalling that the current operators differ from~\eqref{eq:J3su-app} by a factor of $i$).  One can define a set of ``bosonic'' $\sutwo$ level $\nfivetil=\nfive-2$ currents $j_\su^a$ that commute with the fermions,
\be
j_\su^a = J_\su^a + \frac i2\epsilon_{abc}\psi_\su^b \psi_\su^c  ~.
\ee
The primary fields $\Psihat_{\jsu \msu \bmsu}$ of the current algebra have conformal dimensions
\be
h = \bar h = \frac{\jsu(\jsu+1)}{\nfive} ~;
\ee
unitarity restricts the allowed spins $\jsu$ of the underlying bosonic current algebra to the range
\be
\label{su2 reps app}
\jsu = 0,\frac12,\dots,\frac{\nfive}{2} -1 ~.
\ee

The primary operators of the supersymmetric theory are built by combining primaries $\Psi_{\jsu \msu \bmsu}$ of the level $\nfivetil$ bosonic current algebra with level two primaries of the fermions (namely the identity operator $\One$, spin operator $\Sigma$, and the fermions themselves).  The highest weight fields are of three types:
\be
\label{su2hwr}
\Psihat_{jjj} = \Psi_{jjj}
~~,~~~~
\Psihat_{j+\half,j+\half,j+\half} = \Psi_{jjj}\Sigma_{++}
~~,~~~~
\Psihat_{j+1,j+1,j+1} = \Psi_{jjj}\psi^+\bar\psi^+
~~,
\ee
with the remaining operators of the zero mode multiplet obtained through the action of the zero modes of the total current $J^-_\su$.  In building massless string states, one uses the purely bosonic highest weight operator for NS sector states whose polarization does not lie along $\sutwo$, and the highest weight operator with a fermion attached when the polarization does lie along $\sutwo$; Ramond sector operators involve $\Sigma$, with the various spinor polarizations reached through the action of the zero modes of $\psi^-_\su,\bar\psi^-_\su$.  In the type II string, the choice of fermion decoration is independent on left and right, and there is a chiral GSO projection onto odd total fermion number in the matter sector.

These operators have a {\it parafermion} decomposition%
\footnote{Our notation here largely follows~\cite{Martinec:2001cf}, see also~\cite{Giveon:2015raa}, except that we work in conventions where $\alpha'=1$, so that T-duality is $R\to 1/R$, instead of the convention $\alpha'=2$ of those works.}
under the current $J_\su^3$%
~\cite{Fateev:1985mm,Gepner:1986hr,Gepner:1987qi}
obtained by extracting the dependence on $J^3_\su, \bar J^3_\su$.  To this end, one bosonizes the currents
\begin{align}
j^3_\su=i\sqrt{\nfivetil}\,\partial Y_\su 
~&,~~~~
\psi^+_\su \psi^-_\su = i\sqrt2\,\partial H_\su ~,
\nn\\
J^3_\su=i\sqrt{\nfive}\,\partial \cY _\su
~&,~~~~
J_\cR^\su = \frac\nfivetil\nfive \psi^+_\su\psi^-_\su - \frac2\nfive j^3_\su
= i\sqrt{\frac{2\nfivetil}{\nfive}} \, \partial \cH_\su ~,
\end{align}
and similarly for the right-movers.  The current $J^3_\su$ forms a $U(1)$ supermultiplet with the fermion $\psi^3_\su$, and every operator in the super-WZW model can be written as a product of a parafermion operator and an operator from the super-$U(1)$ theory.  
In particular, the $\sutwo$ primary field $\Psihat_{\jsu \msu \bmsu}$ can be decomposed as
\be
\label{supf app}
\Psihat_{\jsu \msu \bmsu} = \medhat\Lambda_{\jsu \msu \bmsu}
\,\exp\Bigl[i\frac2{\sqrt\nfive}\Bigl(\msu \cY_{\!\su}+\bmsu \bar\cY_{\!\su}\Bigr)\Bigr] ~.
\ee
The conformal dimension of the $\sutwo$ primary $\Lambda_{\jsu \msu \bmsu}$ decomposes as
\be
\label{supfspec-app}
h\big[\medhat\Lambda_{\jsu \msu \bmsu}\big] = \frac{\jsu (\jsu+1)-\msu^2}{\nfive} 
~~,~~~~
\bar h\big[\medhat\Lambda_{\jsu \msu \bmsu}\big] = \frac{\jsu (\jsu+1)-\bmsu^2}{\nfive} 
\ee
with the rest made up by the dimension of the $\cY,\bar\cY$ exponentials.  

The shift $\msu \to (\msu \!+\!\frac12\nfive \wsu)$ in the $\cY$ exponential in $\Psihat_{\jsu \msu \bmsu}$ in~\eqref{supf app} defines the left spectral flow of interest to us here.  The states flowed in this way have a shifted exponential but the same underlying superparafermion state; their conformal dimensions are
\be
h\big[\Psihat^{(\wsu,\bwsu)}_{\jsu \msu \bmsu}\big] = \frac{\jsu
  (\jsu+1)}{\nfive} + \msu \wsu +\frac{\nfive}{4} \wsu^2
\ee
and similarly for the right-handed spectral flow. 


\refstepcounter{subsection}
\subsection*{\thesubsection \quad SL(2)} \label{sec:sltwoapp}

We parametrize $\sltwo$ as $SU(1,1)$ via
\be
g_{\sl} \;=\; e^{\frac{i}{2}(\tau-\sigma)\sigma_3}e^{\rho \sigma_1}e^{\frac{i}{2}(\tau + \sigma)\sigma_3} \,.
\ee
Again these are the same conventions used in~\cite{Martinec:2018nco} (related to those of~\cite{Martinec:2017ztd} by $\sigma \to - \sigma$).\\
The generators and structure constants of the Lie algebra $\su(1,1)$ are 
\be
T^1_{\sl} \,=\, \frac{i}{2}\sigma_1 \,, \quad~ T^2_{\sl} \,=\, \frac{i}{2}\sigma_2 \,, \quad~ T^3_{\sl} \,=\, \frac{1}{2}\sigma_3 \;; \qquad~~%
 \T{(f_{\sl})}{12}{3} \,=\, -i \,, \quad \T{(f_{\sl})}{23}{1} \,=\, \T{(f_{\sl})}{31}{2} \,=\,  i \,.
\ee
We define the left- and right-invariant one-forms $\jj^a_{\sl}$, $\bar{\jj}^a_{\sl}$ as follows:
\bea 
\label{eq:J3sl}
{\jj}^1_{\sl} &=& \nfive \tr \left[ (-i T_\sl^1 )\partial g_{\sl} \;\! g_{\sl}^{-1}  \right] ~=~  
\nfive\,\big[ \cos(\tau\tight-\sigma)\, \partial\rho + \hf \sinh2\rho \,\sin(\tau\tight-\sigma) \big(\partial\tau+\partial\sigma\big)\big]
\nn\\[3pt]
{\jj}^2_{\sl} &=& \nfive \tr \left[ ( -i T_\sl^2 )\partial g_{\sl} \;\! g_{\sl}^{-1}  \right] ~=~  
\nfive\,\big[ -\sin(\tau\tight-\sigma)\, \partial\rho + \hf \sinh2\rho \,\cos(\tau\tight-\sigma) \big(\partial\tau+\partial\sigma\big)\big]
\nn\\[3pt]
{\jj}^3_{\sl} &=& \nfive \tr \left[ (-i T_\sl^3 )\partial g_{\sl} \;\! g_{\sl}^{-1}  \right] ~=~  \nfive\bigl[ \cosh^2\!\rho \, \partial \tau + \sinh^2 \!\rho \,\partial \sigma \bigr]
\\[3pt]
\bar{\jj}^1_{\sl} &=&\nfive\tr \big[ ( -i T_\sl^1) g_{\sl}^{-1} \bar\partial g_{\sl} \;\!   \big] ~=~ 
\nfive\,\big[ \cos(\tau\tight+\sigma)\, \bar\partial\rho + \hf \sinh2\rho \,\sin(\tau\tight+\sigma) \big(\bar\partial\tau-\bar\partial\sigma\big)\big]
\nn\\[3pt]
\bar{\jj}^2_{\sl} &=&\nfive\tr \big[ ( -i T_\sl^2) g_{\sl}^{-1} \bar\partial g_{\sl} \;\!   \big] ~=~ 
\nfive\,\big[ \sin(\tau\tight+\sigma)\, \bar\partial\rho - \hf \sinh2\rho \,\cos(\tau\tight+\sigma) \big(\bar\partial\tau-\bar\partial\sigma\big)\big]
\nn\\[3pt]
\bar{\jj}^3_{\sl} &=&\nfive\tr \big[ (-i T_\sl^3) g_{\sl}^{-1} \bar\partial g_{\sl} \;\!   \big] ~=~ \nfive\bigl[ \cosh^2\!\rho \,\bar\partial \tau - \sinh^2 \!\rho \, \bar\partial \sigma \bigr] ~.
\nn
\eea
Once again one has the relations
\be
\label{SL(2) one-forms to ops}
J^a_\sl = i\, \jj_\sl^a  ~~,~~~~  \bar J^a_\sl = -i \;\! C \;\! \bar \jj_\sl^a C^{-1}  
\ee
(again $J_\sl^a$ and $\bar J_\sl^a$ are related by flipping the sign of $\sigma$).
Corresponding to the ${J}^a_{\sl}$ are vector fields $\sfV^a_{\sl}$ which are right-invariant in SU(1,1), and corresponding to the $\bar{J}^a_{\sl}$ are vector fields $\bar{\sfV}_{\sl}^a$ which are left-invariant in SU(1,1). In our conventions these satisfy
\be
[\sfV_{\sl}^a,\sfV_{\sl}^b] =  \T{(f_{\sl})}{ab}{c} \;\! \sfV_{\sl}^c \;, 
\qquad [\bar{\sfV}_{\sl}^a,\bar{\sfV}_{\sl}^b] = \T{(f_{\sl})}{ab}{c} \;\!  \bar{\sfV}_{\sl}^c ~~,
\ee
(note that the components of $J_\sl^a$ and $\sfV_\sl^a$ are related by 
$(J_\sl^{a})_i=-2 G_{ij}(\sfV_\sl^a)^{j} $, similarly $(\bar J_\sl^{a})_i=-2 G_{ij}(\bar\sfV_\sl^a)^{j} $ due to the fact that the metric $-\frac{n_5}{2}\tr((g^{-1}_\sl dg_\sl)^2)$ differs from the $AdS_3$ metric by an overall sign).
One can check that the operators $J_\sl^a,\bar J_\sl^a$ have the correct algebra acting on center of mass wavefunctions, \eg\
\be
\label{sl2rep}
\Phi_{j,j,j} = \frac{e^{-i2j\tau}}{\cosh^{2j}\!\rho}
~~,~~~~
\Phi_{j,-j,-j} = \frac{e^{i2j\tau}}{\cosh^{2j}\!\rho} ~.
\ee

The supersymmetric $\sltwo$ level $\nfive$ current algebra consists of currents $J_\sl^a$ and their fermionic superpartners $\psi_\sl^a$ having the OPE structure
\begin{align}
J_\sl^a(z)\,J_\sl^b(0) &\sim \frac{\frac12 \nfive \,h^{ab}}{z^2} +  \frac{\T{(f_{\sl})}{ab}{c} J_\sl^c(0)}{z}
\nn\\
J_\sl^a(z)\, \psi_\sl^b(0) &\sim \T{(f_{\sl})}{ab}{c}\frac{\psi^c_\sl(0)}{z}
\\
\psi_\sl^a(z)\, \psi_\sl^b(0) &\sim \frac{h^{ab}}{z}
\nn
\end{align}
with the Killing metric $h^{ab} = {\rm diag}(+1,+1,-1)$.  One can similarly define a set of ``bosonic'' $\sltwo$ level $\nfivehat\!=\!\nfive\!+\!2$ currents $j_\sl^a$ that commute with the fermions,
\be
j_\sl^a = J_\sl^a + \frac 12 \T{(f_{\sl})}{a}{bc}\psi_\sl^b \psi_\sl^c  ~.
\ee
The primary fields $\Phihat_{\jsl \msl \bmsl}$ of the current algebra have conformal dimensions
\be
h = \bar h = -\frac{\jsl (\jsl-1)}{\nfive} ~.
\ee

As before the primary operators of the supersymmetric theory are built by combining primaries $\Phi_{\jsl \msl \bmsl}$ of the level $\nfivehat$ bosonic current algebra with level minus two primaries of the fermions $\One,\Sigma,\psi$.  The highest weight fields are similarly built by tensoring highest weight fields in the bosonic theory with one of these three, with the remaining operators of the zero mode multiplet obtained through the action of the zero modes of the total current $J^-_\sl$.  In building massless string states, one again uses the purely bosonic highest weight operator for NS sector states whose polarization does not lie along $\sltwo$, and the highest weight operator with a fermion attached when the polarization does lie along $\sltwo$; and Ramond sector operators involve $\Sigma$, with the various spinor polarizations reached through the action of the zero modes of $\psi^-_\sl,\bar\psi^-_\sl$. Again, in the type II string, the choice of fermion decoration is independent on left and right, and there is a chiral GSO projection onto odd total fermion number in the matter sector.

These operators also have a superparafermion decomposition under the current $J_\sl^3$%
~\cite{Dixon:1989cg,Griffin:1990fg,Dijkgraaf:1991ba}%
\footnote{Again our notation here largely follows~\cite{Martinec:2001cf}, see also~\cite{Giveon:2015raa}.}
obtained by extracting the dependence on $J^3_\sl, \bar J^3_\sl$.  To this end, one bosonizes the currents
\begin{align}
j^3_\sl=-i\sqrt{\nfivehat}\,\partial Y_\sl 
~&,~~~~
\psi^+_\sl \psi^-_\sl = i\sqrt2\,\partial H_\sl ~,
\nn\\
J^3_\sl=-i\sqrt{\nfive}\,\partial \cY _\sl
~&,~~~~
J_\cR^\sl = \frac\nfivehat\nfive \psi^+_\sl\psi^-_\sl + \frac2\nfive j^3_\sl
= i\sqrt{\frac{2\nfivehat}{\nfive}} \, \partial \cH_\sl ~,
\end{align}
and similarly for the right-movers.  Note that the bosons $\cY,\bar\cY$ and $Y,\bar Y$ are timelike.  The $\sltwo$ primary field $\Phihat_{jm\mbar}$ can then be decomposed as
\begin{equation}
\label{slpf app}
\Phihat_{\jsl \msl \bmsl} = \medhat V_{\jsl \msl \bmsl }
\,\exp\Bigl[i\frac2{\sqrt\nfive}\Bigl(\msl \cY_{\!\sl}+\bmsl\bar\cY_{\!\sl}\Bigr)\Bigr] ~.
\end{equation}
The conformal dimension of the $\sltwo$ primary $\Phihat_{\jsl \msl \bmsl}^{\sl}$ decomposes as
\begin{equation}
\label{slpfspec-app}
h\big[\medhat V_{\jsl \msl \bmsl}\big] = \frac{-\jsl (\jsl-1)+\msl^2}{\nfive} 
~,~~~~
\bar h\big[\medhat V_{\jsl \msl \bmsl}\big] = \frac{-\jsl (\jsl-1)+\bmsl^2}{\nfive} ~,
\end{equation}
with the rest made up by the dimension of the $\cY,\bar\cY$
exponentials.  Again the fields $V_{\jsl \msl \bmsl}$ commute with the current $J^3_\sl$, and so are the natural building blocks for representations of the gauged theory.
The shift of the $J^3_\sl$ charge $\msl \to (\msl \!+\!\frac12 \nfive \wsl)$ leads to the flowed conformal dimension 
\be
h\bigl[V^{(\wsl,\bwsl)}_{\jsl \msl \bmsl}\bigr] = -\frac{\jsl
  (\jsl-1)}{\nfive} - \msl \wsl - \frac{\nfive}{4}  \wsl^2 \;.
\ee
Because we are working on the universal cover of $\sltwo$, the $\tau$ direction is non-compact and so there is no independent left and right spectral flow, but rather a simultaneous spectral flow that shifts $\msl,\bmsl$ by the same amount.

Unitary representations of bosonic $\sltwo$ current algebra are as follows. One has the principal discrete series (on both left and right)
\begin{equation}
\cD_j^+ = \bigl\{ \ket{j,m}~\bigl| ~  j\in\IR_+\, ;~~ m\!=\! j+n\, ,~~n\in\IN \bigr\}
\end{equation}
and its conjugate 
\begin{equation}
\cD_j^- = \bigl\{ \ket{j,m}~\bigl| ~  j\in\IR_+\, ;~~ m\!=\! -(j+ n)\,
,~~ n\in\IN \bigr\} \,,
\end{equation}
restricted to the range
\be
\frac 12 < j < \frac{\nfive+1}{2} ~;
\end{equation}
in addition one has the continuous series representations $\cC_j^\alpha$ (again on both left and right)
\begin{equation} \label{eq:cont-series-app}
\cC_j^\alpha = \bigl\{ \ket{j,m}~\bigl| ~  j\!=\!
\coeff12(1+i\nu)\, ,~~\nu \in\IR\, ;~~m\!=\! \alpha+n\, ,~~n\in\IZ\, ,~~0\!\le\alpha\!<1\in\IR \bigr\} ~.
\end{equation}


\section{Spacetime supersymmetry}
\label{app:spacetimesusy}

Spacetime supersymmetry is perhaps simplest to discuss in a framework where the worldsheet fermions are bosonized as
\begin{align}
\label{eq:bosoniz}
\psi^+_\sl\psi^-_\sl = i\sqrt2\partial H_\sl
~,~~~~
\psi^+_\su\psi^-_\su &= i\sqrt2\partial H_\su
~,~~~~
\psi^3_\sl\psi^3_\su = i\sqrt2\partial H_3~,
\nn\\
\psi^t\psi^y = i\sqrt2\partial H_{ty}
~,~~~~~
i\chi^6\chi^7 &= i\sqrt2\partial H_{67}
~,~~~~~
i\chi^8\chi^9 = i\sqrt2\partial H_{89}~,
\end{align}
where $\chi^i$ are the worldsheet superpartners of the $\bT^4$ coordinates $X^i$, $i=6,7,8,9$.  
The operators
\be
\label{O(10,2) spinfield}
S_\varepsilon = \exp\Big[ \frac{i}{\sqrt2}\Big(\varepsilon_\sl H_\sl + \varepsilon_\su H_\su + \varepsilon_3 H_3 + \varepsilon_{ty} H_{ty} + \varepsilon_{67} H_{67} + \varepsilon_{89} H_{89} \Big) \Big] ~,
\ee
where each $\varepsilon_a=\pm1$ are the Dynkin indices of the spinor polarization, transform in the spinor representation of $O(10,2)$. 
One can impose the Majorana-Weyl condition $\prod_a \varepsilon_a=\pm1$ independently on left- and right-movers to describe the (10+2)-dimensional analog of either type IIA or type IIB, which gauge down to the usual type IIA and IIB theories respectively.
The spacetime supersymmetry operators in the $-\half$ picture (for $\varphi$) are then%
\footnote{As discussed in~\cite{Giveon:1998ns}, there are two canonical constructions of supercharges in $AdS_3\times\bS^3$~-- the standard spin field of the orthogonal group current algebra of the fermions, dressed by worldsheet ghosts; and the spin field generated by spectral flow in a worldsheet $\cN=2$ $\cR$-symmetry~\cite{Banks:1987cy}.  In the case of Coulomb branch NS5-brane backgrounds, these two constructions differ by a null gauge transformation.  The supersymmetry currents built from~\eqref{O(10,2) spinfield} have suitable OPE's with all vertex operators, while the supersymmetry current built using $\cN=2$ spectral flow may have $\bZ_{\nfive}$ branch singularities with some vertex operators outside the physical spectrum.
}
\be
\label{susy charge}
Q_\varepsilon^{ (-\half)} = \oint e^{-\half(\varphi-\tilde\varphi)} S_\varepsilon ~,
\ee
where we have bosonized the superghosts $\beta,\gamma$ and spinor ghosts for null gauging $\tilde\beta,\tilde\gamma$ according to~\cite{Martinec:1988bg,Takama:1988it,Horowitz:1988xf,Horowitz:1988ip}
\vspace{-4mm}
\begin{align}
\begin{aligned}
\beta\gamma= -\partial\varphi
~~,~~~~
\gamma &=  e^\varphi \eta
~~,~~~~
\beta = e^{-\varphi} \partial\xi
\cr
\tilde\beta\tilde\gamma = -\partial\tilde\varphi
~~,~~~~
\tilde\gamma &=  e^{\tilde\varphi} \tilde\eta
~~,~~~~
\tilde\beta = e^{-\tilde\varphi} \partial\tilde\xi
\end{aligned}
\end{align}
in order to construct the ghost spin field $e^{-\half(\varphi-\tilde\varphi)}$ that intertwines NS and R sector ground states for the ghosts. Note that the conformal dimension $h[e^{\pm\tilde\varphi/2}]=-\frac18$  compensates the added $h=+\frac18$ of the spin field $S_\beta$ of $SO(10,2)$ relative to that of $SO(9,1)$.  There are thus ``pictures'' of physical state vertex operators for both the $\beta,\gamma$ and $\tilde\beta,\tilde\gamma$ systems.

The BRST charge $\cQ$ of equation~\eqref{BRST charges} imposes several constraints on the supersymmetry charge~\eqref{susy charge}.  In the supercurrent term $\gamma G$, the $f_{abc}\psi^a\psi^b\psi^c$ terms in the $\sltwo$ and $\sutwo$ WZW models impose the constraints%
\footnote{Our conventions agree with those of~\cite{Itzhaki:2005tu} Appendix A, apart from our convention $l_1=l_2=1$ for the circular Coulomb branch array, and our bosonization conventions which are related by $(\varepsilon_3,{H}_3)\to  (-\varepsilon_3,-{H}_3)$.}
\be
\varepsilon_\sl\,\varepsilon_\su\,\varepsilon_3 = -1
~,~~~~~
\varepsilon_{ty}\,\varepsilon_{67}\,\varepsilon_{89} = -1  
\ee
where we have fixed an overall spinor parity in $O(10,2)$.
The $\tilde c\cJ$ term imposes the constraint
\be
\label{eq:J-constraint}
l_1\,\varepsilon_\sl + l_2\,\varepsilon_\su = 0 ~.
\ee
Finally, the fermionic null constraint $\tilde\gamma\lamb$ involves
\begin{align}
\label{supernull current}
\lamb &= 
l_1 \psi^3_\sl+ l_2\psi^3_\su + l_3\psi^t +l_4 \psi^y 
\\[.2cm]
&= 
\frac{e^{i\sqrt{2} H_3}}{\sqrt{2}}\big(l_1+l_2\big) + \frac{e^{-i\sqrt{2} H_3}}{\sqrt{2}}\big(-l_1+l_2\big) +
\frac{e^{i\sqrt{2} H_{ty}}}{\sqrt{2}}\big(l_3+l_4\big) + \frac{e^{-i\sqrt{2} H_{ty}}}{\sqrt{2}}\big(-l_3+l_4\big) \,.
\nn
\end{align}
Recall that for Coulomb branch configurations, $l_1=l_2$ while $l_3=l_4=0$ (see Eqs.~\eqref{eq:kv-1},~\eqref{eq:CB-params}). For supertube configurations, $l_1=l_2$ while $l_3=-l_4\neq 0$ (see Eq.~\eqref{eq:circ-sup-params}). In general, this constraint imposes the restriction that the spinor polarization be transverse to the null vector; acting on the supercharge~\eqref{susy charge}, the result is that $l_1=l_2\neq 0$ implies $\varepsilon_3=+1$ always.  For supertubes, additionally $l_3=-l_4\neq 0$ implies $\varepsilon_{ty}=-1$.  

For the spectrally flowed three-charge 1/8-BPS supertubes~\cite{Giusto:2004id,Giusto:2004ip,Giusto:2012yz} and non-BPS JMaRT backgrounds~\cite{Jejjala:2005yu} described in our formalism in~\cite{Martinec:2017ztd,Martinec:2018nco}, the null constraint coefficients have $|l_1|\ne |l_2|$ and $|l_3|\ne|l_4|$, and there are no solutions to the fermionic null constraint for candidate supercharges, and thus no spacetime supersymmetry coming from this worldsheet chirality. For the 1/8-BPS spectrally flowed supertubes, there remain four right-moving BRST-invariant spacetime supercharges, while for the JMaRT backgrounds, one additionally has $|r_1|\ne |r_2|$ and $|r_3|\ne|r_4|$, and all spacetime supersymmetry is broken.

Returning to the Coulomb branch and supertube configurations that are our main focus, one finds that the supercharge $Q_\beta$ lies in the BRST cohomology for
\be
\label{spinor charges}
\big(q_\varphi,q_{\tilde\varphi};\varepsilon_\sl,\varepsilon_\su,\varepsilon_3 \big) =  \big( -\hf,+\hf; \:\! \varepsilon,\;\!-\varepsilon,+1 \big) 
~~,~~~~
\varepsilon_{ty}\, \varepsilon_{67}\, \varepsilon_{89} = -1 
~~,~~~~
\varepsilon =\pm1 ~,
\ee
with in addition $\varepsilon_{ty}=-1$ for supertubes.
Together with the analogous right-moving counterparts one has (8,8) supersymmetries for the Coulomb branch configurations and (4,4) spacetime supersymmetry for supertubes, as expected.
As usual, picture changing~\cite{Friedan:1985ge} yields versions of these supersymmetry charges in other superselection sectors of the ghost numbers measured by the currents $\beta\gamma$ and $\tilde\beta\tilde\gamma$.
The end result is a perturbative string S-matrix that satisfies the constraints of spacetime supersymmetry.

One can then check that the operators~\eqref{BPSverts} form a chiral supermultiplet under this supersymmetry.  
This fact follows from the general structure that relates worldsheet $\cN=2$ supersymmetry to spacetime supersymmetry~\cite{Banks:1987cy}, and the fact that the $\sltwo\times\sutwo$ WZW model admits an $\cN=2$ structure~\cite{Rastelli:2005ph,Spindel:1988sr} under which the operators~\eqref{BPSverts} are worldsheet $\cN=2$ chiral superfields with integer $\cR$ charge.  The latter property guarantees that they belong to BPS supermultiplets in spacetime.  For supertubes, the spinor charge assignments of the holomorphic component of these operators, together with the supersymmetry operator, are shown in Table~2.
%
%
\begin{equation}
\arraycolsep=1.4pt\def\arraystretch{1.1}
~\begin{array}{|c||c|c||c|c|c|c|c|c|}
\hline
 ~ & ~q_\varphi ~& ~q_{\tilde\varphi}~ &  ~\varepsilon_\sl ~&~ \varepsilon_\su ~&~ \varepsilon_3 ~&~ \varepsilon_{ty} ~&~ \varepsilon_{67} ~&~ \varepsilon_{89} ~\\[2pt]
\hline
\hline
~\cV_j^+ ~&-1 & 0 & -2 & 0 & 0 & 0 & 0 & 0  \\[2pt]
\hline
\cS_j^A &-\hf & -\hf  & -1 & - 1 & -1 & +1 & -A  & -A  \\[2pt]
\hline
\cV_j^- &-1 & 0 & 0 & -2 & 0 & 0 & 0 & 0  \\[2pt]
\hline
\hline
Q_A 			&-\hf & +\hf  & -1 & +1 & +1 & -1  &  A &  A \\[2pt]
\hline
\widetilde Q_A 	&-\hf & +\hf  & +1 & -1 & +1 & -1 &  A &  A \\[2pt]
\hline
\end{array}
\nn\\
\end{equation}
\nopagebreak
\begin{center}
{{Table 2. }\textit{Charges of left-moving massless vertex operators and supercharges; $A=\pm1$.}}
\end{center}
%

\vskip 1cm


\section{Determination of $\H,\A,\K$ for the elliptical supertube}
\label{app:ellpert}

In this appendix we sketch a derivation of the harmonic functions $\H,\A,\K$ that appear in the geometry~\eqref{LMgeom} for a fivebrane source distributed along an ellipse.  The strategy will be to carry out a perturbation expansion in the deformation away from a circular source, and use the coefficients in that expansion to pin down the exact expression in terms of the functions $\bA_m,\bB_m,f$ in~\eqref{AmBm}.
We have the source located at
\be
\F(v) = \bigl( a\sqrt{1+\epsilon}\,\cos v, a\sqrt{1-\epsilon}\,\sin v,\,0,\,0\bigr) ~,
\ee
and in elliptical bipolar coordinates~\eqref{ellbip} the denominator in the Green's function~\eqref{greensfn} evaluates to
\begin{align}
|\xx-\F(v)|^2 &= r^2+a^2 +a^2\sin^2\theta + \epsilon \,a^2\bigl[\hf(z^2\tight+z^{-2})+\sin^2\theta\cos 2\phi\bigr]
- w z - \bar w z^{-1}
\end{align}
where $z=e^{iv}$, and
\begin{align}
w &= \sqrt{a^2(1+\epsilon)}\sqrt{r^2+a^2(1+\epsilon)}\,\sin\theta \, \cos\phi
\nn\\
&\hskip 1cm -i \sqrt{a^2(1-\epsilon)}\sqrt{r^2+a^2(1-\epsilon)}\,\sin\theta \, \sin\phi ~.
\end{align}
Expanding in $\epsilon$, at leading order one has
\be
|\xx-\F(v)|^2 =  \Bigl({a\sin\theta}\,z- {\sqrt{r^2+a^2}\,e^{i\phi}}\Bigr)\Bigl({a\sin\theta}\,z^{-1} - {\sqrt{r^2+a^2}\,e^{-i\phi}}\Bigr) + O(\epsilon)~.
\ee
In the expansion of the Green's function, one then has a series of terms of the form
\be
\oint \frac{dz}{z} \frac{P_n(z)}{\bigl({a\sin\theta}\,z- {\sqrt{r^2\tight+a^2}\,e^{i\phi}}\bigr)^{n+1}\bigl({a\sin\theta}\,z^{-1} - {\sqrt{r^2\tight+a^2}\,e^{-i\phi}}\bigr)^{n+1}}
\ee
where $P_n(z)$ is a Laurent polynomial with powers of $z$ up to $\pm2n$, $\pm(2n+1)$, or $\pm(2n+2)$ in the expansion of $\H$, $\A$, or $\K$ respectively.  Evaluating the integrals via residues, at each order in $\epsilon$ one finds a series of terms involving various inverse powers of the unperturbed harmonic function denominator $\Sigma_0=r^2+a^2\cos^2\theta$.  Powers of $\Sigma_0$ beyond the leading order must come from the $\epsilon$ expansion of the denominator, and one can resum to any given order to arrive at an expression with only a single pole at a shifted location.  In the case of $\H$, all powers of $\epsilon$ come from this denominator expansion, and one has 
\be
\H = \frac{1}{\Sigma_0 + \epsilon \Sigma_1 + \epsilon^2 \Sigma_2 + \dots} ~,
\ee
while for $\A$ and $\K  $ one expects
\be
\A = \H \tilde \A ~~,~~~~ \K = \H\tilde \K
\ee
where $\tilde \A$, $\tilde \K$ have Taylor series expansions in $\epsilon$.  The first few orders in the expansion are sufficient to pin down a guess for $\H,\tilde \A,\tilde \K$, namely~\eqref{LMellipse}.  One then checks that this guess yields an exact harmonic $\H,\A,\K$, which because it was arrived at from an evaluation of the Green's function integrals, is guaranteed to have the appropriate source distribution.  In evaluating the various gradients involved, the fact that $\bA_1=1$ in~\eqref{AmBm} implies
\be
\partial_i r^2 = \frac{2x^i}{(r^2+a_i^2)\bA_2} ~,
\ee
leading to relations among the gradients of $\bA_m,\bB_m,f$:
\begin{align}
\partial_i\bA_m &= \frac{2x^i}{(r^2+a_i^2)^m} - \frac{2mx^i\bA_{m+1} }{(r^2+a_i^2)\bA_2}
\nn\\[.3cm]
\partial_i\bB_m &= - \frac{2mx^i\bB_{m+1} }{(r^2+a_i^2)\bA_2}
\\[.3cm]
\partial_i f &= f\,\frac{2x^i\bB_1}{\bA_2} ~.
\nn
\end{align}
These relations can then be used to systematically evaluate the Laplacians of the candidate expressions for $\H,\A,\K$ in terms of these functions, and verify that they vanish away from the source.

Note that the series expansion of $\H$ in powers of $\epsilon$ (and similarly for $\K$ \etc.) leads to expressions which are ever more singular at higher orders, diverging at the supertube as $\Sigma_0^{-n}$.  While na\"ively one might take this as an indication that focussing effects may be leading to a gravitational singularity, that would be the wrong lesson to extract.  Instead, rather than summing up to something more singular than the original supertube, we see that the terms sum up nicely to a smooth perturbed profile for the supertube.



\newpage
\vskip 1cm

\bibliographystyle{JHEP}      

\bibliography{microstates}


\end{document}